\documentclass{svmult}
\usepackage{amssymb}
\usepackage{graphicx,setspace}
\usepackage{amsmath}
\usepackage{epsfig}
\usepackage{amssymb}
\usepackage{gensymb}

\usepackage{mathptmx}
\usepackage{helvet}
\usepackage{courier}
\usepackage{makeidx}
\usepackage{multicol}
\usepackage{footmisc}

\newcommand{\eqname}[1]{\label{eq:#1}}
\newcommand{\bgar}{\begin{eqnarray}}
\newcommand{\enar}[1]{\label{eq:#1}\end{eqnarray}}

\newcommand{\kk}{ {\bf k}}
\newcommand{\vv}{ {\bf v}}

\newcommand{\rr}{ {\bf r}}
\newcommand{\qq}{ {\bf q}}

\newcommand{\nh}{ {\hat {\bf n}}}

\newcommand{\eq}[1]{(\ref{eq:#1})}

\newcommand{\phit}{\tilde\phi}

\begin{document}
\title*{The \u Cerenkov effect revisited: from swimming ducks to zero modes in gravitational analogs}
\titlerunning{The \u Cerenkov effect revisited: from swimming ducks to zero modes} 
\author{Iacopo Carusotto \and Germain Rousseaux}
\institute{Iacopo Carusotto \at INO-CNR BEC Center and Dipartimento di Fisica, Universit\`a di Trento, I-38123 Povo, Italy, European Union. \email{carusott@science.unitn.it} \and
Germain Rousseaux \at Universit\'{e} de Nice-Sophia Antipolis, Laboratoire J.-A. Dieudonn\'{e}, UMR CNRS-UNS 6621, Parc Valrose, 06108 Nice Cedex 02, France, European Union. \email{Germain.Rousseaux@unice.fr} }


\maketitle

\abstract{
We present an interdisciplinary review of the generalized \u Cerenkov emission of radiation from uniformly moving sources in the different contexts of classical electromagnetism, superfluid hydrodynamics, and classical hydrodynamics. The details of each specific physical systems enter our theory via the dispersion law of the excitations. A geometrical recipe to obtain the emission patterns in both real and wavevector space from the geometrical shape of the dispersion law is discussed and applied to a number of cases of current experimental interest. Some consequences of these emission processes onto the stability of condensed-matter analogs of gravitational systems are finally illustrated.
}


\section{Introduction}

The emission of radiation by a uniformly moving source is a widely used paradigm in field theories to describe a number of very different effects, from the wake generated by a swimming duck on the surface of a quiet lake~\cite{whitham,lighthill,Mei,Darrigol,Sorensen,Yih,Torsvik,Soomere,Raphael_deGennes,Chevy}, to the \u Cerenkov emission by a charged particle relativistically moving through a dielectric medium~\cite{Jelley}, to the sound waves emitted by an object traveling across a fluid or a superfluid at super-sonic speed~\cite{IC_CC_Superfl,AmoBramati_NPhys,Cornell_exp,Carusotto:PRL2006}. 
Of course, the radiated field has a different physical nature in each case, consisting e.g. of gravity or capillary waves at the water/air interface, or electromagnetic waves in a dielectric medium, or Bogoliubov excitations in the superfluid. In spite of this, the basic qualitative features of the emission process are very similar in all cases and a unitary discussion is possible.

In the present chapter, we shall present an interdisciplinary review of this {\em generalized \u Cerenkov effect} from the various points of view of classical electromagnetism, superfluid hydrodynamics, and classical hydrodynamics. The details of each specific physical systems enter our theory via the dispersion law $\Omega(\kk)$ of its excitations. In particular, the emission patterns in both real and wavevector space can be extracted from the geometrical shape of the intersection of the $\Omega(\kk)$ dispersion law with the $\Omega=\kk\cdot \vv$ hyper-plane that encodes energy-momentum conservation. Once the dispersion law of a generic system is known, our geometrical algorithm provides an efficient tool to obtain the most significant qualitative features of wake pattern in a straightforward and physically transparent way.

In the last years, the interest of the scientific community on this classical problem of wave theory has been revived by several experiments which have started exploring the peculiar features that appear in new configurations made accessible by the last technological developments, e.g. the \u Cerenkov emission of electromagnetic radiation in resonant media~\cite{resonant_dispersive,dipolar_nonlopt_cherenk} and the Bogoliubov-\u Cerenkov emission of sound waves in bulk dilute superfluids~\cite{AmoBramati_NPhys,Cornell_exp,Carusotto:PRL2006}. 
Another reason for this renewed interest comes from the condensed-matter models of gravitational systems that are the central subject of the present book. In many of such analog models, the presence of a horizon may be responsible for the  emission of waves from the horizon by generalized \u Cerenkov processes. A full understanding of these classical effects is then required if one is to isolate quantum features such as the analogs of Hawking radiation, dynamical Casimir emission and anomalous Doppler effect. 

\vspace{0.5cm}

The structure of the chapter is the following. 
In Sec.\ref{sec:theory}, we shall introduce the general field-theoretical formalism to calculate the real and momentum space emission patterns and the geometrical construction to obtain qualitative information on them. These methods will then be applied in the following sections to a few different systems of current interest. 
As a first example, in Sec.\ref{sec:Cerenkov} we will review the main features of the \u Cerenkov emission  of electromagnetic waves from relativistically moving charges in a dielectric medium. We shall restrict our attention to the simplest case of a non-dispersive dielectric with frequency-independent refractive index $n$, where the phase and group velocities are equal and constant. In this case, a \u Cerenkov emission takes place as soon as the charge speed exceeds the velocity of light in the medium $c=c_0/n$. Modern developments for the case of a strongly dispersive media~\cite{resonant_dispersive,dipolar_nonlopt_cherenk,noi_cerenkov}, photonic crystals, and left-handed metamaterials~\cite{Cerenk_PC} will be briefly mentioned.
In Sec.\ref{sec:superfluid} we shall review the emission of sound waves by a supersonically moving impurity in the bulk of a dilute superfluid. In addition to the Mach cone that appears in the wake behind the object, the presence of single-particle excitations in the excitation spectrum is responsible for the appearance of a series of curved wavefronts ahead of the impurity. On the other hand, a subsonically moving impurity will produce no propagating wave and the perturbation will remain localized in its vicinity: the resulting frictionless motion is one of the clearest examples of the class of phenomena that go under the name of superfluidity~\cite{LPSS_book,PinesNozieres,Leggett}.
This physics is currently of high experimental relevance, as first real-space images of the density perturbation pattern induced by a moving impurity have been recently obtained using Bose-Einstein condensates of ultracold atoms~\cite{Cornell_exp} and of exciton-polaritons in semiconductor microcavities~\cite{AmoBramati_NPhys}.
The case of a parabolic dispersion will be presented in Sec.\ref{sec:parabolic}: this specific functional form allows for an elementary analytical treatment of the wake pattern in both real and momentum space. On one hand, this discussion provides an useful guideline to understand the qualitative shape of the wake in superfluids and in surface waves. On the other hand, it is of central importance in view of the experimental realization of gravitational analogs based on magnon excitations in magnetic solids~\cite{magnons}.
The physics of a material object such as a boat, a duck or a fishing line creating surface waves on the air/water interface of a lake will be considered in Sec.\ref{sec:surface}: not only does this example provide the most intuitive example of the generalized \u Cerenkov effect, but is perhaps also the richest one in terms of different behaviors that can be observed depending on the system parameters, e.g. the velocity of the object with respect to the fluid, the depth of the water, the surface tension of the fluid~\cite{whitham,lighthill,Mei,Darrigol,Sorensen,Yih,Torsvik,Soomere,Raphael_deGennes,Chevy}. 
The concepts that have been laid down so far are finally applied in Sec.\ref{sec:gravity} to analog models of gravity based on flowing superfluids or surface waves on flowing water. In both these cases, classical \u Cerenkov emission into the so-called {\em zero modes} at the horizon may disturb detection of the analog Hawking radiation as well as affect the dynamical stability of the analog black/white hole~\cite{noiWH}.  Conclusions are finally drawn in Sec.\ref{sec:conclu}.

\section{Generic model}
\label{sec:theory}

In this section, we introduce the generic model that will be used to study the different physical systems in the following sections. The model is based on a linear partial differential equation for a scalar $\mathbb{C}$-number field $\phi(\rr,t)$: in most relevant cases, the multi-component physical field (i.e. the vector electromagnetic field or the Bogoliubov spinor) can in fact be reduced to a single scalar field upon straightforward algebraic manipulations under controlled approximations. We are also assuming that quantum fluctuations of the field $\phi(\rr,t)$ can be fully neglected. The geometry under investigations consists of a spatially homogeneous system interacting with a spatially localized moving source describing the moving electric charge, or the interaction potential of the moving impurity with the fluid, or the extra pressure exerted on the fluid surface by the moving object. In this geometry, the microscopic information on the field dynamics is summarized in the dispersion law relating the frequency $\Omega$ of a plane wave to its wavevector $\kk$: different forms of dispersion laws corresponding to first- or higher-order partial differential equations are discussed in the subsections Secs.\ref{sec:first} and \ref{sec:PDEgeneral}.
The geometric construction of the wake pattern starting from the dispersion law $\Omega(\kk)$ is discussed in Sec.\ref{sec:wake}.

\subsection{The wave equation and the source term}
\label{sec:first}

We start by considering a generic, $d$-dimensional classical complex field $\phi(\rr,t)$ ($d=2$ in the figures) that evolves according to the generic linear, first-order in time, partial differential equation:
\begin{equation}
\eqname{wave_eq}
i\partial_t \phi(\rr,t) =\Omega(-i\nabla_\rr)\phi(\rr,t)+S(\rr,t)
\end{equation}
with a source term $S(\rr,t)$. 

The function $\Omega(\kk)$ defines the so-called dispersion law for free field propagation, that is the frequency of the plane wave solutions
\begin{equation}
\phi(\rr,t)=\phi_0\,e^{i\kk\rr}\,e^{-i\Omega(\kk)\, t}.
\end{equation}
as a function of the wavevector $\kk$ in the absence of sources, $S(\rr,t)=0$.

Throughout this chapter we shall consider a spatially localized and uniformly moving source term of the form
\begin{equation}
S(\rr,t)=S_0(\rr-\vv\,t),
\eqname{source}
\end{equation}
with a spatial profile $S_0(\rr)$ concentrated in the vicinity of $\rr=0$ and moving at a speed $\vv$.

Thanks to the translational invariance of the free field problem in both space and time, solution of the full wave equation \eq{wave_eq} in the presence of the source term is easily obtained in Fourier space with respect to both space and time. 
Defining the Fourier transform in the usual way
\begin{equation}
\phit(\kk,\omega)=\int\!dt\int\!d^d\rr \,\phi(\rr,t)\,e^{-i\kk\cdot\rr}\,e^{i\omega t},
\eqname{phi_k}
\end{equation}
the source term in Fourier space has the simple form
\begin{equation}
\tilde{S}(\kk,\omega)=2\pi \tilde{S}_0(\kk)\,\delta(\omega-\kk\cdot \vv)
\end{equation}
in terms of the structure factor $\tilde{S}_0(\kk)$ defined as the Fourier transform of the source shape $S_0(\rr)$.

In Fourier space, the solution of \eq{wave_eq} is then
\begin{equation}
\phit(\kk,\omega)=\frac{2\pi S_0(\kk)\, \delta(\omega-\kk\cdot \vv)}{\omega-\Omega(\kk)+i\,0^+},
\eqname{solut_wave}
\end{equation}
where an infinitesimally small imaginary part is introduced in the denominator of \eq{solut_wave}  to specify the integration contour to be followed around the poles and, in this way, ensure causality of the solution. This trick dates back to Rayleigh~\cite{Darrigol} and is equivalent to a infinitesimal shift of the dispersion law into the lower half-space, $\Omega(\kk)\rightarrow \Omega(\kk)-i0^+$. Physically, it corresponds to introducing a very weak damping of the plane wave solutions in time, 
\begin{equation}
\phi(\rr,t)=\phi_0\,e^{i\kk\rr}\,e^{-i\Omega(\kk)\, t}\,e^{-0^+ t}.
\end{equation}
or to assume that the source term is slowly switched on in time~\cite{Raphael_deGennes}.

The real-space pattern is obtained by an inverse Fourier transform of \eq{solut_wave},
\begin{equation}
\phi(\rr,t)=
-\int\!\frac{d^d\kk}{(2\pi)^d}\,\frac{\tilde{S}_0(\kk)\,e^{i\kk(\rr-\vv\,t)}}{\Omega(\kk)-\kk \cdot \vv -i\,0^+}=\phi(\rr-\vv\,t).
\eqname{field_real}
\end{equation}
Thanks to the $\delta(\omega -\kk\cdot \vv)$ factor in \eq{solut_wave}, this expression only depends on the combination $\rr'=\rr-\vv\,t$: as expected, the wake pattern is rigidly moving at the speed of the source. Within Galilean invariance, the $\rr'$ coordinate corresponds to the spatial coordinate in the reference frame of the source in motion at velocity $\vv$.

Evaluation of \eq{field_real} can be performed with standard numerical tools. The result for some most interesting cases will be presented in the next sections. Now, we shall rather proceed with some analytical manipulations of \eq{field_real} that allow to extract qualitative information on the emitted field pattern from the dispersion law $\Omega(\kk)$. The first step in this direction is to note that the integral in  \eq{field_real} is dominated by those $\kk$ values for which the resonant denominator vanishes, that is 
\begin{equation}
\Omega(\kk)=\kk\cdot \vv.
\eqname{cerenkov}
\end{equation}
This equation recovers the standard \u Cerenkov condition for emission of radiation~\cite{Jelley} and can be geometrically interpreted as the intersection of the dispersion surface $\Omega(\kk)$ with the $\Omega=\kk\cdot \vv$ plane. In a quantum description of the \u Cerenkov emission by a massive charged particle, the condition \eq{cerenkov} naturally appears when energy-momentum conservation is imposed to the photon emission process~\cite{Jelley}. When reformulated in the reference frame of the moving source, the condition \eq{cerenkov} reduces to $\Omega'=\gamma(\Omega-\kk\cdot \vv)=0$, meaning that the perturbation pattern around the source at rest is stationary in time in the moving reference frame.

The locus $\Sigma$ of $\kk \neq 0$ modes that satisfy \eq{cerenkov} is a central object in all the following discussion as it defines the modes in $\kk$ space into which the \u Cerenkov emission will be peaked~\footnote{The $\kk=0$ mode corresponds to a spatially constant modulation that does not transport energy nor momentum. As discussed in~\cite{Lock}, many other interesting features of wave propagation can be graphically studied starting from isofrequency surfaces analogous to the locus $\Sigma$.}. In particular, no emission of propagating waves takes place if the locus $\Sigma$ is empty; the non-resonant contributions to \eq{field_real} only provide a non-radiative perturbation that remains spatially localized in the close vicinity of the source and is not able to transport energy away. In spite of this, the momentum and energy that are stored in the localized moving pattern of the field $\phi$ are responsible for a sizable renormalization of the particle mass~\cite{Raphael_deGennes,Pomeau_ell,astra}.

\subsection{Qualitative geometrical study of the wake pattern}
\label{sec:wake}

\begin{figure}[!htbp]
\begin{center}
\includegraphics[width=0.5\columnwidth]{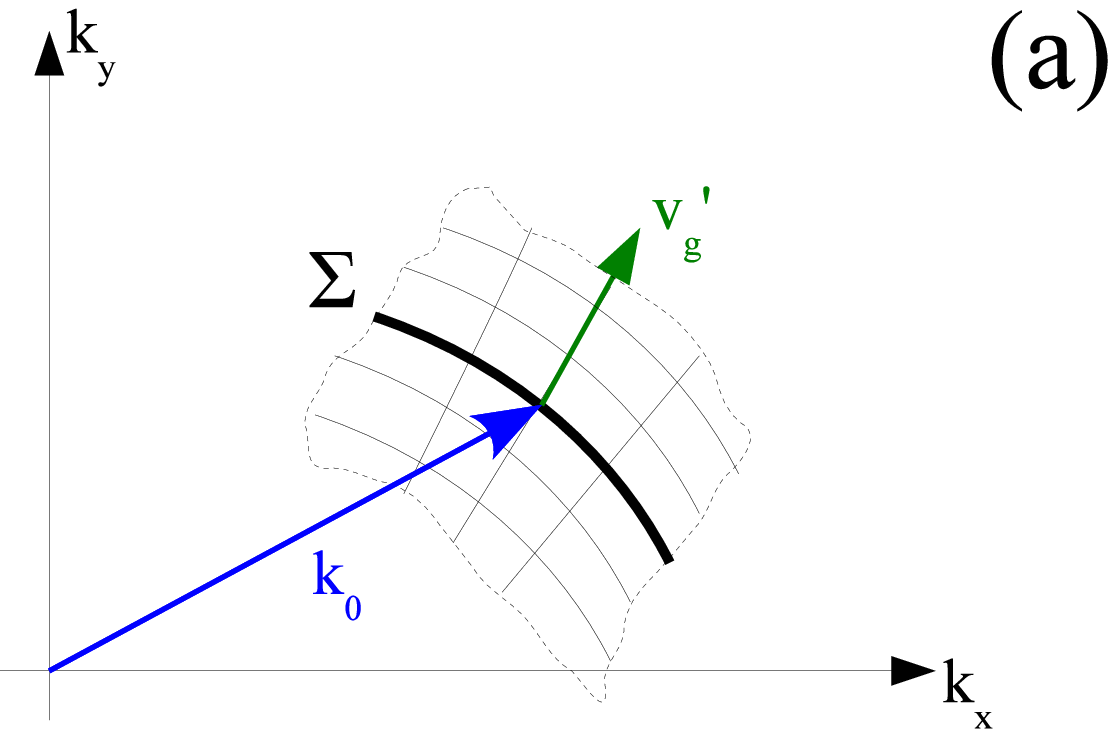}
\hspace{1cm}
\includegraphics[height=0.35\columnwidth]{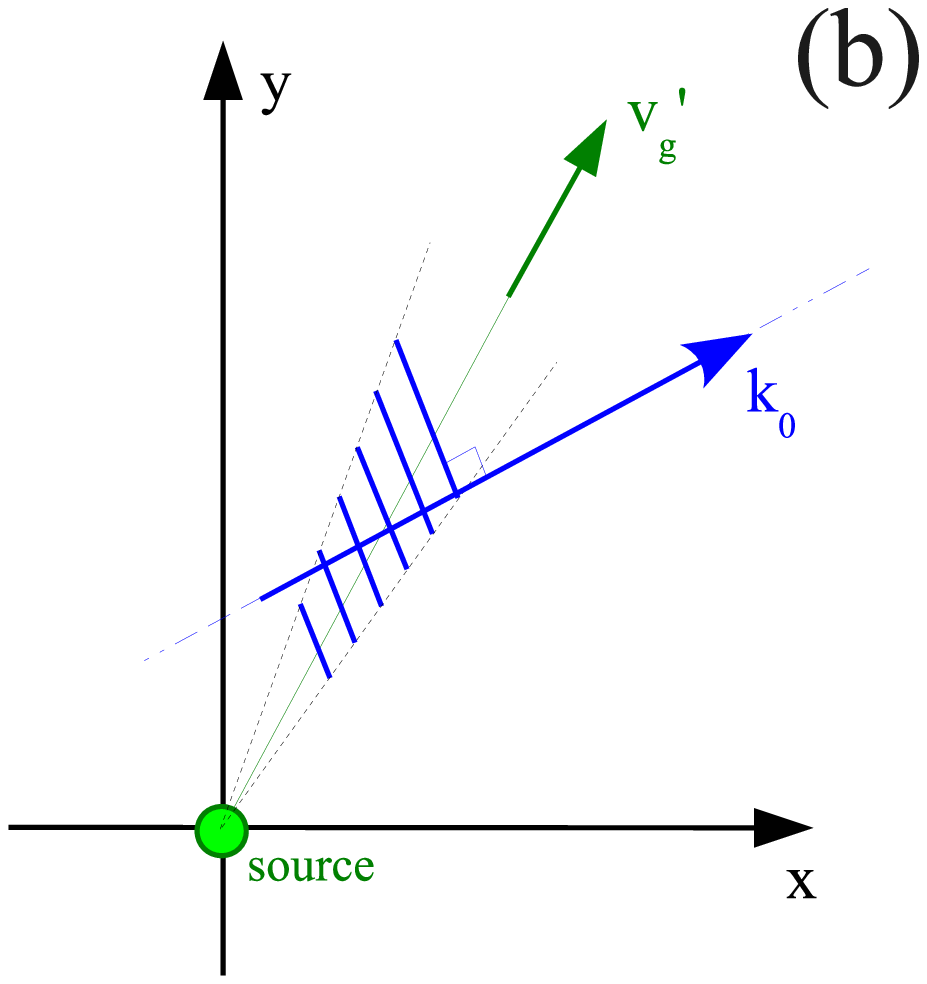}
\end{center}
\caption{Left (a) panel: $\kk$-space sketch of a patch of the locus $\Sigma$ around the wavevector $\kk_0$. The grid shows the $(\qq_\parallel,q_n)$ coordinate system used in the geometrical construction of the wake pattern. Right (b) panel: sketch of the region of the wake pattern generated by the emission in the neighborhood of $\kk_0$: the blue fringes have wavevector $\kk_0$, the direction of propagation $\vv'_g$ is determined by the normal to the locus $\Sigma$ at the point $\kk_0$.}
\label{fig:coord}
\end{figure}

Let us consider a generic point $\kk_0\in\Sigma$. Within a neighborhood of $\kk_0$, we introduce a new set of $\kk$-space coordinates defined as follows: for each point $\qq$, $q_n$ is the distance of $\kk$ from the $\Sigma$ surface and the position of the closest point on the surface $\Sigma$ is parametrized by the $(d-1)$-dimensional $\qq_\parallel$ curvilinear coordinate system. A sketch of this coordinate system is indicated as a grid in Fig.\ref{fig:coord}(a).

In this new coordinate system, the Fourier integral giving the emitted field pattern in real space can be approximately rewritten as:
\begin{equation}
\phi(\rr')=-\tilde{S_0}(\kk_0) \int_\Sigma\!\frac{d^{d-1}\qq_\parallel}{(2\pi)^{d-1}}\,e^{i\bar{\kk}(\qq_\parallel)\cdot\rr'} \int\!\frac{dq_n}{2\pi}\,\frac{e^{iq_n\nh\cdot\rr'}}{v'_g\,q_n-i\,0^+},
\eqname{field_real_approx}
\end{equation}
where $\bar{\kk}(\qq_\parallel)$ is the position of the point on the surface $\Sigma$ corresponding to the value $\qq_\parallel$ of the $(d-1)$-dimensional coordinate and $\nh$ is the unit vector normal to the surface at $\kk_0$ in the direction of growing $\Omega(\kk)-\kk\cdot \vv$. 
As the surface $\Sigma$ is defined by the zeros of $\Omega(\kk)-\kk\cdot \vv$, the velocity 
\begin{equation}
\vv'_g=v'_g\,\nh= \nabla_\kk\big[\Omega(\kk)-\kk\cdot\vv\big]=\vv_g-\vv
\end{equation}
is directed along the normal $\nh$ and corresponds to the group velocity of the wave, as measured relative to the moving source at $\vv$. For a non-relativistic source speed $v\ll c_0$, it can be interpreted as the group velocity observed from the source reference frame.

The integral over $q_n$ can be performed by closing the contour on the complex plane. The only pole is located slightly above the real axis. Depending on the sign of $\nh\cdot \rr'$, the contour has to be closed in the upper or lower half plane, which gives
\begin{equation}
\phi(\rr')=-i S_0(\kk_0) \int_\Sigma\!\frac{d^{d-1}\qq_\parallel}{(2\pi)^{d-1}}\,\frac{e^{i\bar{\kk}(\qq_\parallel)\cdot\rr'}}{v'_g}\Theta[\vv'_g\cdot \rr'].
\eqname{field_real_approx_2}
\end{equation}

The expression \eq{field_real_approx_2} can be further simplified by performing the so-called stationary phase approximation, as first proposed by Thomson~\cite{Darrigol}. For each value $\rr'$ of the relative coordinate, the integral over $\qq_\parallel$ is dominated by those points for which the phase is stationary, i.e. the variation of $\bar{\kk}(\qq_\parallel)$ on $\qq_\parallel$ is orthogonal to $\rr'$. In combination with the Heaviside-$\Theta$ function in \eq{field_real_approx_2}, this is equivalent to requiring that the vector $\rr'$ is parallel to the normal $\nh$ to the surface $\Sigma$ at point $\kk_0$ in the direction of growing $\Omega(\kk)-\kk\cdot \vv$ , i.e. parallel to the relative group velocity $\vv'_g$.

For a generic relative position $\bar{\rr}'$, there are only a few discrete points $\kk_j$ on $\Sigma$ such that this condition is met. As a consequence, for generic values of $\rr'$ in a neighborhood of $\bar{\rr}'$, one can approximately write
\begin{equation}
\phi(\rr')\approx-i\frac{S_0(\kk_j)}{(2\pi)^{d-1}} \sum_j \frac{\Delta k_j^2}{v'_{g,j}}\,e^{i\kk_j\cdot\rr'},
\eqname{field_real_approx_fin}
\end{equation}
where the sum is over the allowed $\kk_j$ vectors: the numerical coefficient $\Delta k_j^2$ is inversely proportional to the curvature of $\Sigma$ at $\kk_j$ and $\vv'_{g,j}$ is the group velocity of the $\kk_j$ mode. 

A physical understanding of this result can be easily obtained by looking at the diagram of Fig.\ref{fig:coord}(b). Within Galilean invariance, sitting in the moving reference frame of the source may facilitate building an intuitive picture of the emission process: every point on the surface $\Sigma$ corresponds to a continuous plane wave of wavevector $\kk_0$ that is emitted from the source and propagates away from it at a group velocity $\vv'_g$ (indicated by the green arrow in the figure). As a result, it is able to reach all points $\rr'$ that lie in the vicinity of the straight line of direction $\vv'_g$. While the group velocity $\vv'_g$ is always along the radial direction, the wavevector $\kk_0$ (blue arrow in the figure) can have arbitrary direction: as a result, the wavefronts (indicated by the blue fringes) are generally tilted and the emission pattern does not necessarily resemble to a spherical wave. Of course, all this reasoning can be performed equally well in the laboratory frame if $\vv'_g$ is interpreted as the relative group velocity of the wave with respect to the moving source. 

\subsection{Generalization to higher-order wave equations}
\label{sec:PDEgeneral}

The geometrical framework introduced in the previous subsections is not restricted to partial differential equations that are of first-order in time, but can be extended to more general wave equations of the form
\begin{equation}
P[i\partial_t,-i\nabla_\rr]\phi(\rr,t)=S(\rr,t),
\eqname{higher_wave}
\end{equation}
where $P$ is an arbitrary polynomial in two variables, a scalar variable and a $d$-component vectorial variable. The degree of the polynomial $P$ corresponds to the order of the partial differential equation for $\phi(\rr,t)$: in the case of electromagnetic waves in a non-dispersive medium, it is of second order in both variables; in the case of Bogoliubov excitations in a superfluid, it is of second order in time and of fourth order in space; in the case of surface waves, it is of second order in time, but it involves arbitrarily high derivatives in the spatial coordinates.
The different branches $\Omega(\kk)$ of the dispersion law are then defined by the roots of $P$ via the equation
\begin{equation}
P[\Omega(\kk),\kk]=0.
\eqname{dispersion_2}
\end{equation}

In the presence of a source term of the form \eq{source}, the solution of \eq{higher_wave} has the form
\begin{equation}
\phit(\kk,\omega)=\frac{2\pi\,\tilde{S}_0(\kk)}{P(\kk\cdot \vv+i0^+,\kk)},
\eqname{field_fourier2}
\end{equation}
where the infinitesimally small imaginary part has been again added in order to enforce causality by shifting the real roots $\Omega$ of the dispersion law \eq{dispersion_2} into the lower half of the complex-plane.

The reasoning to extract from \eq{field_fourier2} the qualitative shape of the real-space pattern is then the same as before, the locus $\Sigma$ in $\kk$-space being now defined by the zeros of the polynomial equation
\begin{equation}
P(\kk\cdot \vv,\kk)=0
\end{equation}
with $\kk\neq 0$.
In the next sections, we shall discuss in full detail a few physical examples illustrating how the geometrical structure of $\Sigma$ determines the shape of the emission pattern in both real and momentum spaces.

\section{\u Cerenkov emission by uniformly moving charges}
\label{sec:Cerenkov}

As a first application of the theory, in this section we shall review the basic features of the emission of electromagnetic radiation by a charged particle relativistically moving through a dielectric medium at speed higher than the phase velocity of light in the medium. This is the well-known \u Cerenkov effect (or, more precisely, Vavilov-\u Cerenkov effect) first observed by Marie Curie, then experimentally characterized by Vavilov and \u Cerenkov~\cite{cerenkov_basics} and finally theoretically understood by Frank and Tamm~\cite{cerenkov_basics_th}.

\subsection{Non-dispersive dielectric}
\label{sec:non-disp}

\begin{figure}[!htbp]
\begin{center}
\includegraphics[width=0.49\columnwidth]{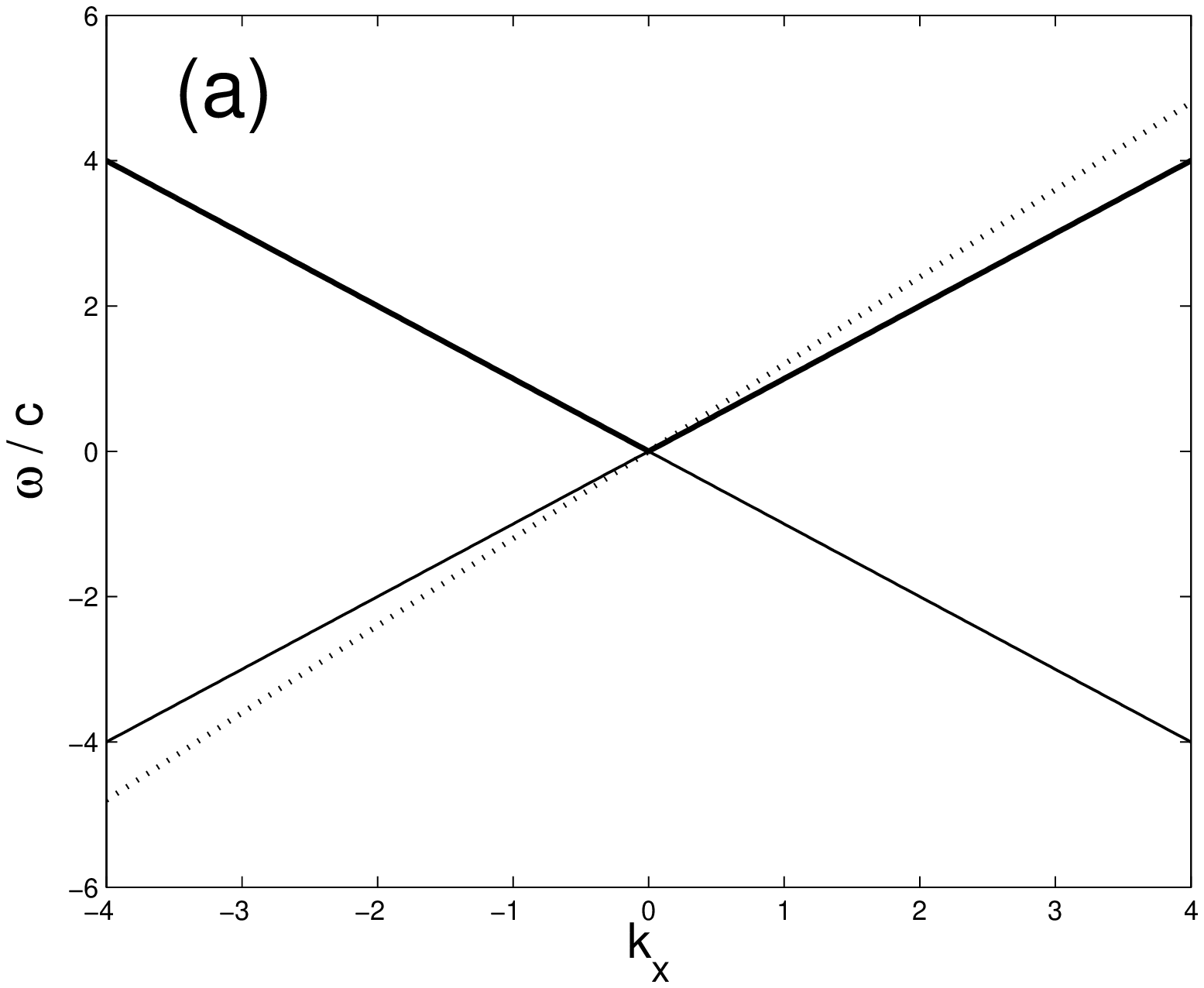} \\
\includegraphics[width=0.49\columnwidth]{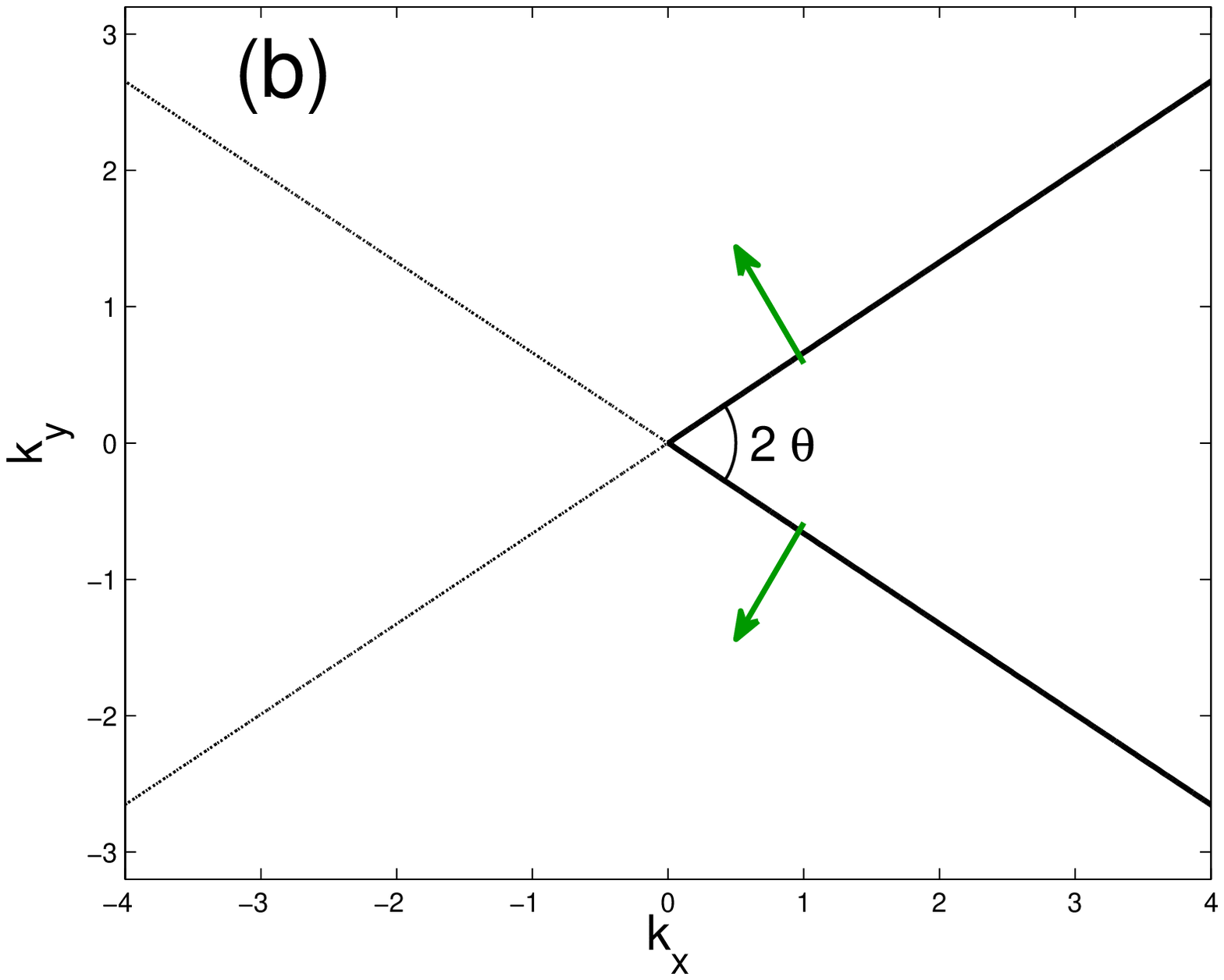} \\
\includegraphics[width=0.49\columnwidth]{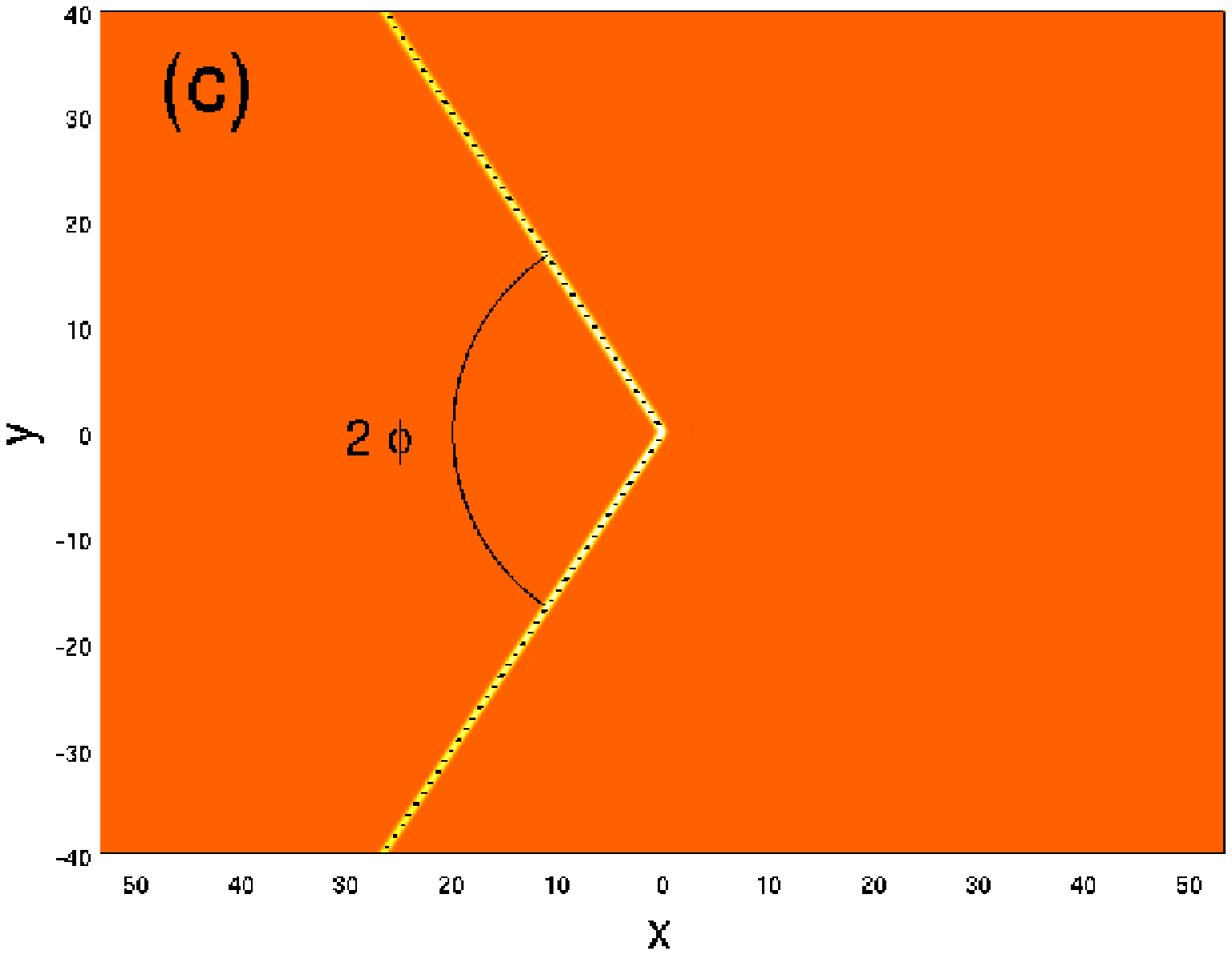}
\end{center}
\caption{Panel (a): cut along $k_y=0$ of the photon dispersion in a non-dispersive medium of frequency-independent refractive index $n$. The dashed line indicates the $\Omega=\kk\cdot \vv$ plane for a super-luminal charge speed $v>c$. Panel (b): $\kk$-space locus $\Sigma$ of resonant modes into which the \u Cerenkov emission occurs, the so-called \u Cerenkov cone; the green arrows indicate the normal to the  $\Sigma$ locus, that is the direction of the relative group velocity $\vv_g'=\nabla_\kk\big[\Omega(\kk)-\kk\cdot\vv\big]$. Panel (c): real-space pattern of the electric field amplitude in the wake of the charge; the pattern is numerically obtained as the fast Fourier transform of the $\kk$-space perturbation \eq{field_fourier2}. The Mach cone around the negative $x$ axis is apparent with aperture $\phi$.}
\label{fig:cerenkov}
\end{figure}

In the simplest case of a non-dispersive dielectric with a frequency-independent refractive index $n$, the dispersion law satisfies the second-order equation 
\begin{equation}
\Omega^2=\frac{c_0^2}{n^2}\,k^2:
\end{equation}
in the $(\Omega,\kk)$ space, the dispersion $\Omega(\kk)$ corresponds to a conical surface with vertex in $\Omega=k=0$. A cut of this cone along the $k_y=0$ line is shown in Fig.\ref{fig:cerenkov}(a): the thick lines indicate the positive frequency part of the conical surface, the thin line indicate the negative frequency part. In the absence of dispersion, the phase and group velocities coincide and are equal to $c=c_0/n$.

The shape of the locus $\Sigma$ of $\kk\neq 0$ points satisfying $\Omega(\kk)-\kk\cdot\vv=0$ crucially depends on whether the charge is moving at a sub-luminal $v<c$ or super-luminal $v>c$ speed. In the former case, the locus $\Sigma$ is empty and no radiation is emitted. The localized, non-radiative perturbation that is present around the charge due to the non-resonant excitation of the field modes contributes to the (velocity-dependent) Coulomb field around the charge. 

The locus $\Sigma$ in the case of a super-luminally moving charge in the positive $x$ direction is illustrated in Fig.\ref{fig:cerenkov}(b): it has the analytic form
\begin{equation}
k_y^2=k_x^2\,\left(\frac{v^2}{c^2} -1\right)
\eqname{kxky}
\end{equation}
and consists of a pair of half straight lines originating from $\kk=0$ and symmetrically located with respect to the $k_x$ axis at an angle $\theta$ such that $\cos\theta=c/v$. The higher the particle speed $v/c$, the wider the angle $\theta$ made by the direction of the \u Cerenkov emission with the direction of the charge motion. 


The most peculiar feature of the locus $\Sigma$ is that the normal vector to $\Sigma$ indicating the direction of the relative group velocity $\vv'_g=\nabla_\kk[\Omega(\kk)-\kk\cdot\vv]=\vv_g-\vv$ [indicated by the green arrows on Fig.\ref{fig:cerenkov}(b)] is constant for all points $\kk$ lying on each of the two straight lines forming $\Sigma$ and points in the backward direction. This last feature is a straightforward consequence of the fact that the charge velocity is larger than the speed of light $c$ in the medium.  
As a result, all modes on $\Sigma$ propagate (as seen from the charge reference frame) into the same direction and the electromagnetic field radiated by the charge is spatially concentrated around the direction of $\vv'_g$. 
This defines a single-sheet conical surface in real space, i.e. a pair of half straight lines in the two dimensional geometry considered here,
\begin{equation}
y^2= \frac{c^2\,x^2}{{v^2}-c^2}\, \hspace{0.3cm} \textrm{with} \hspace{0.3cm} x<0
\eqname{xy}
\end{equation} 
Its aperture $\phi$ around the negative $x$ axis~\footnote{The coefficients of the analytical form \eq{xy} can be understood from the Fourier transform of a delta function peaked on the conically-shaped locus $\Sigma$ of equation \eq{kxky}.} is determined by the condition $\sin\phi=c/v$: the faster the charge speed, the narrower the cone behind the charge. In the analogy with the conical sonic wake generated by a super-sonically moving bullet in a bulk fluid, we will refer to this real space cone as the {\em Mach cone}.
The very thin shape of the Mach cone results from the interference of the continuum of points on the $\Sigma$ locus. For each $\kk_0\in \Sigma$, the fringes are orthogonal to the Mach cone and have different spacing: the interference is everywhere destructive but for the thin surface of the Mach cone. If the correct form of the structure factor $S_0(\kk)$ is included, the usual $\delta$-shape for the Mach cone is recovered~\cite{Jelley}.

In view of the discussion of the next sections, it is crucial to clearly keep in mind the conceptual distinction between the Mach cone in real space on which the electric field intensity is spatially concentrated and the $\kk$-space {\em \u Cerenkov cone} defining the directions into which the radiation does occur. The former was experimentally detected and characterized in~\cite{dipolar_nonlopt_cherenk,fs_nonlopt_cherenk} by looking at the spatial profile of the electric field wake behind the charge~\footnote{It is interesting to note that in both these experiments the moving charge responsible for the \u Cerenkov emission did not consist of a charged physical particle travelling through the medium, but rather consisted of a moving bullet of nonlinear optical polarization generated by a femtosecond optical pulse via the so-called inverse electro-optic effect.}.
The latter is observed in any standard \u Cerenkov radiation experiment measuring the far-field angular distribution of the radiation, that turns out to be concentrated in the forward direction on a conical surface making a \u Cerenkov radiation angle $\theta$ with the charge velocity.

The conceptual distinction between the \u Cerenkov and the Mach cones is related to the distinction between the so-called {\em phase} and {\em group cones}, first pointed out in the context of the \u Cerenkov emission in dispersive media in~\cite{groupcone}. Restricting for a moment our attention to a given emission frequency, the {\em wave cone} is defined as the real space conical wavefront passing through the source and orthogonal to the direction of the far-field emission in $\kk$ space: its aperture $\phi_{ph}$ around the negative $x$ axis is determined by the {\em phase} velocity as $\sin\phi_{ph}=v_{\rm ph}/v$ and is related to the aperture of the \u Cerenkov cone by $\phi_{ph}=\pi/2-\theta$. With some caveats, it can be interpreted as the wavefront on which the \u Cerenkov emission has a constant phase.
On the other hand, the {\em group cone} is defined as the Mach cone for the given frequency and describes the spatial points on which the (spectrally filtered) electric field intensity is peaked. Its aperture  $\phi$ depends on the {\em group} velocity of light $v_{\rm gr}$ as $\sin\phi=v_{\rm gr}/v$. The distinction between the phase and group cones has been anticipated in~\cite{noi_cerenkov} to be most striking in the case of ultra-slow light media where $v_{\rm gr}$ is reduced to the m/s range while $v_{\rm ph}$ remains of the order of the speed of light in vacuo $c_0\simeq 3\cdot10^8$~m/s~\cite{slowlight}.

The study of the \u Cerenkov effect in strongly dispersive media where the refractive index $n(\omega)$ has a strong dependence on the frequency and/or the medium exhibits a non-trivial spatial patterning is still a very active domain of research from both the theoretical and the experimental points of view. For instance, the consequences of a strong resonance in $n(\omega)$ were theoretically investigated in~\cite{resonant_dispersive}:  the sub-linear dispersion of the photon in a resonant medium is responsible for the disappearance of the threshold velocity for the \u Cerenkov emission and for a non-trivial spatial patterning of the electric field wake behind the charge. These striking results were experimentally confirmed in~\cite{dipolar_nonlopt_cherenk} and bear a close resemblance to the surface wave pattern in the wake of a duck swimming on shallow water that will be discussed in Sec.\ref{sec:shallow}. Another active and promising research line is addressing those new features of \u Cerenkov radiation that follow from the peculiar band dispersion of photons in spatially periodic media~\cite{Cerenk_PC}  and in negative refractive index metamaterials, the so-called left-handed media~\cite{Cerenk_LH}.

\section{Moving impurities in a superfluid}
\label{sec:superfluid}

A central concept in the theory of superfluids~\cite{LPSS_book,PinesNozieres,Leggett} is the so-called Landau criterion for superfluidity, that determines the maximum speed at which a weak impurity can freely travel across a superfluid without experiencing any friction force and without generating any propagating perturbation in the fluid. In terms of the dispersion $\Omega(\kk)$ of the excitations in the superfluid, the Landau critical velocity has the form
\begin{equation}
v_{\rm cr}=\min_\kk \left[\frac{\Omega(\kk)}{k}\right].
\eqname{Landau}
\end{equation}
This cornerstone of our theoretical understanding of quantum liquids has a simple interpretation in terms of the theory of the generalized \u Cerenkov effect reviewed in Sec.\ref{sec:theory}: the friction force experienced by the moving impurity is due to the emission of elementary excitations in the fluid by a mechanism that is a quantum fluid analog of \u Cerenkov emission. The $v<v_{\rm cr}$ condition for superfluidity corresponds to imposing that the locus $\Sigma$ of excited modes is empty, while for faster impurities a characteristic wake pattern is generated around the impurity.

\begin{figure}[!htbp]
\begin{center}
\includegraphics[width=0.2\columnwidth,clip]{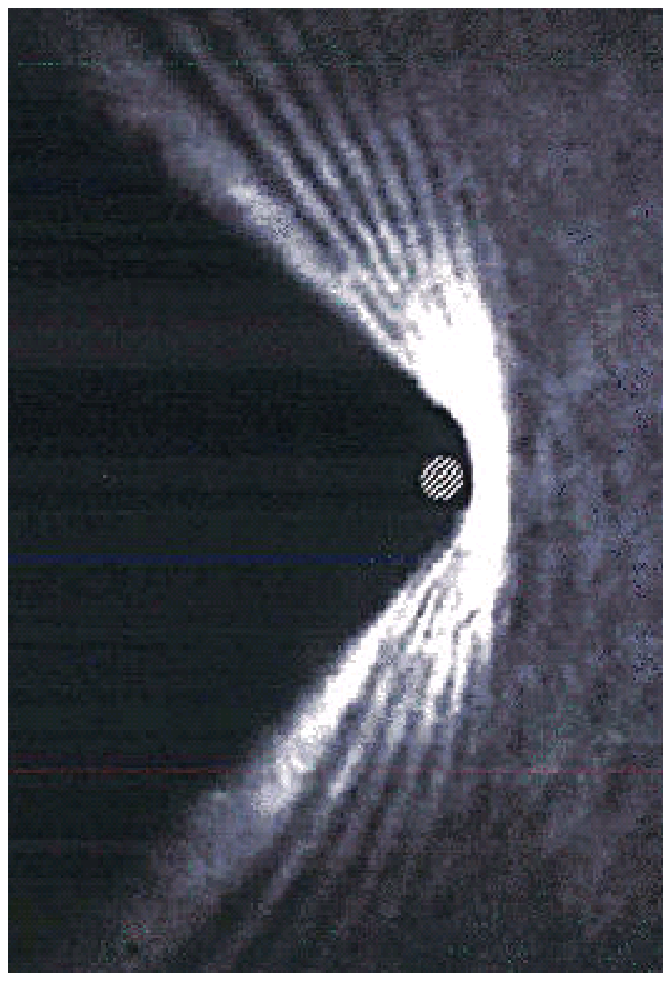}
\includegraphics[width=0.3\columnwidth,clip]{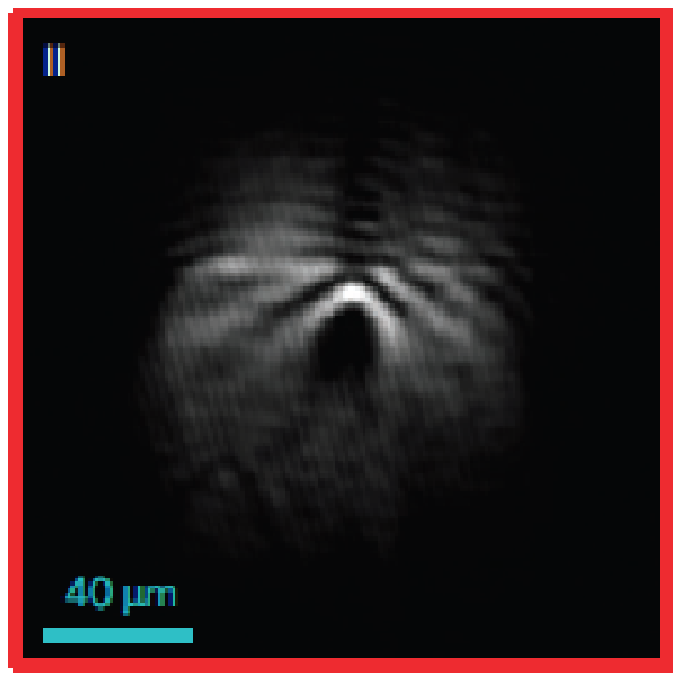}
\includegraphics[width=0.35\columnwidth,clip]{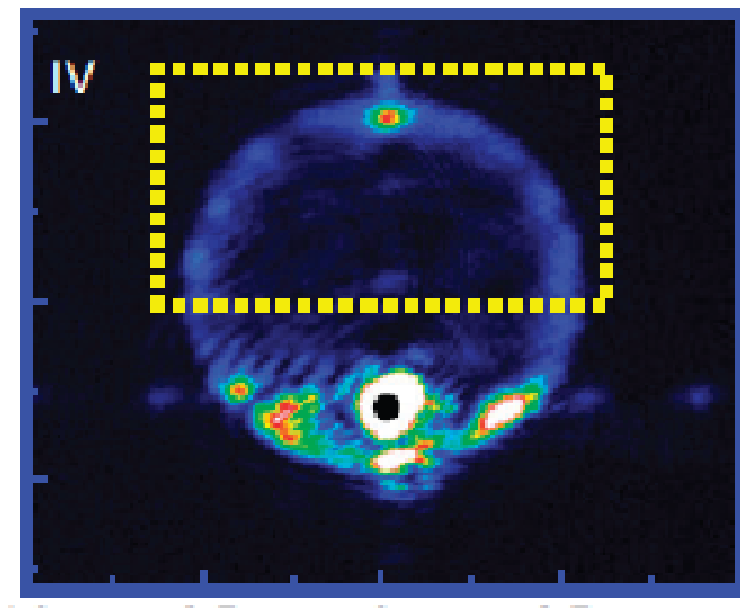}
\end{center}
\caption{Left panel: experimental image of the real-space wake pattern that appears in a Bose-Einstein condensate of ultracold atoms hitting the repulsive potential of a blue-detuned laser beam. The condensate motion is from right to left. Figure taken from~\cite{Cornell_exp}, as published in~\cite{Carusotto:PRL2006}.
Middle and right panels: experimental images of the real-space wake pattern (middle) and the momentum distribution (right) for a Bose-Einstein condensate of exciton-polaritons hitting a fabrication defect in the planar microcavity. The polariton flow is from top to bottom. The value of the density in the right panel is very small and interactions negligible.
Figures taken from~\cite{AmoBramati_NPhys}.
}
\label{fig:exp_qfluid}
\end{figure}

An experimental image of this wake using a dilute Bose-Einstein condensate of ultracold atoms hitting\footnote{Needless to say that the configuration of a moving superfluid hitting an impurity at rest is fully equivalent modulo a Galilean transformation to the case of a moving impurity crossing a superfluid at rest.} the repulsive potential of a blue-detuned laser is reproduced in the left panel of Fig.\ref{fig:exp_qfluid}; an analogous image for a condensate of exciton-polaritons in a semiconductor microcavity is reproduced in the middle panel. In both cases, the density wake extends both behind and ahead of the impurity. The geometrical shape of the $\Sigma$ locus is instead clearly visible in the momentum distribution pattern shown in the right panel.

The situation is of course more complex when stronger impurities are considered, e.g. a finite-sized impenetrable object: in this case, the critical speed for frictionless flow was predicted in~\cite{Pomeau_vc} to be limited by the nucleation of pairs of quantized vortices at the surface of the object, and therefore to be significantly lower than the speed of sound. This mechanism was recently confirmed in experiments for with atomic~\cite{atom_vort} and polariton~\cite{polar_vort} condensates. Furthermore, it is worth reminding that all our reasonings are based on a mean-field description of the condensate that neglects quantum fluctuations: more sophisticated Bethe ansatz calculations for a strongly interacting one-dimensional Bose gas~\cite{astra} have anticipated the appearance of a finite drag force also at sub-sonic speed. Including higher order terms of the Bogoliubov theory led the authors of~\cite{Pomeau_vc_qf} to a similar claim for a three-dimensional condensate.

\begin{figure}[!htbp]
\begin{center}
\includegraphics[width=0.49\columnwidth]{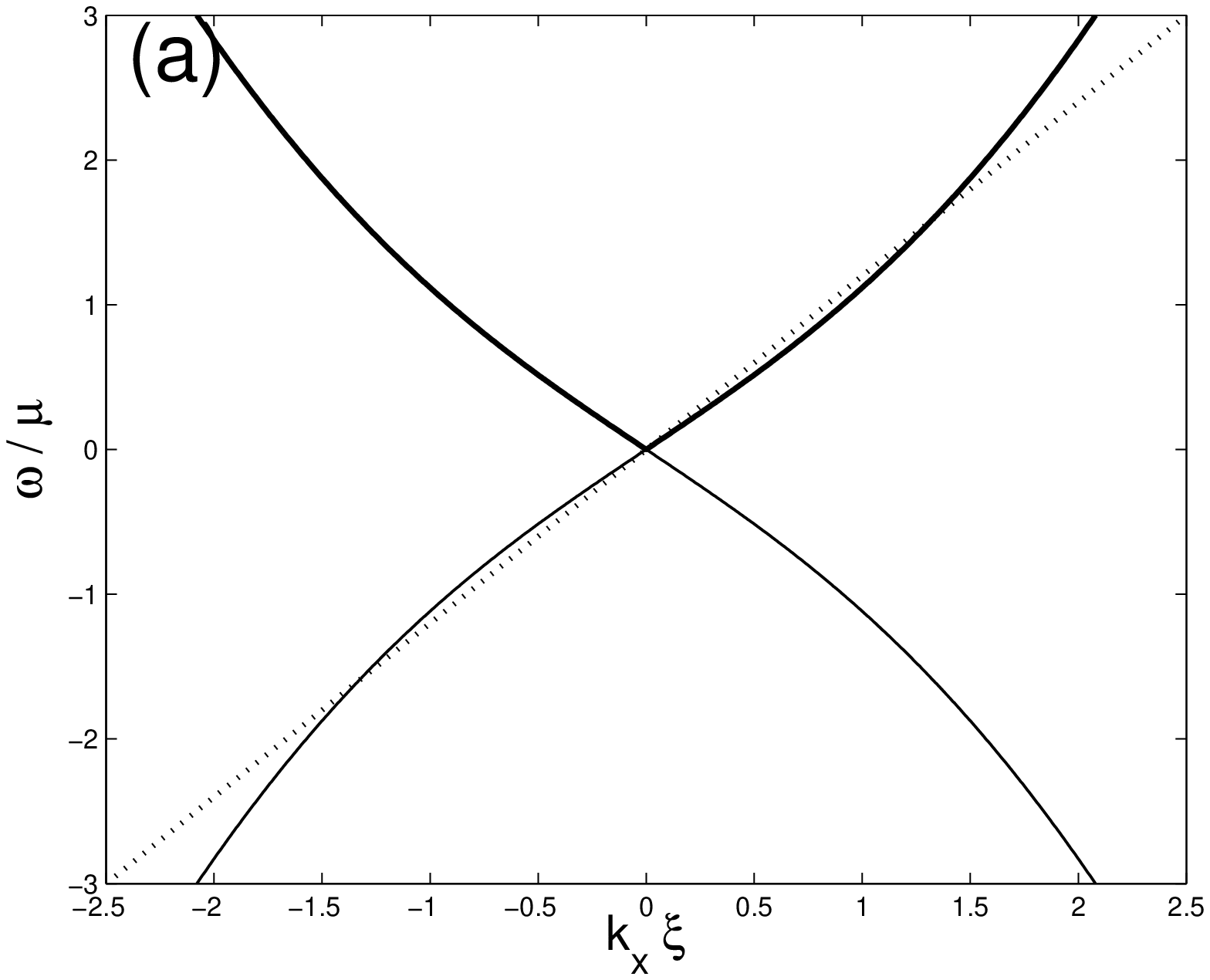}
\includegraphics[width=0.49\columnwidth]{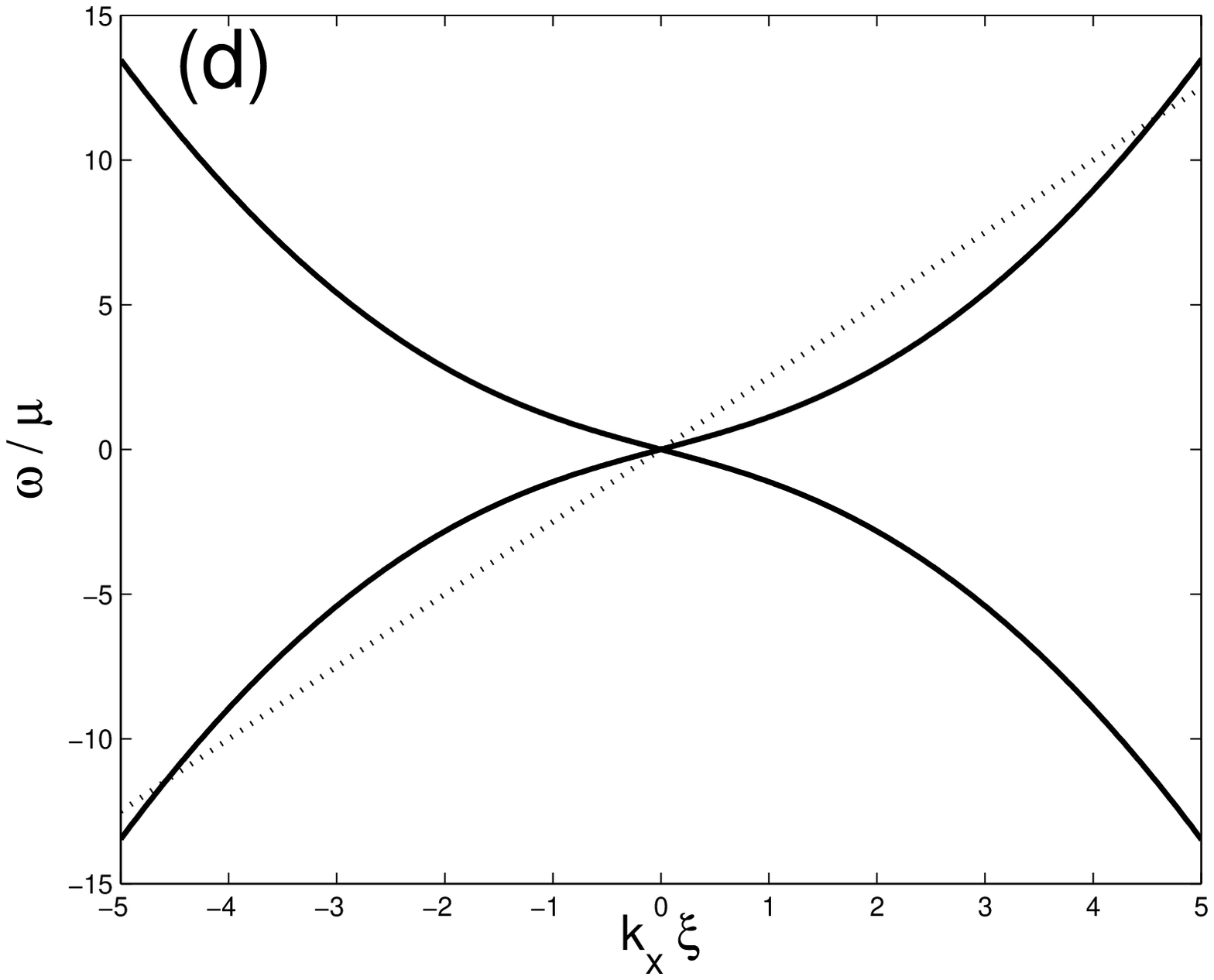}\\
\includegraphics[width=0.49\columnwidth]{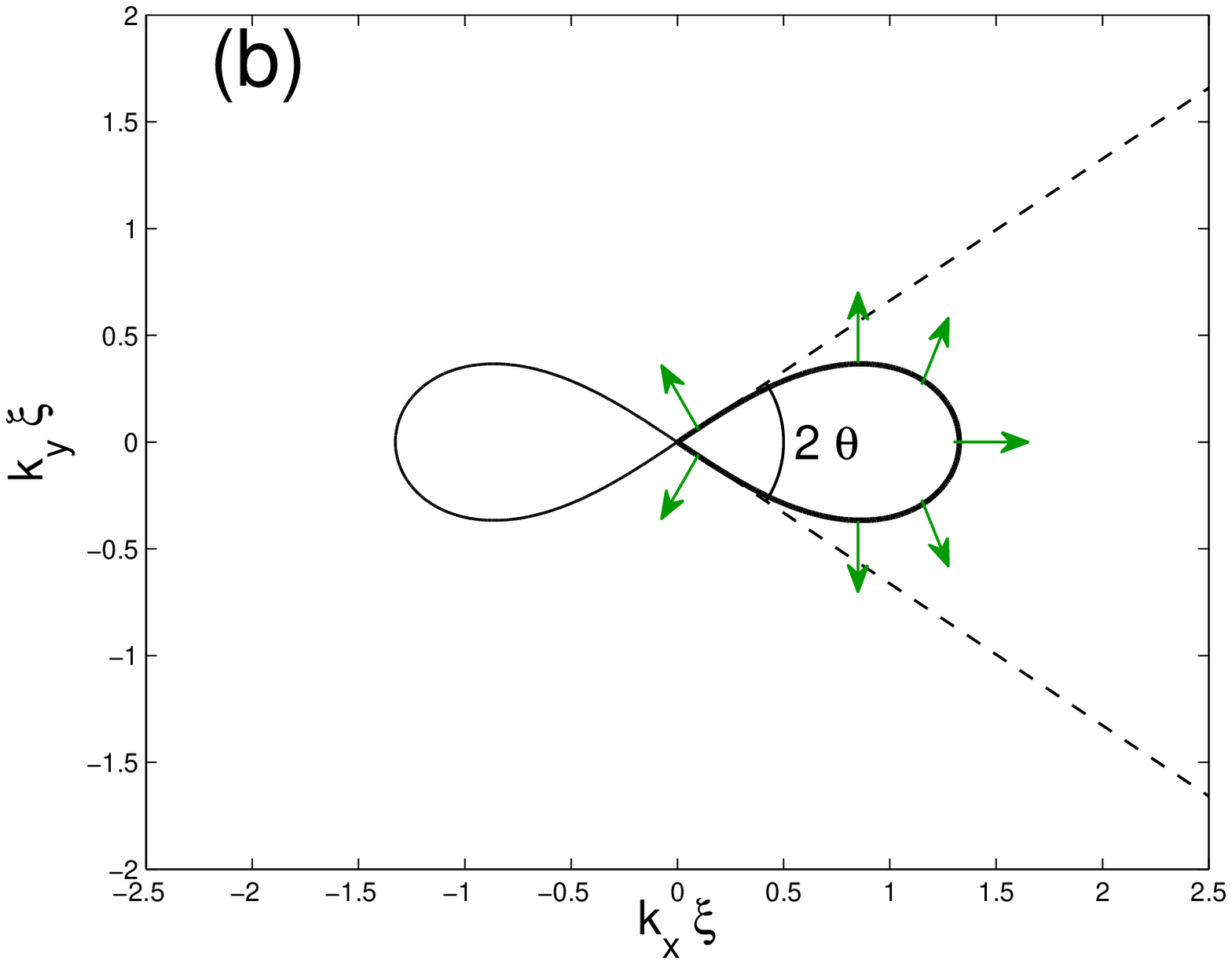}
\includegraphics[width=0.49\columnwidth]{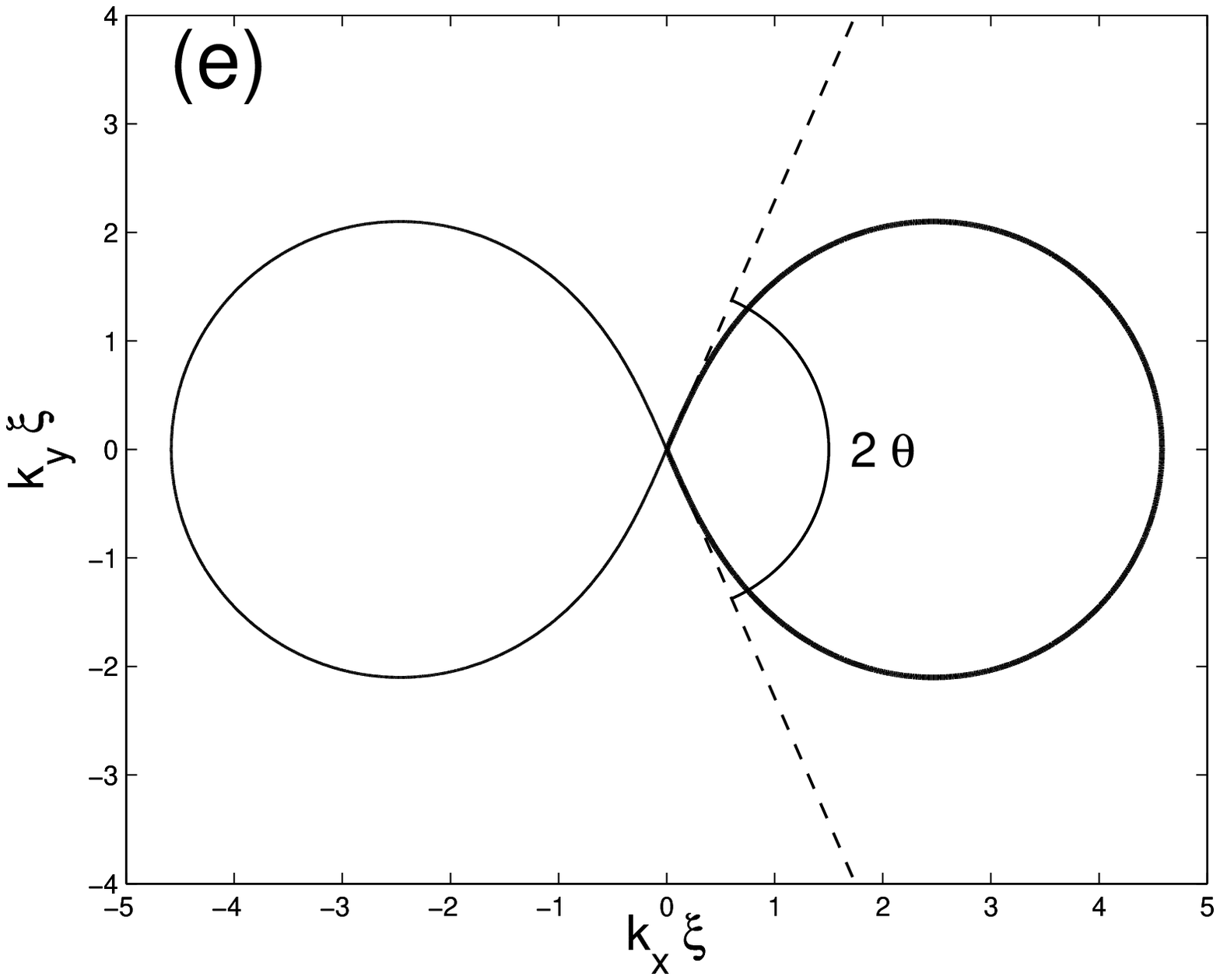} \\
\includegraphics[width=0.49\columnwidth]{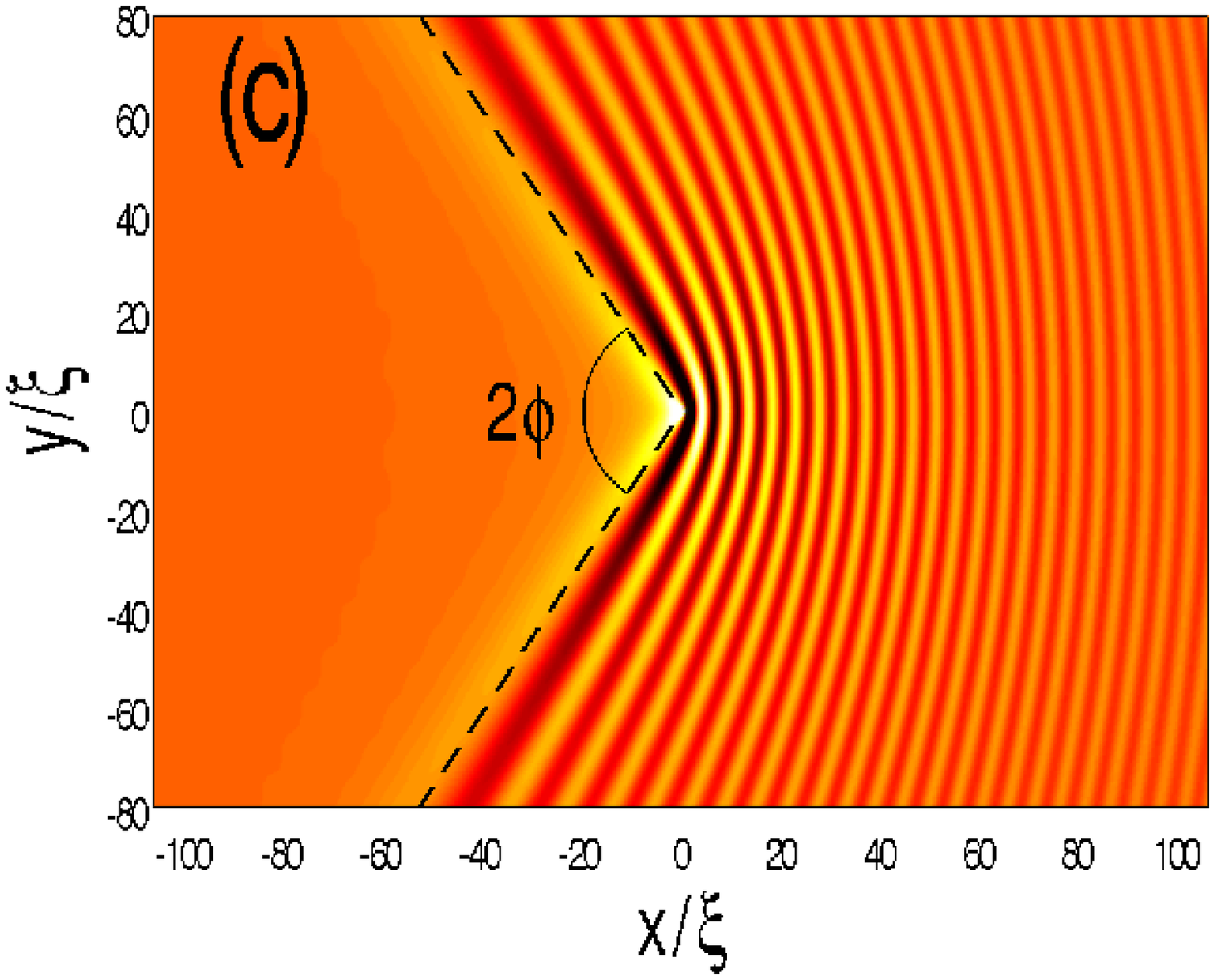}
\includegraphics[width=0.49\columnwidth]{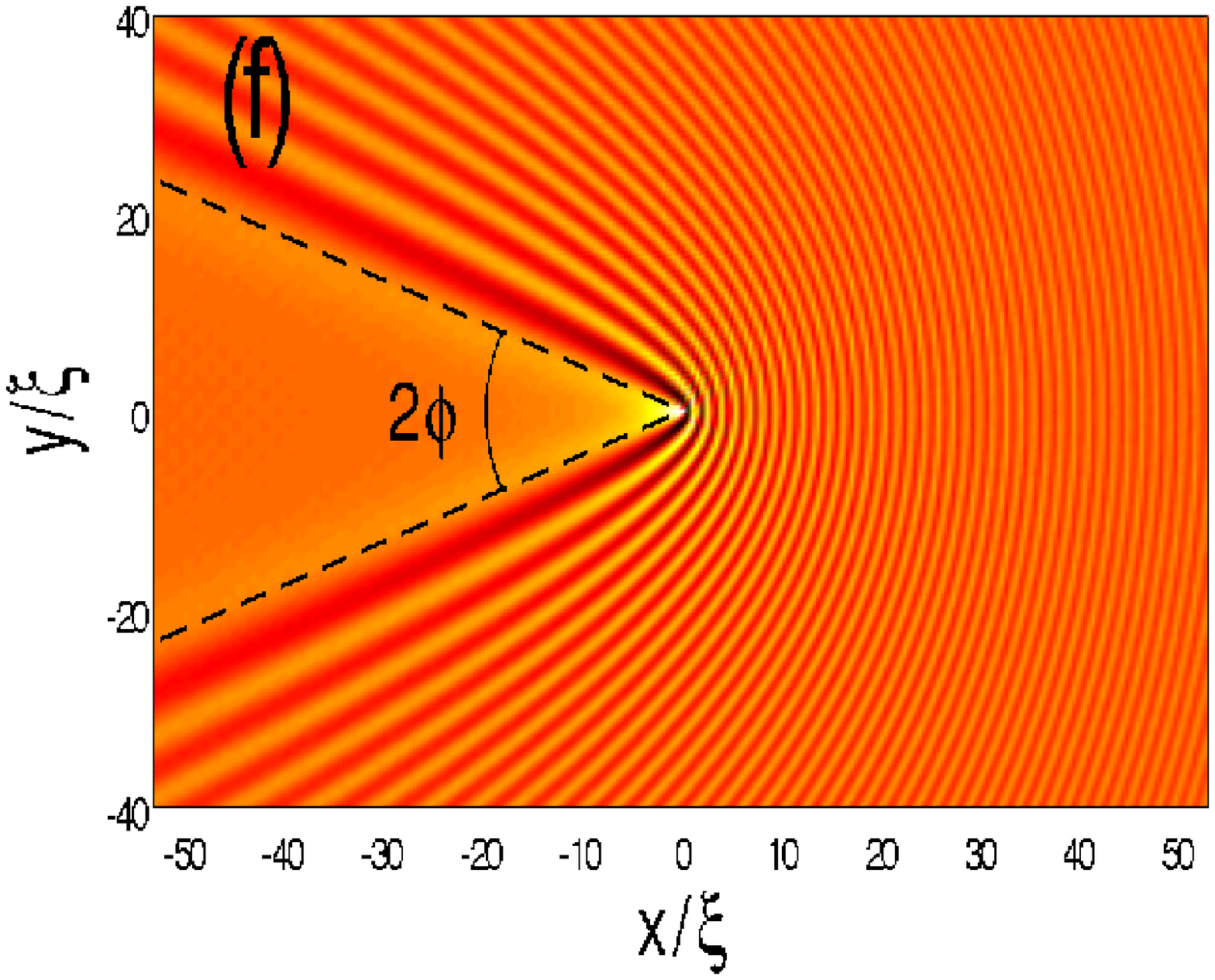}
\end{center}
\caption{Top row: Bogoliubov dispersion of excitations in a dilute Bose-Einstein condensate. The dashed line indicates the $\Omega=\kk\cdot \vv$ plane for two different impurity speeds $v/c_s=1.2$ [panel (a)] and $v/c_s=2.5$ [panel (d)].
Middle row: shape of the corresponding $\kk$-space locus $\Sigma$ of resonantly excited modes. The dashed lines indicate the \u Cerenkov cone in the low wavevector region $k\xi \ll 1$; the green arrows indicate the normal to the  $\Sigma$ locus, that is the direction of the relative group velocity $\vv_g'$.
Bottom row: real space pattern of the density modulation. All patterns are numerically obtained performing the integral via a fast Fourier transform of the $\kk$-space perturbation \eq{field_fourier2}.
The black dashed lines indicate the Mach cone.
}
\label{fig:Bogo}
\end{figure}

\subsection{The Bogoliubov dispersion of excitations}

The theoretical description of superfluids is simplest in the case of a dilute Bose gas below the transition temperature $T_{BEC}$ for Bose-Einstein condensation~\cite{LPSS_book}. For $T\ll T_{BEC}$ and sufficiently weak interactions, most of the atoms are accumulated in the same one-particle orbital, the so-called Bose-Einstein condensate. The elementary excitations in a dilute Bose-Einstein condensate are characterized by the Bogoliubov dispersion~\cite{LPSS_book}
\begin{equation}
\eqname{Bogo_disp}
\hbar^2\Omega^2=\frac{\hbar^2k^2}{2m}\left(\frac{\hbar^2k^2}{2m}+2\mu\right),
\end{equation}
where $m$ is the mass of the constituent particles and the chemical potential $\mu$ is given (at zero temperature) by
\begin{equation}
\mu=\frac{4\pi\hbar^2 a_0 n}{m}
\end{equation}
in terms of the particle-particle low-energy collisional scattering length $a_0$ and the particle density $n$. In the standard three-dimensional case, the weak interaction (or diluteness) condition requires that $na_0^3\ll1$ 

The characteristic shape of the Bogoliubov dispersion \eq{Bogo_disp} is illustrated in the (a,d) panels of Fig.\ref{fig:Bogo}.
For small momenta $k\xi\ll 1$ (the so-called healing length $\xi$ being defined as $\hbar^2/m\xi^2=\mu$), the dispersion has a sonic behavior
\begin{equation}
\Omega^2\simeq c_s^2\,k^2
\eqname{sonic}
\end{equation}
with a sound speed $c_s=\sqrt{\mu/m}$, while at high wavevectors $k\xi\gg 1$ it grows at a super-sonic rate and eventually recovers the parabolic behavior of single particles,
\begin{equation}
\Omega\simeq \pm \left[\frac{\hbar k^2}{2m} + \mu\right].
\eqname{1part}
\end{equation}
An explicit calculation from \eq{Bogo_disp} shows that the Landau critical velocity \eq{Landau} in the dilute Bose gas is determined by the speed of sound $v_{\rm cr}=c_s$. It is worth reminding that this is no longer true in more complex superfluids with strong interparticle interactions as liquid He-II, where $v_{\rm cr}$ is determined by the roton branch of the elementary excitations~\cite{LPSS_book,PinesNozieres,Leggett,Liq_Hel}. Remarkably, super-linear dispersions in the form \eq{Bogo_disp} also appear in the theory of surface waves on shallow fluids when the fluid depth is lower than the capillary length, see Eqs.\ref{eq:water_Bogo} in Sec.\ref{sec:surface}.

The effect of the moving impurity onto the superfluid can be described by a time-dependent external potential of the form $V(\rr,t)=V_0(\rr-\vv t)$ coupled to the particles forming the superfluid. Inclusion of this external potential in the Bogoliubov theory requires including a classical source term in the Bogoliubov equations of motion for the two-component spinor describing the quantum field of the non-condensed particles: a complete theoretical discussion along these lines can be found in the recent works~\cite{IC_CC_Superfl,Carusotto:PRL2006}. Here we shall use an approximate, yet qualitatively accurate model based on the simplified scalar theory of Sec.\ref{sec:theory}: the real and imaginary parts of the field $\phi(\rr,t)$ correspond to the density and phase modulation of the condensate.

\subsection{Superfluidity vs. Bogoliubov-\u Cerenkov wake}

As we have already mentioned, the locus $\Sigma$ is empty for sub-sonic impurity speeds $v<c_s$: the impurity is able to cross the superfluid without resonantly exciting any propagating mode of the fluid. As a result, within mean-field theory it is not expected to experience any friction force. Modulo a Galilean transformation, this effect is equivalent to a frictionless flow along a containing pipe in spite of the roughness of the walls, which is one of the clearest signatures of superfluid behavior~\cite{LPSS_book,PinesNozieres,Leggett}.

Still, the non-resonant excitation of the Bogoliubov modes by the moving impurity is responsible for a sizable density modulation in the vicinity of the impurity, that quickly decays to zero in space with an exponential law. An important consequence of this localized density perturbation is a sizable renormalization of the mass of the object~\cite{astra}: the linear momentum that is associated to the moving impurity gets in fact a contribution from the portion of fluid that is displaced by it.

For super-sonic motion, the locus $\Sigma$ consists of a conical region at small $k \xi \ll 1$ analogous  to what was found in Sec.\ref{sec:non-disp} for a purely linear dispersion: as in that case, the aperture angle $\theta$ of the $\kk$-space \u Cerenkov cone [dashed lines in Fig.\ref{fig:Bogo}(b,e)] defining the far-field angle at which phonons are emitted by the impurity is defined by the condition $\cos\theta=c_s/v$.

Correspondingly, the aperture $\phi$ of the Mach cone that is visible in the real-space density modulation pattern behind the impurity is defined by $\sin\phi=c_s/v$ (dashed black lines in [Fig.\ref{fig:Bogo}(c,f)]. This cone is the superfluid analog of the Mach cone that is created in a generic fluid by a super-sonically moving object, e.g. an aircraft or a bullet. An experimental image of a Mach cone in a superfluid of exciton-polaritons is shown in the central panel of Fig.\ref{fig:exp_qfluid}. As usual, the faster the impurity, the narrower the Mach cone. 

Differently from sound waves in an ordinary fluid, the Bogoliubov dispersion of the excitations in a superfluid is characterized by a parabolic shape at large wavevectors $k\xi \gg 1$ according to \eq{1part}. This region of the Bogoliubov spectrum is responsible for the smooth arc in the high wavevector region of $\Sigma$ that connects the two straight lines emerging from the origin  $\kk=0$. In experiments, the shape of $\Sigma$ can be inferred following the peak of the momentum distribution of the particles in the superfluid: an example of experimental image using exciton-polaritons in the low-density regime is reproduced in the right panel of Fig.\ref{fig:exp_qfluid}. Analogous images for atomic gases can be found e.g. in~\cite{ketterle}.

In the low wavevector region, the relative group velocity $\vv'_g$ is oriented along the edges of the Mach cone. Along the high wavevector part of $\Sigma$, the relative group velocity $\vv'_g$ rotates in a continuous and monotonous way spanning all intermediate directions external to the Mach cone.  As no point on $\Sigma$ corresponds to a relative group velocity oriented in the backward direction inside the Mach cone, the density profile remains unperturbed in this region. On the other hand, the density perturbation shows peculiar features in front of the Mach cone, with a series of curved wavefronts extending all the way ahead of the impurity. These wavefronts are clearly visible in the experimental images that are shown in Fig.\ref{fig:exp_qfluid} for atomic (left panel) and polaritonic (middle panel) superfluids. In the $\kk$-space diagrams of Fig.\ref{fig:Bogo}(b,e), these waves correspond to the regions in the vicinity of the extreme points of $\Sigma$  where $\vv'_g$ is directed in the direction of the impurity speed along the positive $x$ direction. 

Physically, these curved wavefronts in the density modulation pattern can be understood as originating from the interference of the macroscopic coherent wave associated to the Bose-Einstein condensate with the atoms that are coherently scattered by the moving impurity. An analytic discussion of their shape is discussed in detail in~\cite{kamchatnov}; their one-dimensional restriction was first mentioned in~\cite{pavloff}.
In the next section we shall present analytical formulas for an approximated theory where the single particle region of the Bogoliubov dispersion is modeled with the parabolic dispersion of single-particle excitations.

\section{Parabolic dispersion: conics in the wake}
\label{sec:parabolic}

Another example of dispersion that is fully amenable to analytic treatment is the parabolic one,
\begin{equation}
\Omega(\kk)=\frac{\hbar k^2}{2m}+\mu.
\eqname{parabolic}
\end{equation}
In spite of its simplicity, this form of dispersion can be used to model a number of different physical configurations, from the large wavevector $k\xi\gg 1$ region of the Bogoliubov dispersion \eq{Bogo_disp} of superfluids, to the resonant Rayleigh scattering in planar microcavities~\cite{RRS}, to magnons in solid-state materials~\cite{magnons,magnons_doppl}. 
In particular, the results of this section will shine further light on the curved wavefronts observed in Fig.\ref{fig:Bogo}(c,f) ahead of the impurity. 

\begin{figure*}[!htbp]
\begin{center}
\includegraphics[width=0.3\columnwidth]{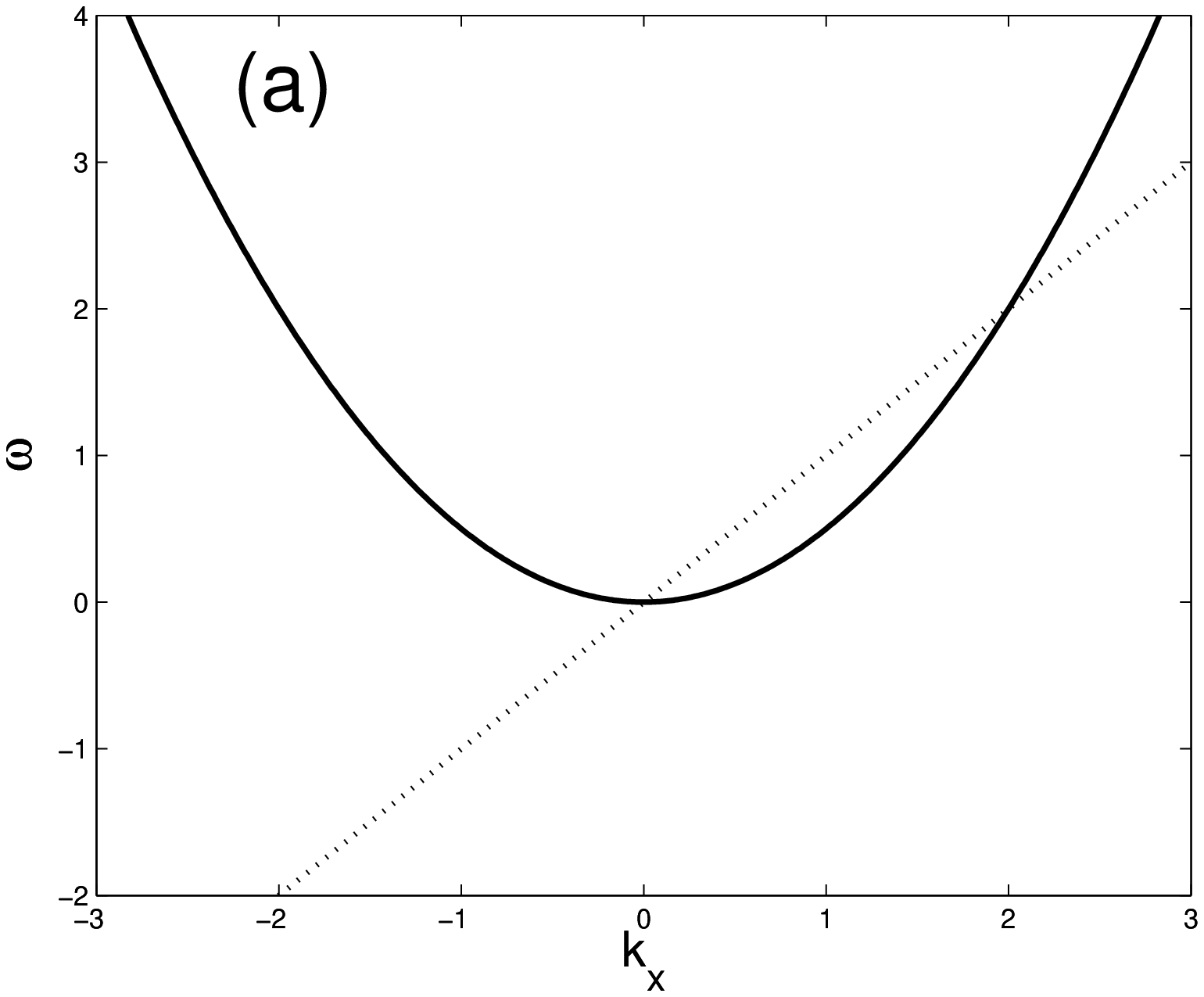}
\includegraphics[width=0.3\columnwidth]{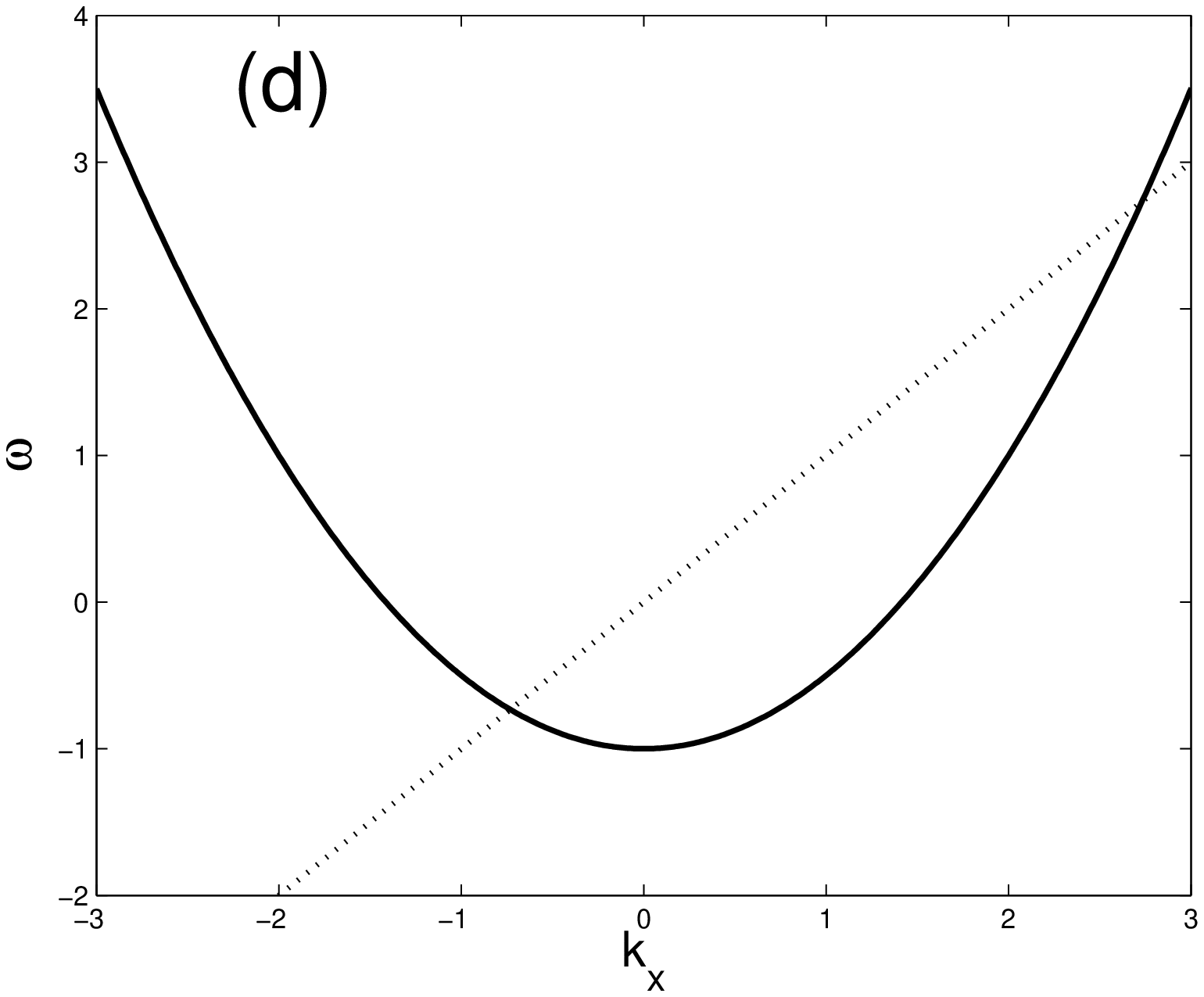}
\includegraphics[width=0.3\columnwidth]{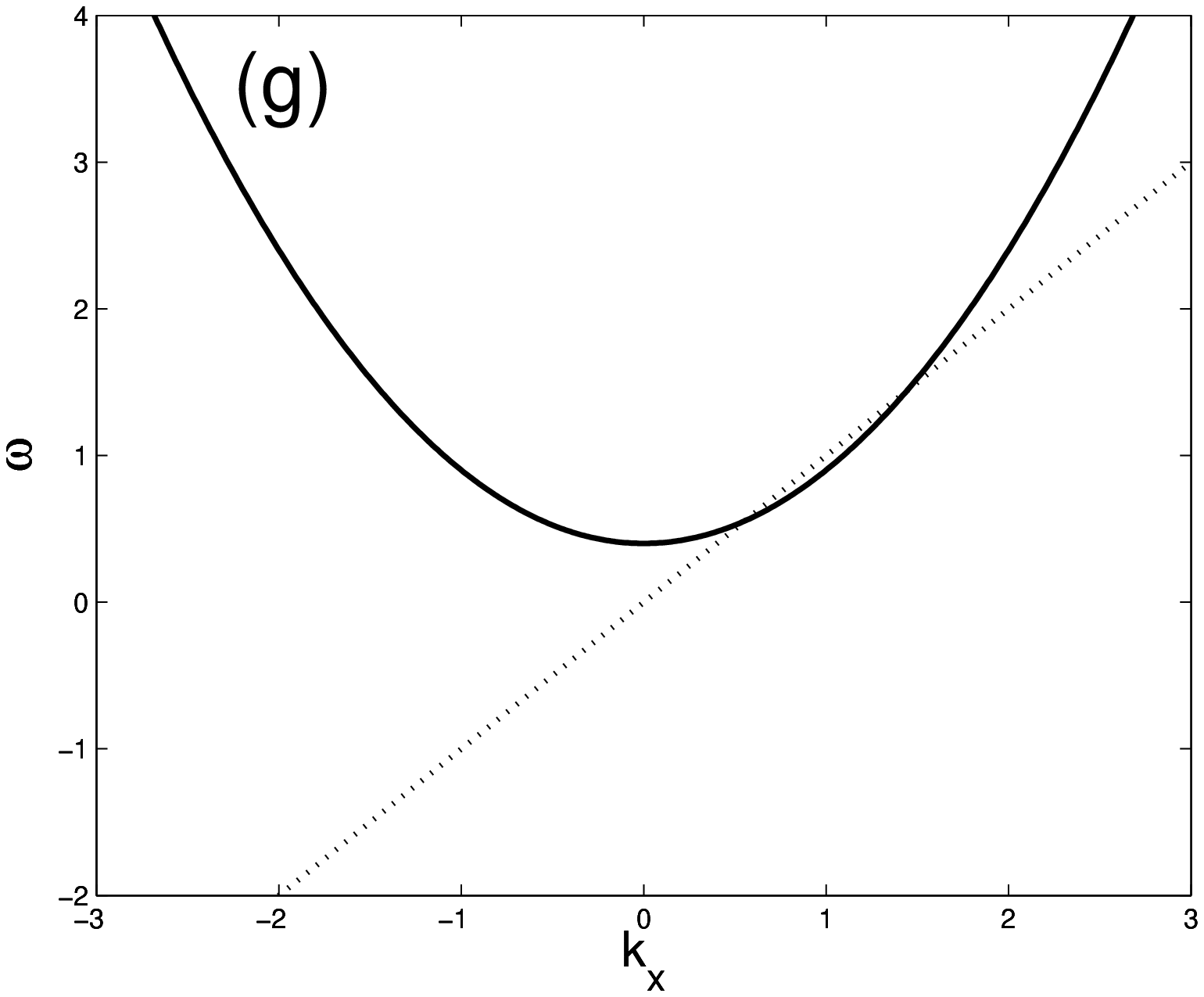} \\
\includegraphics[width=0.3\columnwidth]{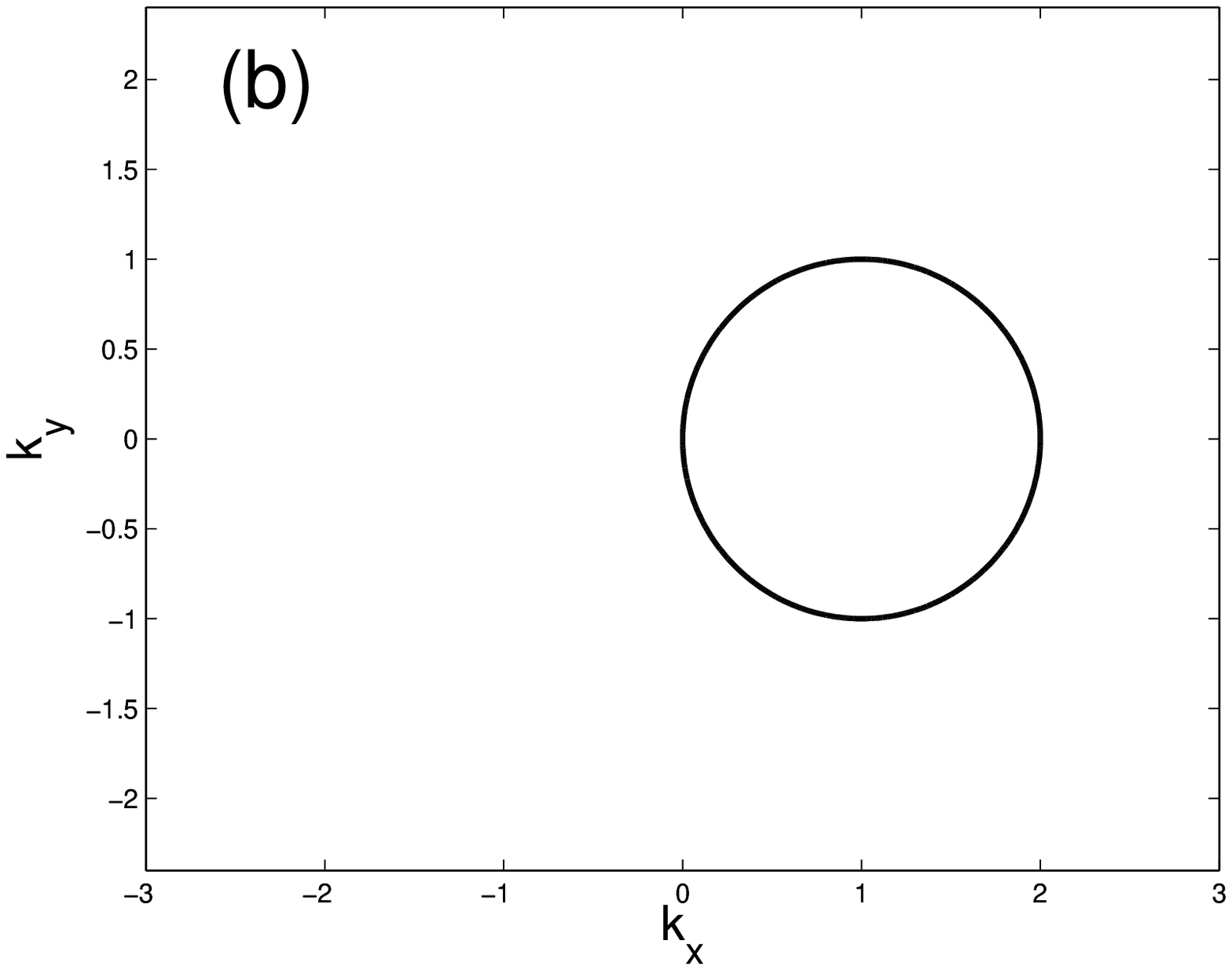}
\includegraphics[width=0.3\columnwidth]{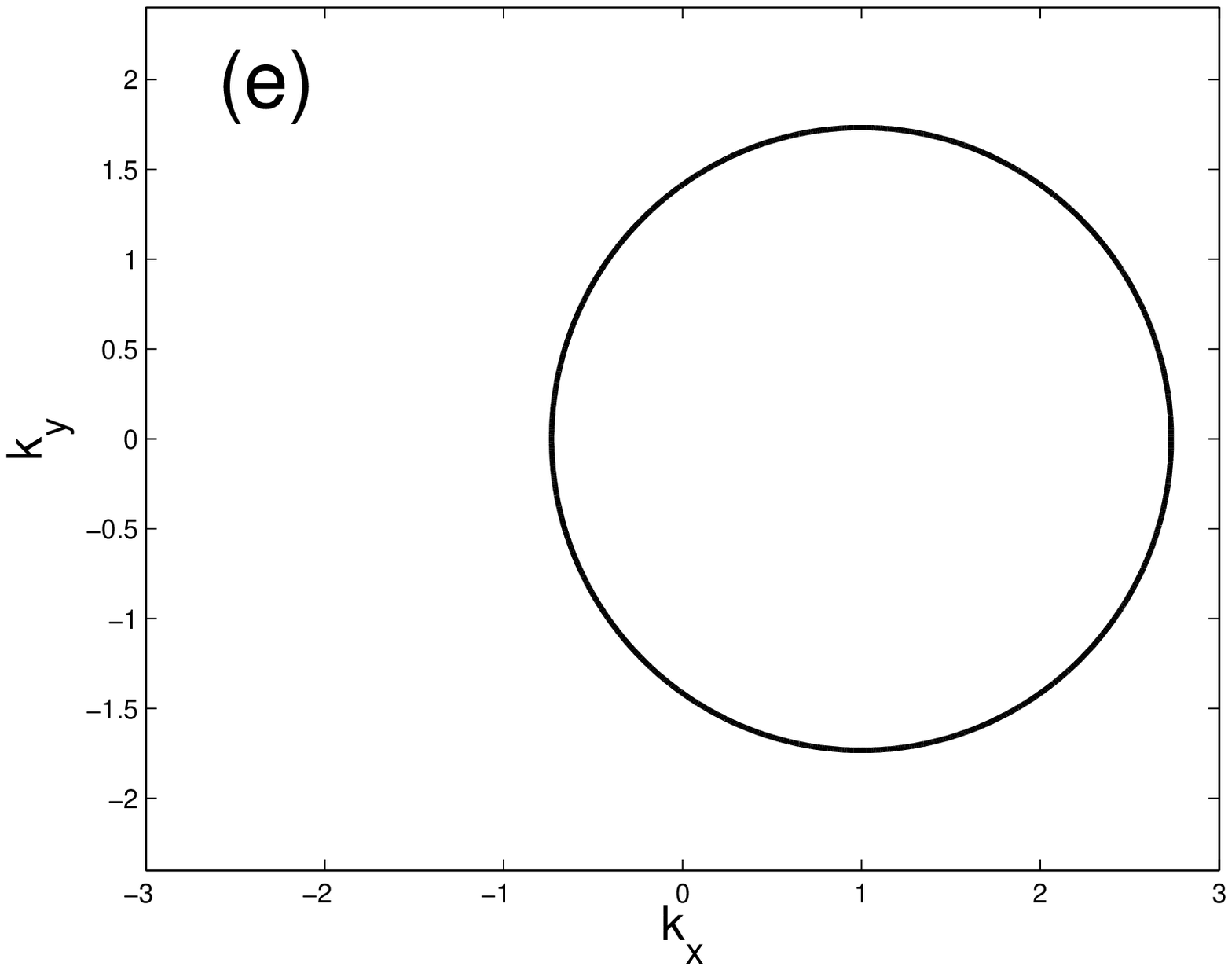}
\includegraphics[width=0.3\columnwidth]{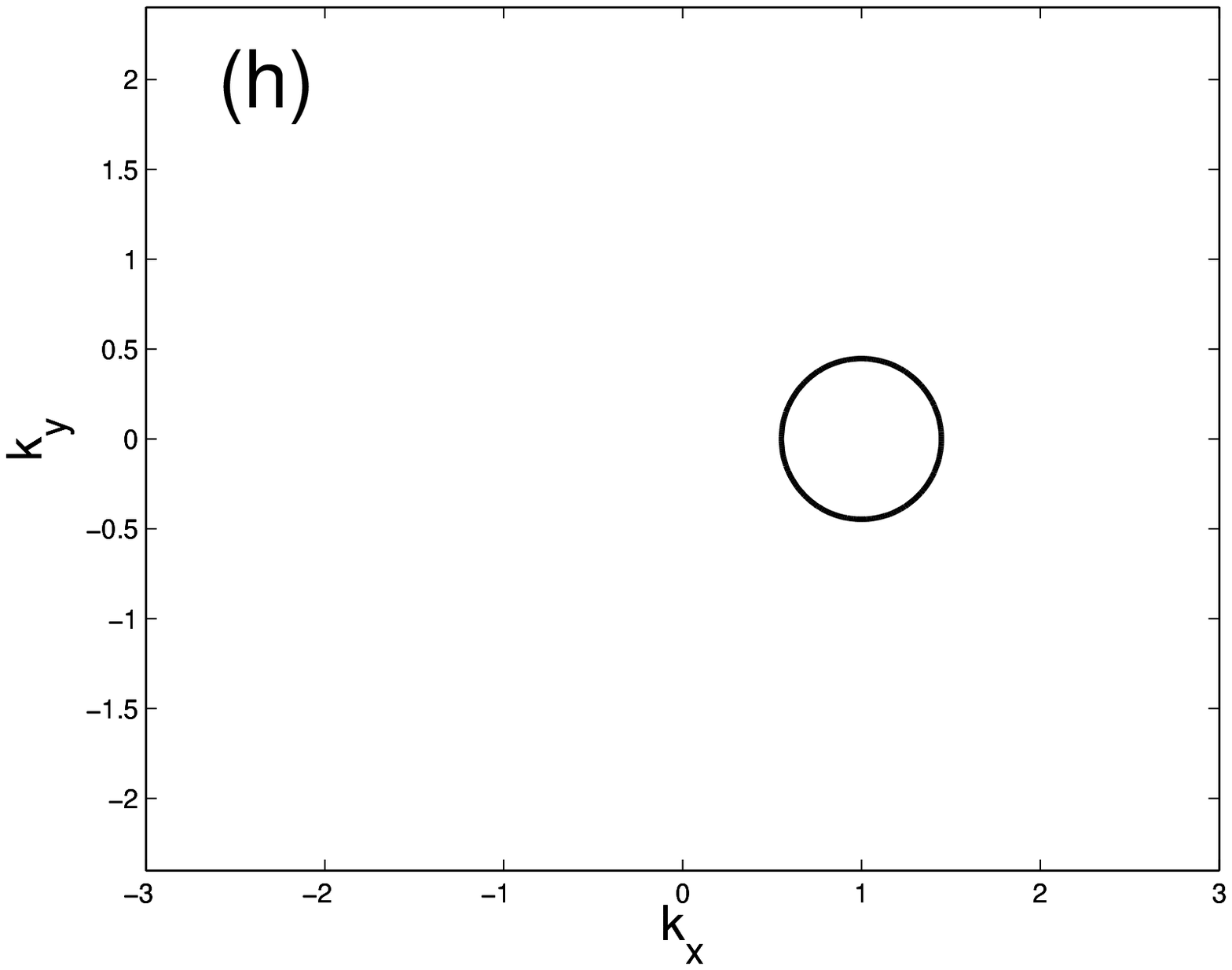}\\
\includegraphics[width=0.3\columnwidth]{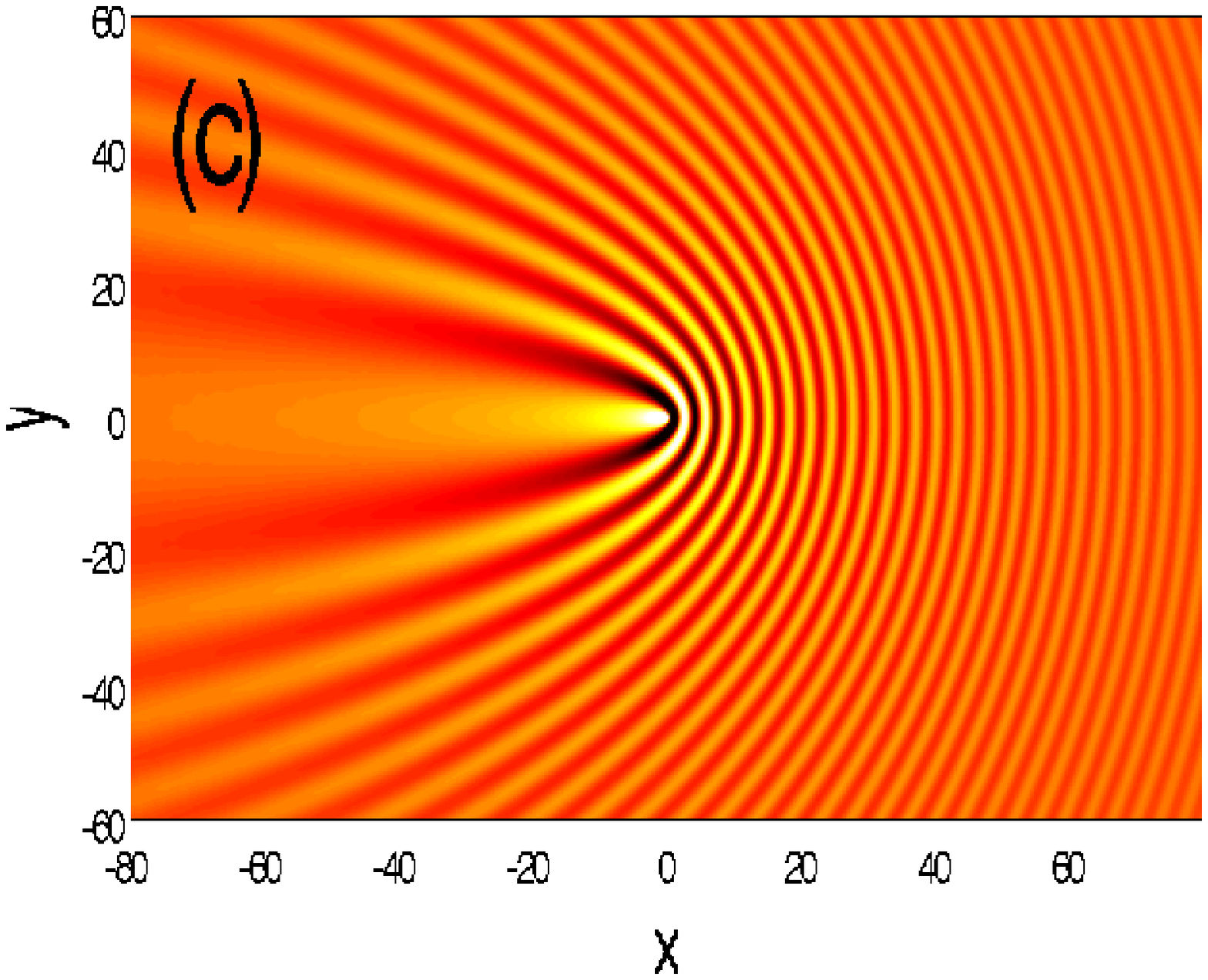}
\includegraphics[width=0.3\columnwidth]{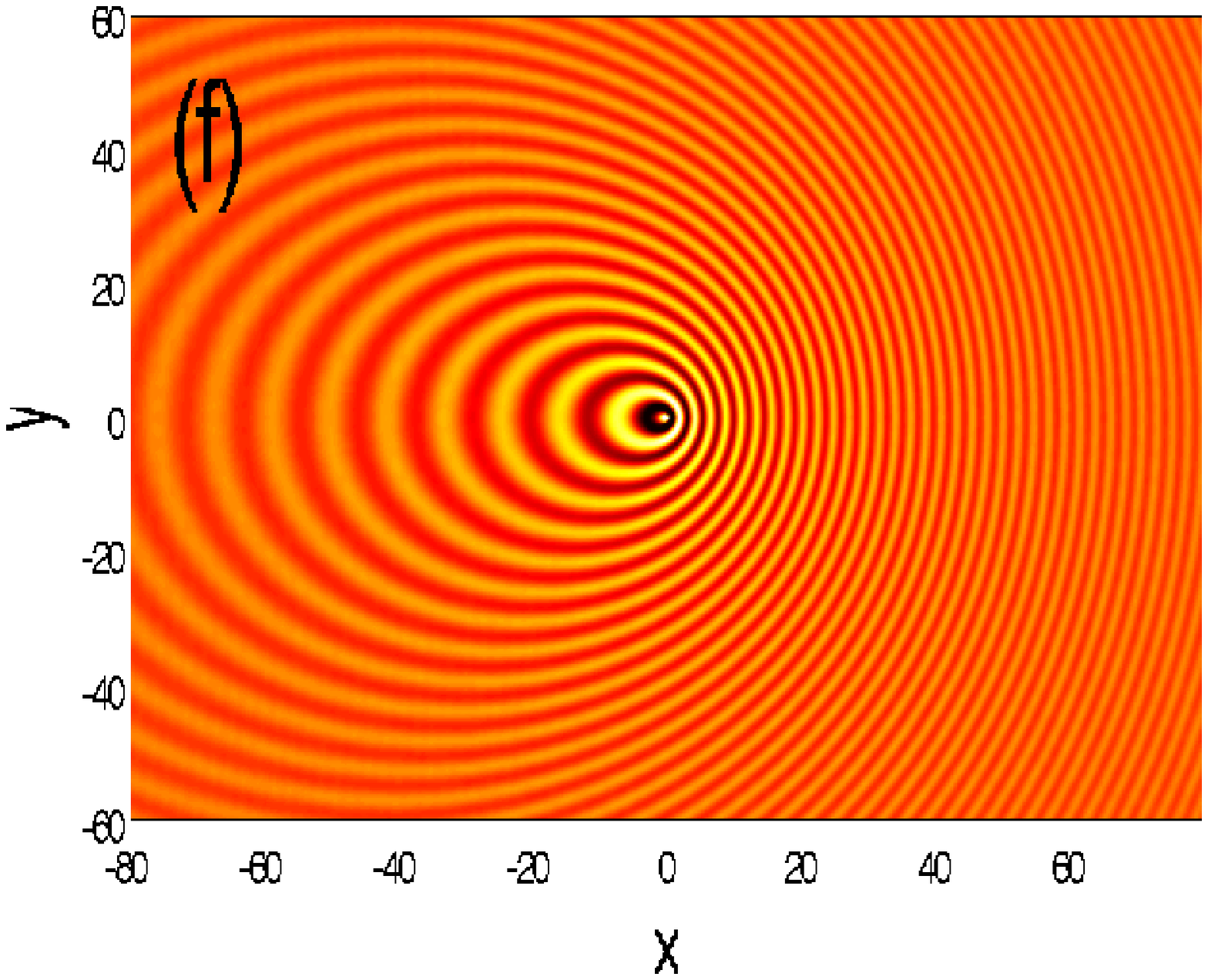}
\includegraphics[width=0.3\columnwidth]{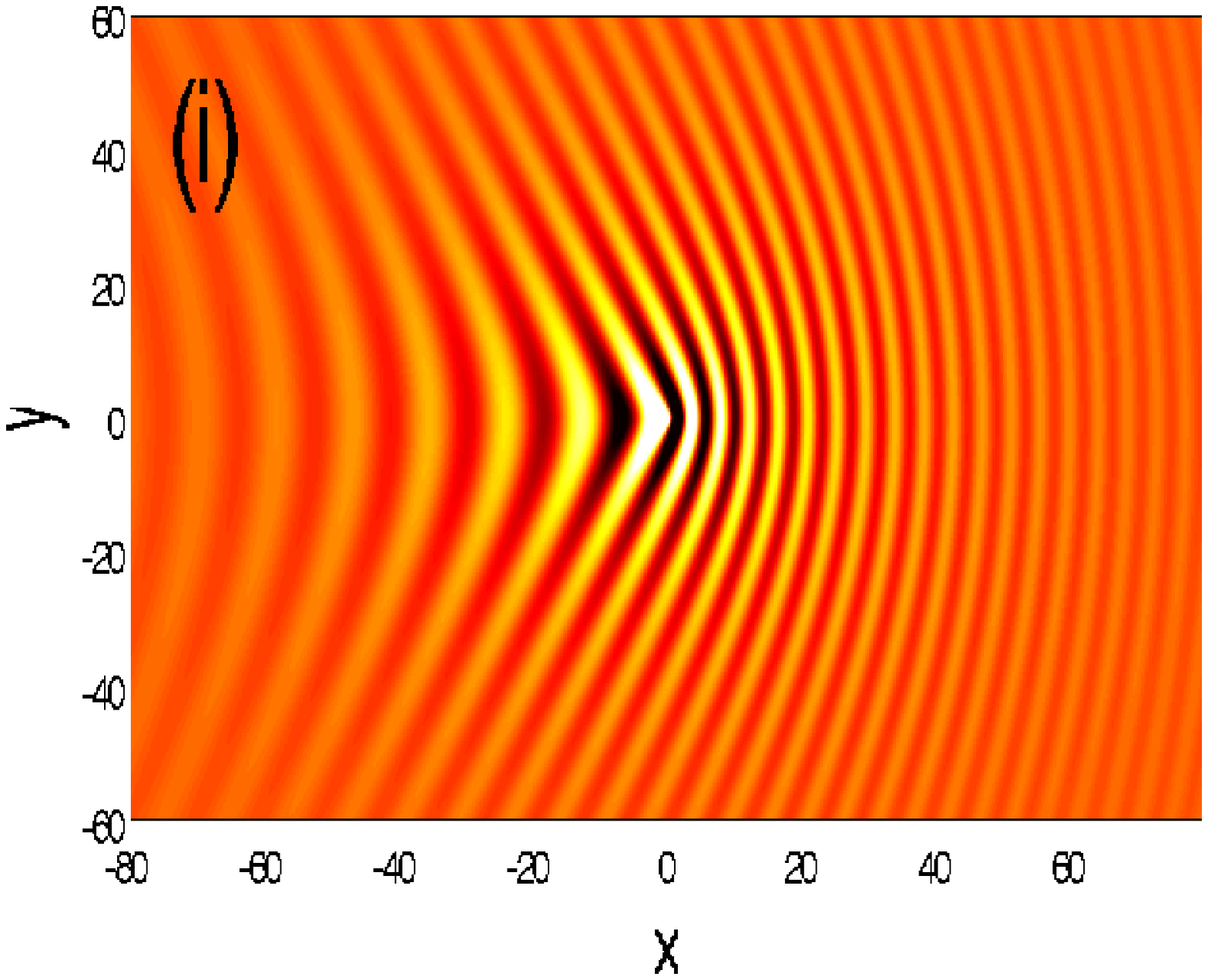}
\end{center}
\caption{Top row: parabolic dispersion of excitations $\omega=k^2/2+\mu$ in the $\mu=0$ (a), $\mu=-1<0$ (b), $\mu=0.4>0$ (c) cases (for notational simplicity, we have set $m=\hbar=1$). The dashed line indicates the $\Omega=\kk\cdot \vv$ plane for a generic particle speed $v=1$ along the positive $x$ direction.
Middle row: circular shape of the $\kk$-space locus $\Sigma$ of resonantly excited modes.
Bottom row: real space patterns of the density modulation.  All patterns are numerically obtained performing the integral in \eq{field_real}  via a fast Fourier transform algorithm.
}
\label{fig:pol}
\end{figure*}

For a generic dispersion of the parabolic form \eq{parabolic}, simple analytical manipulations show that the locus $\Sigma$ has a circular shape as shown in Fig.\ref{fig:pol}(b,e,h). Assuming again that the particle speed $\vv$ is directed along the positive $x$ axis, the center of the circle is located at $k_x=k_o=m v/\hbar$, $k_y=0$ and has a radius $\bar{k}$ such that 
\begin{equation}
\frac{\hbar \bar{k}^2}{2m}=\frac{m v^2}{2\hbar}-\mu.
\eqname{kbar}
\end{equation}
Depending on the relative value of the velocity $v$ and of the $\mu$ parameter, different regimes can be identified.

For positive $\mu$ (but such that the RHS of \eq{kbar} is still positive), the radius $\bar{k}$ is smaller than $k_o$ and the origin $\kk=0$ lies outside the circle. This is the typical case of large $\kk$ excitations in superfluids, whose dispersion is approximated by Eq.\ref{eq:1part}. 
The usual resonant Rayleigh scattering ring~\cite{RRS} passing through the origin $\kk=0$ is recovered in the $\mu=0$ case describing the case of an ideal gas of non-interacting particles: an experimental example of such a ring is visible in the momentum distribution shown in Fig.\ref{fig:exp_qfluid}(c) for a low-density gas of (almost) non-interacting polaritons flowing against a localized impurity potential.
For negative $\mu$, the radius is instead larger $\bar{k}>k_0$ and the origin $\kk=0$ falls inside the circle.

The relative group velocity $\vv'_g$ is directed in the outward radial direction. As a consequence of the smooth shape of $\Sigma$, $\vv'_g$  spans all possible directions and the real-space perturbation shown in Fig.\ref{fig:pol}(c,f,i) extends to the whole plane. However,  the wavefronts can have very different shapes depending on the relative value of $v$ and $\mu$. 

A closed form for the real-space wake pattern is straightforwardly obtained by noting that the integral in the right-hand side of \eq{field_real} is in this case closely related to the retarded Green's function for a free non-relativistic particle~\cite{CCT_MQ},
\begin{equation}
G_{\rm ret}(\rr,\omega)=\int\!\frac{d^d\kk}{(2\pi)^d}\,\frac{e^{i\kk\cdot \rr}}{\hbar \kk^2/2m - \omega-i\, 0^+}.
\end{equation}
In a generic dimension $d$, the asymptotic form of $G_{\rm ret}$ at large $\rr$ has the outgoing spherical wave form
\begin{equation}
G_{\rm ret}(\rr,\omega)= \frac{2\pi i \, C_d}{r^{(d-1)/2}}\,e^{i \bar{k} r}
\end{equation}
with a wavevector $\bar{k}$ such that
\begin{equation}
\frac{\hbar \bar{k}^2}{2m}=\omega.
\end{equation}
Of course, for $\omega>0$ (or $\omega<0$), the solution such that $\bar{k}>0$ (or $\textrm{Im}[\bar{k}]>0$) must be considered. $C_d$ is a dimension- and energy-dependent normalization constant.

Using this result, the expression \eq{field_real} for the wake generated by a point-like source term can be simplified into
\begin{multline}
\phi(\rr')=-\int\!\frac{d^2\kk}{(2\pi)^2}\,\frac{e^{i\kk\cdot\rr'}}{\mu+\frac{\hbar k^2}{2m}-\kk\cdot \vv-i\,0^+} = \\
=- \int \frac{d^2\kk}{(2\pi)^2}\,\frac{e^{i\kk\cdot\rr'}}
{\frac{\hbar k^2}{2m}-\left(\frac{m v^2}{2\hbar}-\mu\right)-i\,0^+}\,e^{i\frac{m}{\hbar}\vv\cdot \rr'}=-\frac{2\pi i\, C_d}{\sqrt{r'}}\,e^{i\bar{k}r'}\,e^{ik_o x'}.
\eqname{parab_real_space}
\end{multline}
The real part of this wave provides the wake pattern plotted in Fig.\ref{fig:pol}(c,f,i):
The different panels correspond to the $\mu=0$ (c),  $\mu<0$ (f) and $\mu>0$ (i) cases, which correspond to $\bar{k}=k_o$ (c), $\bar{k}>k_o$ (f), and $\bar{k}<k_o$ (i), respectively.

The shape of the wavefronts is obtained as the constant phase loci of \eq{parab_real_space}. For instance, the loci of points for which the phase of the field $\phi$ equals an integer  multiple of $2\pi$ are described by
\begin{equation}
\bar{k} \sqrt{x^2+y^2} + k_o x=2\pi M
\eqname{phase_comparison}
\end{equation}
with $M$ a generic integer. 
After moving the $k_o x$ term to the LHS and then taking the square of both members, this equation is straightforwardly rewritten as a quadratic equation in the spatial coordinates. The shape of the wavefronts in the two-dimensional plane is therefore described by conic curves: the specific nature of the conic in the different cases depends on the ratio $\bar{k}/k_o$.

For $k_o=\bar{k}$, the wavefronts have a parabolic shape described by the equation
\begin{equation}
4\pi^2 M^2-4\pi M k_o x= k_o y^2.
\eqname{parab}
\end{equation}
As the square root has by definition a non-negative value, the further condition $2\pi M - k_o x\geq 0$ has to be imposed to ensure that the RHS of \eq{phase_comparison} is non-negative. Combined with \eq{parab}, this condition is equivalent to imposing that  the integer $M\geq 0$. An example of these parabolic wavefronts is shown in Fig.\ref{fig:pol}(c).

For $\bar{k}>k_o$, the wavefronts have an elliptic shape described by the equation
\begin{equation}
\bar{k}^2 y^2 +(\bar{k}^2-k_o^2)\left[x+\frac{2\pi M k_o}{\bar{k}^2-k_o^2} \right]^2=\frac{4\pi^2 M^2 \bar{k}^2}{\bar{k}^2-k_o^2}.
\eqname{elliptic}
\end{equation}
An example of these elliptic wavefronts is shown in Fig.\ref{fig:pol}(f).
The condition on the non-negativity of the RHS of \eq{phase_comparison} imposes that $M\geq 0$. The ellipticity of $\Sigma$ is a function of the ratio $\bar{k}/k_o$: the closer this ratio is to 1, the more elongated the ellipse is. In the limit $\bar{k}/k_o \to 1$, the ellipse tends to a parabola, recovering the case $\bar{k}=k_o$ discussed above. The larger the ratio $\bar{k}/k_o$, the closer the wavefront shape to a series of concentric circles.

Finally, for $\bar{k}<k_o$, the wavefronts have hyperbolic shapes described by the equation
\begin{equation}
(k_o^2-\bar{k}^2)\,\left[x-\frac{2\pi M k_o}{k_o^2-\bar{k}^2} \right]^2-\bar{k}^2 y^2 =
\frac{4\pi^2 M^2 \bar{k}^2}{k_o^2-\bar{k}^2}.
\eqname{hyper}
\end{equation}
In this case, the condition on the RHS of \eq{phase_comparison} does not impose any condition on $M$ that can have arbitrary positive or negative integer values. However, combining this condition with the equation defining the hyperbola, one finds that for each $M$ only the left branch of the hyperbola at lower $x$ has to be retained. Positive vs. negative values of $M$ are responsible for the different periodicities that are visible in Fig.\ref{fig:pol}(i) in the $x>0$ and $x<0$ regions, respectively.

From the $\kk$ space diagrams in Fig.\ref{fig:pol}(b,e,h), it is immediate to see that the points on $\Sigma$ corresponding to the backward and forward propagating waves are the two intersections of the circle with the $x$ axis: the wider spacing of the wavefronts in the backward direction is due to the smaller magnitude of the wavevector at the intersection point at lower $x$. In the parabolic case, this point coincides with the origin, which explains the absence of density oscillations on the negative $x$ axis. 

The characteristic curvature of the forward propagating wavefronts provides a qualitative explanation for the shape of the density modulation experimentally observed ahead of the impurity and illustrated in the left and central panel of Fig.\ref{fig:exp_qfluid}. Of course, the absence of backward propagating waves in the superfluid behind the impurity is due to the $\kk=0$ singularity of the $\Sigma$ locus for the case of the Bogoliubov dispersion. It is worth reminding that, in contrast to previous works, the shape of the forward propagating wavefronts is exactly parabolic only in the $\mu=0$ limit of non-interacting particles. For the generic $\mu>0$ case of Bogoliubov theory, the Hartree potential in \eq{1part} makes their shape to be closer to (part of) an hyperbola.

\section{Surface waves on a liquid}
\label{sec:surface}

\begin{figure}[!htbp]
\begin{center}
\begin{minipage}[c]{0.67\columnwidth}
\begin{flushright}
\includegraphics[width=1\columnwidth,clip]{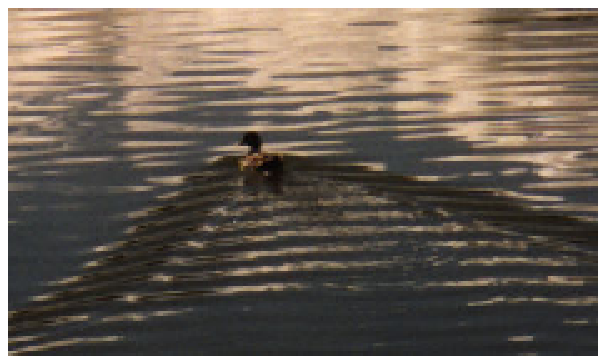}
\end{flushright}
\end{minipage}
\hspace{0.05\columnwidth}
\begin{minipage}[c]{0.24\columnwidth}
\begin{flushright}
\includegraphics[width=0.88\columnwidth,clip]{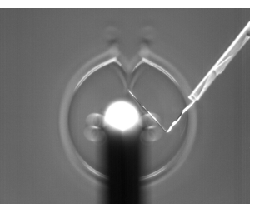} \\
\includegraphics[width=1\columnwidth,clip]{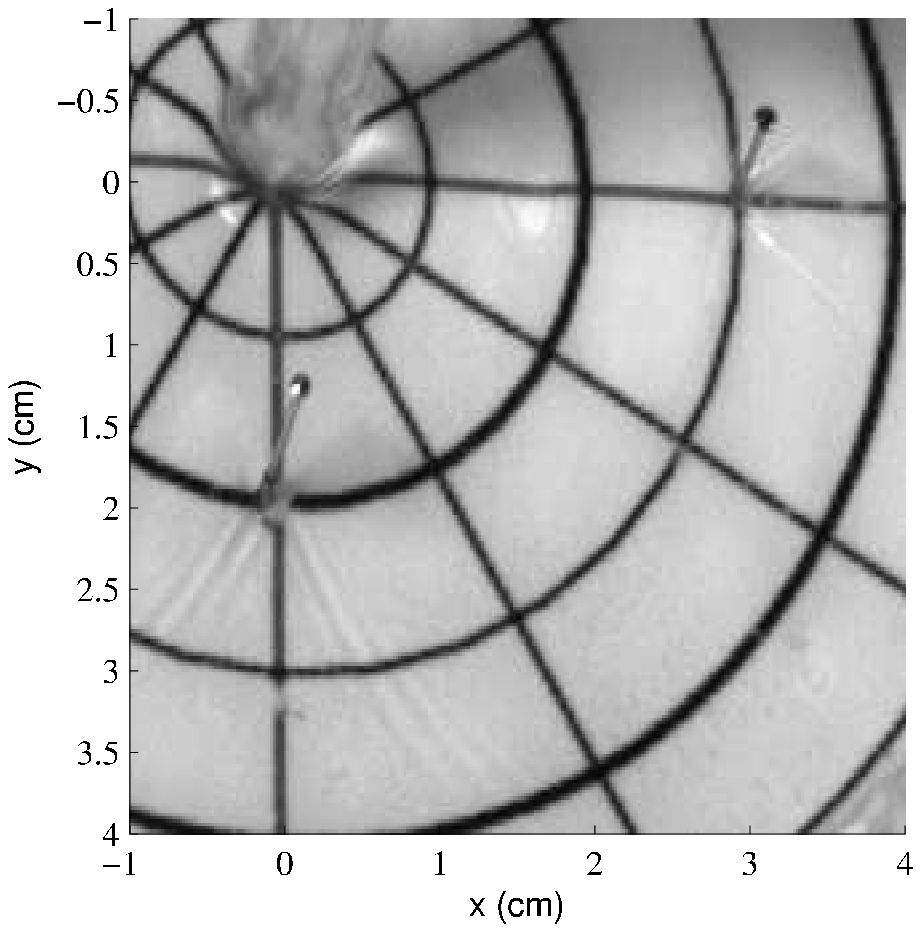} 
\end{flushright}
\end{minipage}
\\
\vspace*{0.05\columnwidth}
\includegraphics[width=0.8\columnwidth,clip]{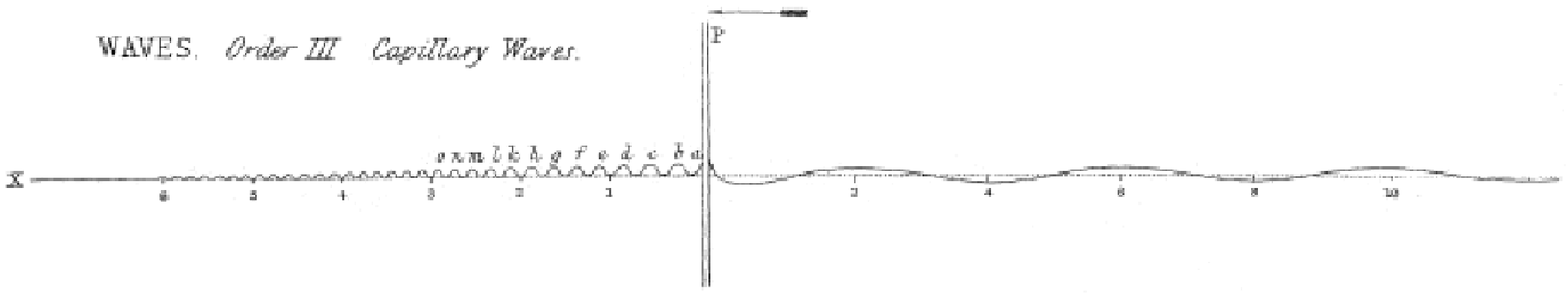} 
\includegraphics[width=0.8\columnwidth,clip]{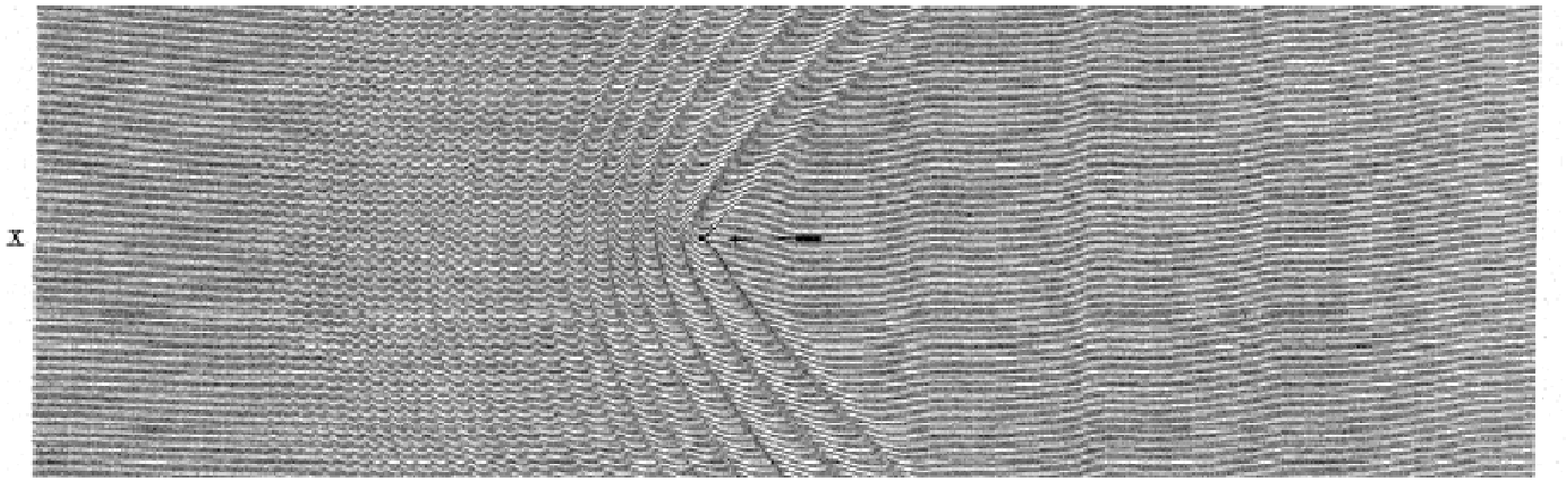} 
\\
\vspace*{0.1\columnwidth}
\includegraphics[width=0.5\columnwidth,clip]{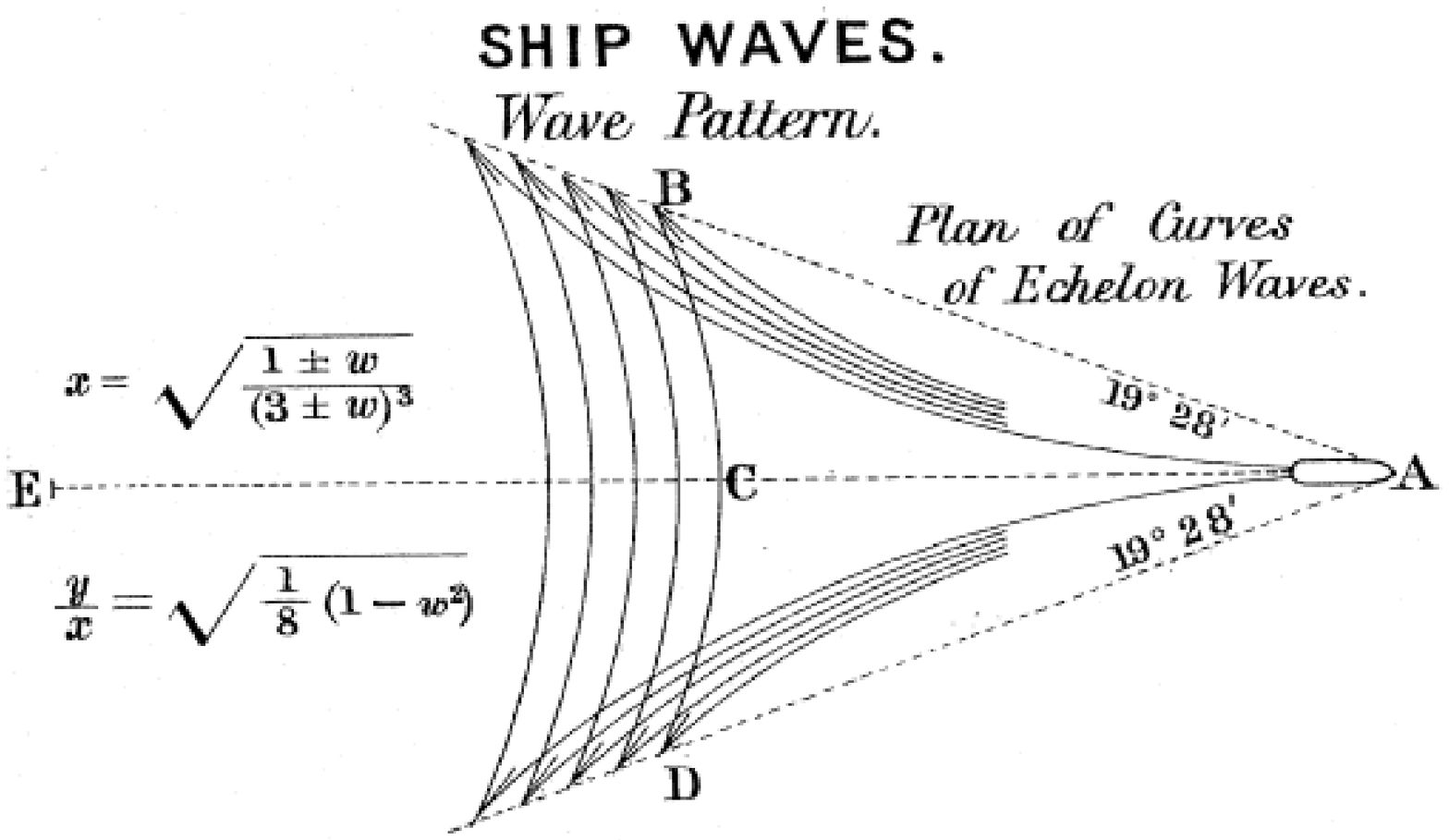} 
\end{center}
\caption{Upper panels:  picture of the Kelvin's ship-wave pattern behind a duck swimming at uniform speed on a quiet lake (left). Photograph courtesy of Fabrice Neyret, ARTIS-CNRS, France. Experimental picture of the Mach cone downstream of a wire immersed in radially flowing silicon oil (upper right). Capillary waves are not visible as they are quickly damped by the larger viscosity of silicon oil. Picture from~\cite{Rousseaux_PRE}.
Experimental picture of the Mach cone downstream of a pin immersed in very shallow flowing water: the surface wave dispersion is supersonic and the height modulation stays outside the Mach cone. Picture courtesy of Silke Weinfurtner (lower right).
Middle panel: Original hand drawing by John Scott Russell~\cite{Russell} of the waves generated by a vertical rod (diameter = 1/16 inch) moving along the water surface with a uniform velocity. The rod moves in the leftward direction: the capillary waves are visible in front of the rod and the gravity waves behind it. A cut of the surface height modulation is shown right above the main drawing.
Lower panel: Original hand drawing by Lord Kelvin of Kelvin's ship-wave pattern~\cite{Kelvin}. The BCD wavefront belongs to the so-called transverse wave pattern. The so-called diverging waves connect the object at A to the dashed lines indicating the edges of the pattern.
}
\label{fig:exp_surface}
\end{figure}

The discussion of the previous sections on the \u Cerenkov effect in classical electromagnetism and on the response of superfluids to moving impurities puts us in the position of getting an easy qualitative understanding of the surface waves that are generated by a duck steadily swimming on the surface of a quiet pond or, equivalently, a fishing line in a uniformly flowing river
\footnote{It is interesting to note that, as in the case of electromagnetic waves, accelerated objects emit surface waves independently from their speed~\cite{Chevy_circ}.}. This system is by far the most accessible from the experimental point of view, but perhaps also the richest one for the variety of different behaviors that can be observed depending on the system parameters. A few examples of experimental pictures are shown in Fig.\ref{fig:exp_surface}.
For the sake of conciseness, we shall restrict ourselves to the case of the water-air interface and restrict to the linear regime of wave propagation described by the model equation \eq{wave_eq}. More complete treatments based on the full hydrodynamic equations including nonlinear effects can be found in the dedicated literature, see e.g.~\cite{whitham,lighthill,Mei,Darrigol,Yih,Torsvik,Soomere,Sorensen,Raphael_deGennes,Chevy}.

\subsection{Dispersion of surface waves}

The dispersion of surface waves on top of a fluid layer of height $h$ and at rest has the form
\begin{equation}
\Omega(k)^2=\left(gk+\frac{\gamma}{\rho} k^3\right)\tanh kh,
\eqname{shallow}
\end{equation}
where $\rho$ is the mass density of the fluid, $g$ is the gravitational acceleration, and $\gamma$ is the surface tension of the fluid-air interface. 

In the simplest case of a {\em deep fluid}, the $\tanh kh$ factor can be approximated with $1$ and the dispersion is characterized by two regions.
For low wavevectors $k\ll k_\gamma$, the dispersion follows the sub-linear square-root behavior
\begin{equation}
\Omega(k)\simeq \pm \sqrt{g k},
\eqname{sqrt}
\end{equation}
of gravity waves, while for large $k\gg k_\gamma$ it is dominated by capillarity effects and has a super-linear growth as
\begin{equation}
\Omega(k) \simeq \pm \sqrt{\frac{\gamma}{\rho}}\, k^{3/2}.
\eqname{capillary}
\end{equation}
The characteristic wavevector scale separating the two regions is fixed by the capillary wavevector
\begin{equation}
k_\gamma = \sqrt{\frac{\rho g}{\gamma}}.
\eqname{k_gamma}
\end{equation}
For the specific case of water/air interface, $k_\gamma\simeq 370\,\textrm{m}^{-1}$, which corresponds to the value  
\begin{equation}
\ell_\gamma=1/k_\gamma= 2.7\cdot 10^{-3}\,\textrm{m}.
\eqname{lambda_gamma}
\end{equation}
for the capillary length.
%

In fluids of finite depth, one can no longer approximate the $\tanh$ in \eq{shallow} with 1. As a result, the dispersion in the low-wavevector region recovers a sonic behavior at low $\kk$'s 
\begin{equation}
\Omega(k)\simeq \pm c_s\,k
\eqname{sonic_2}
\end{equation}
with a speed of sound $c_s = \sqrt{gh}$ proportional to the square root of the fluid depth.
The sign of the first correction to the sonic behavior \eq{sonic_2},
\begin{equation}
\Omega(k)^2\simeq gh k^2 + \left[ \ell_\gamma^2 - \frac{h^2}{3} \right] c_s^2\,k^4
\eqname{water_Bogo}
\end{equation}
critically depends on the depth of the fluid as compared to the capillary length \eq{lambda_gamma}. 
For relatively deep fluids such that $h>\sqrt{3}\ell_\gamma$, the dispersion has a sub-linear behavior, while it recovers a super-linear behavior analogous to the Bogoliubov dispersion \eq{Bogo_disp} for very shallow fluids such that $h<\sqrt{3}\ell_\gamma$.

Independently of the fluid depth $h$, the super-linear behavior of the dispersion $\Omega(k)$ at large $k$ makes the locus $\Sigma$ to be either empty or to consist of a closed curve. 
For very slow sub-sonic motions $v<v_{\rm min}$ (with $v_{\rm min}$ to be defined in the next subsection), the locus $\Sigma$ is empty. For intermediate $c_s>v>v_{\rm min}$ (or infinitely deep fluids, $c_s=\infty$), the locus $\Sigma$ shown in Fig.\ref{fig:surface}(e) consists of a smooth closed curve that does not encircle the origin point $\kk=0$. For super-sonic motions $v>c_s$, the locus $\Sigma$ shown in Fig.\ref{fig:surface2}(b,e) develops a conical \u Cerenkov  singularity at $\kk=0$.

\subsection{Deep fluid}

\begin{figure*}[!htbp]
\begin{center}
\includegraphics[width=0.49\columnwidth]{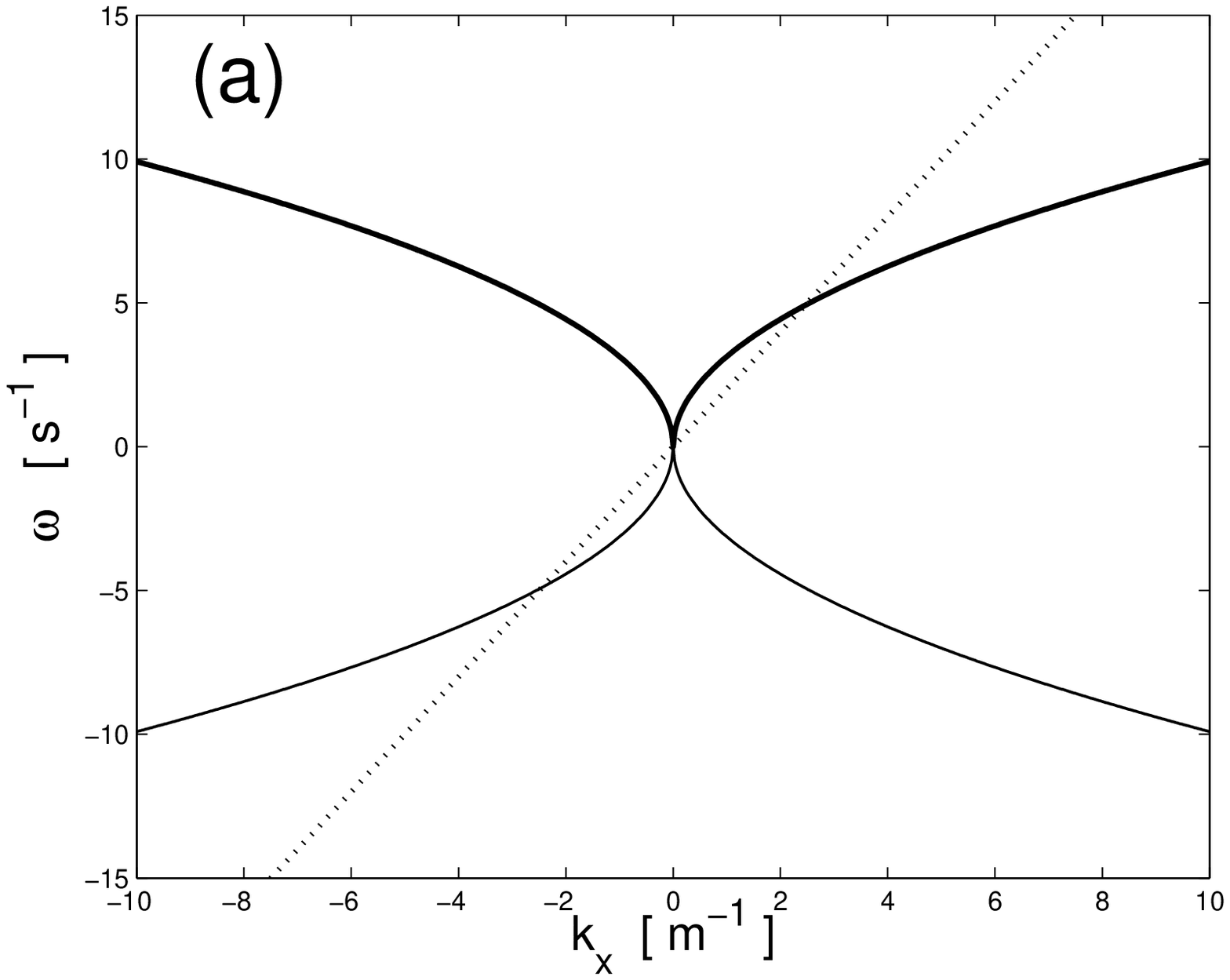}
\includegraphics[width=0.49\columnwidth]{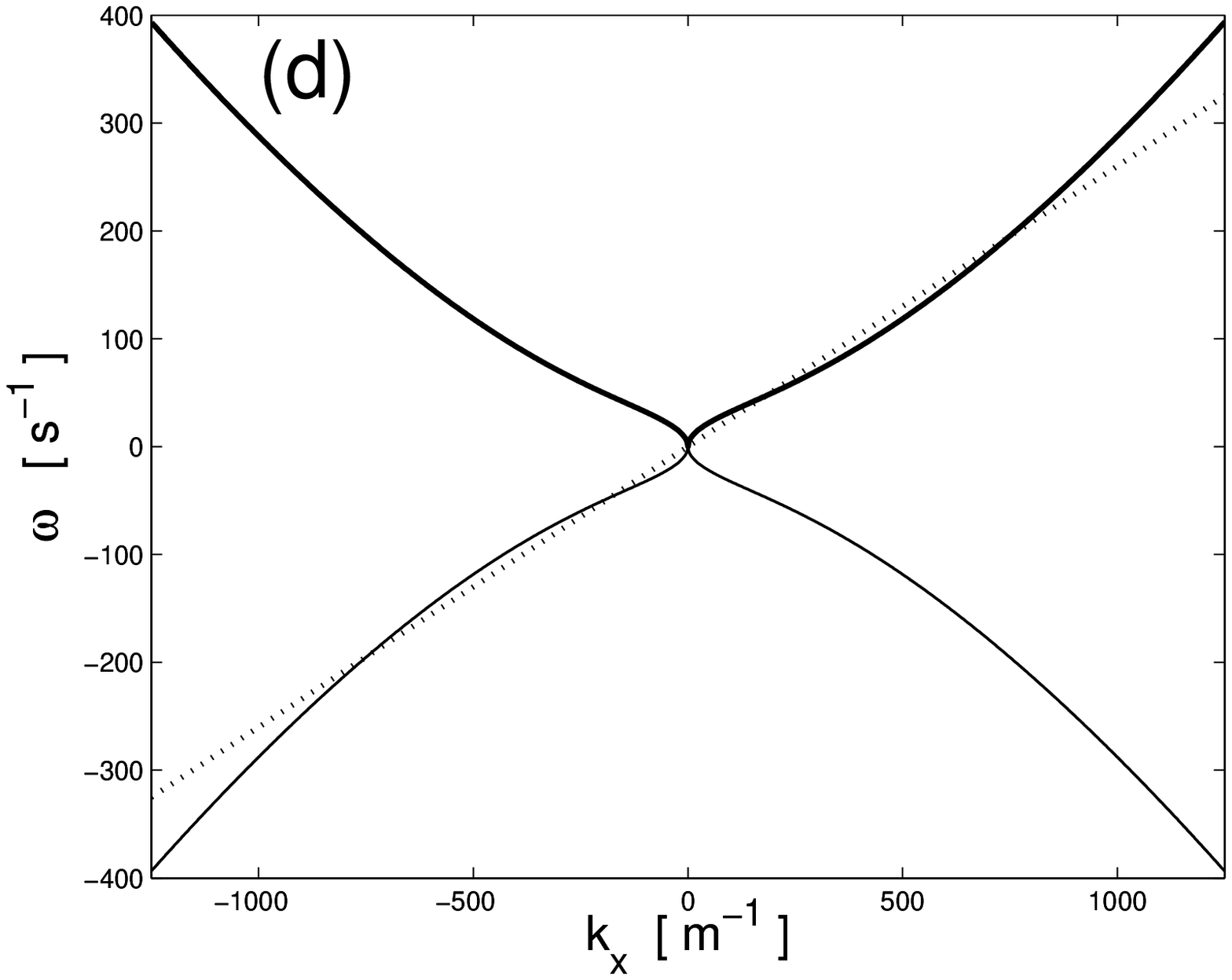}
\\
\includegraphics[width=0.49\columnwidth]{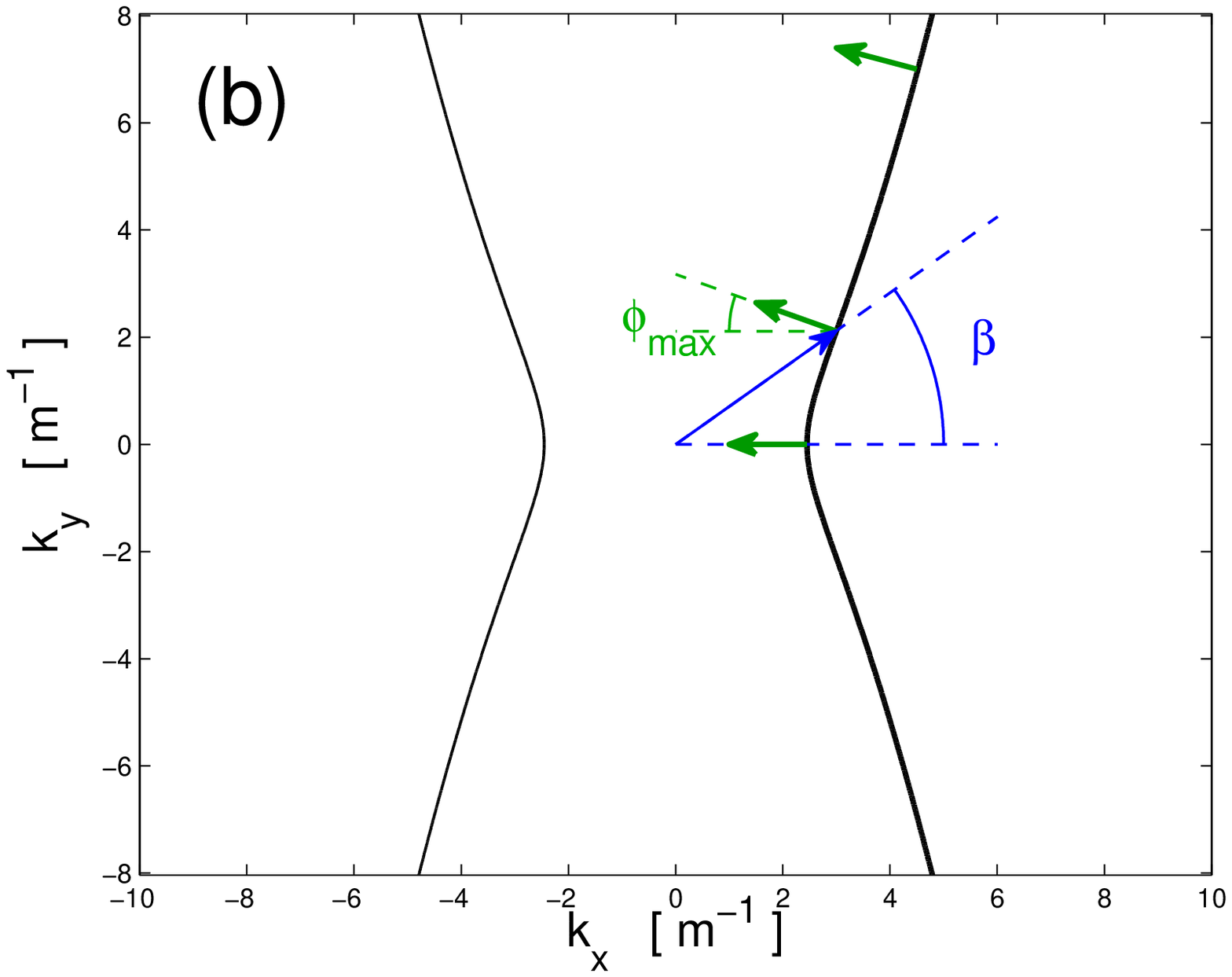}
\includegraphics[width=0.49\columnwidth]{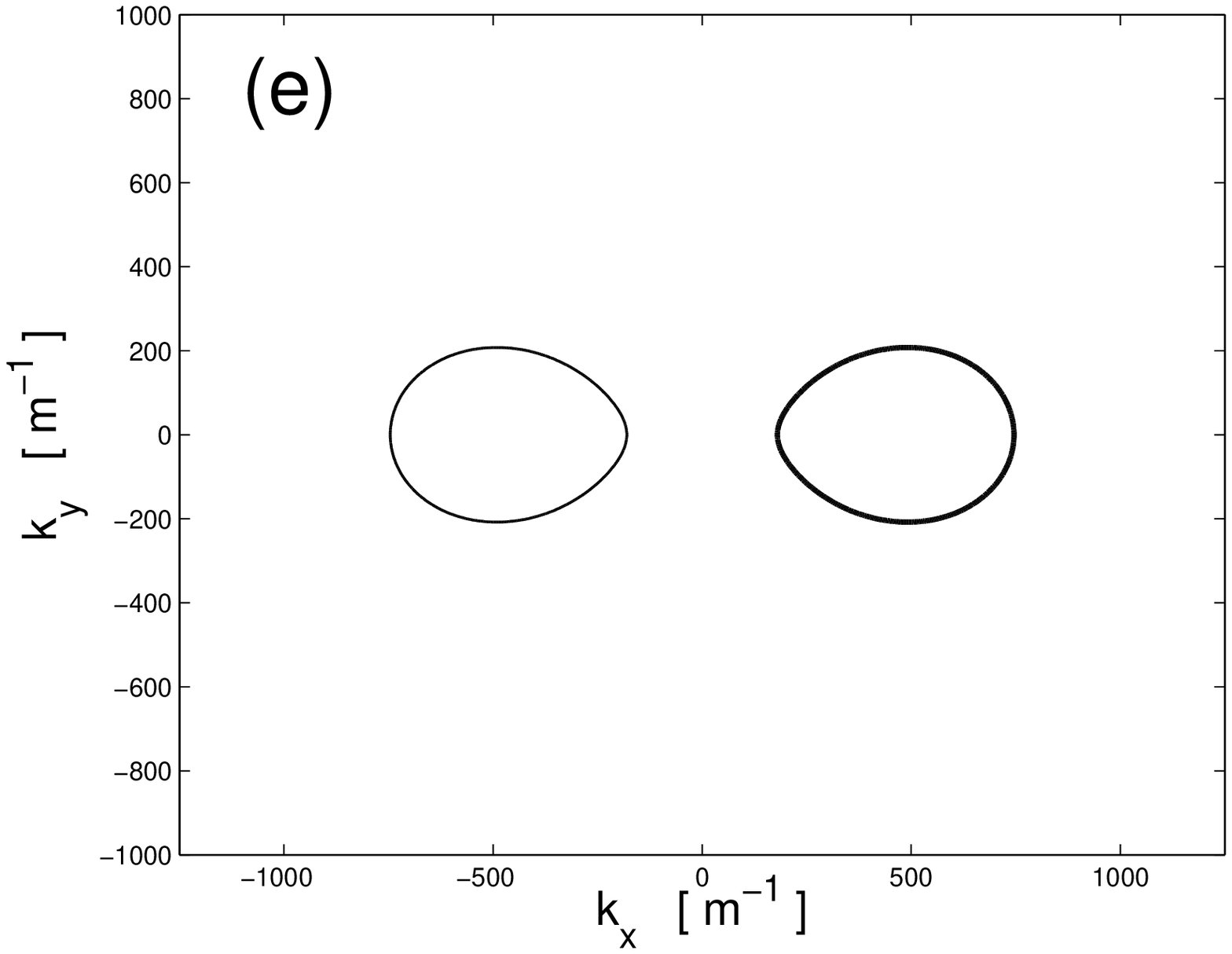}
\\
\includegraphics[width=0.49\columnwidth]{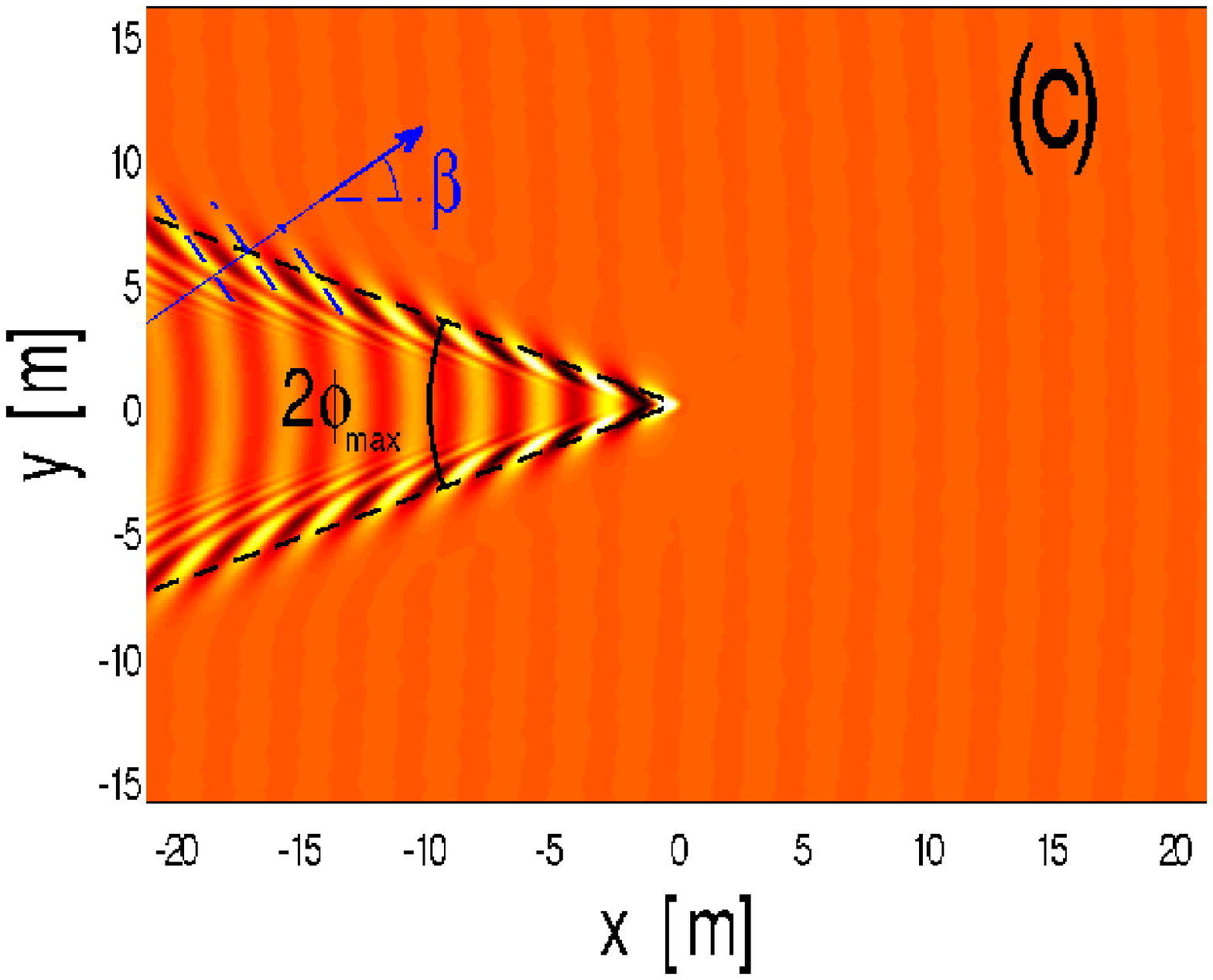} 
\includegraphics[width=0.49\columnwidth]{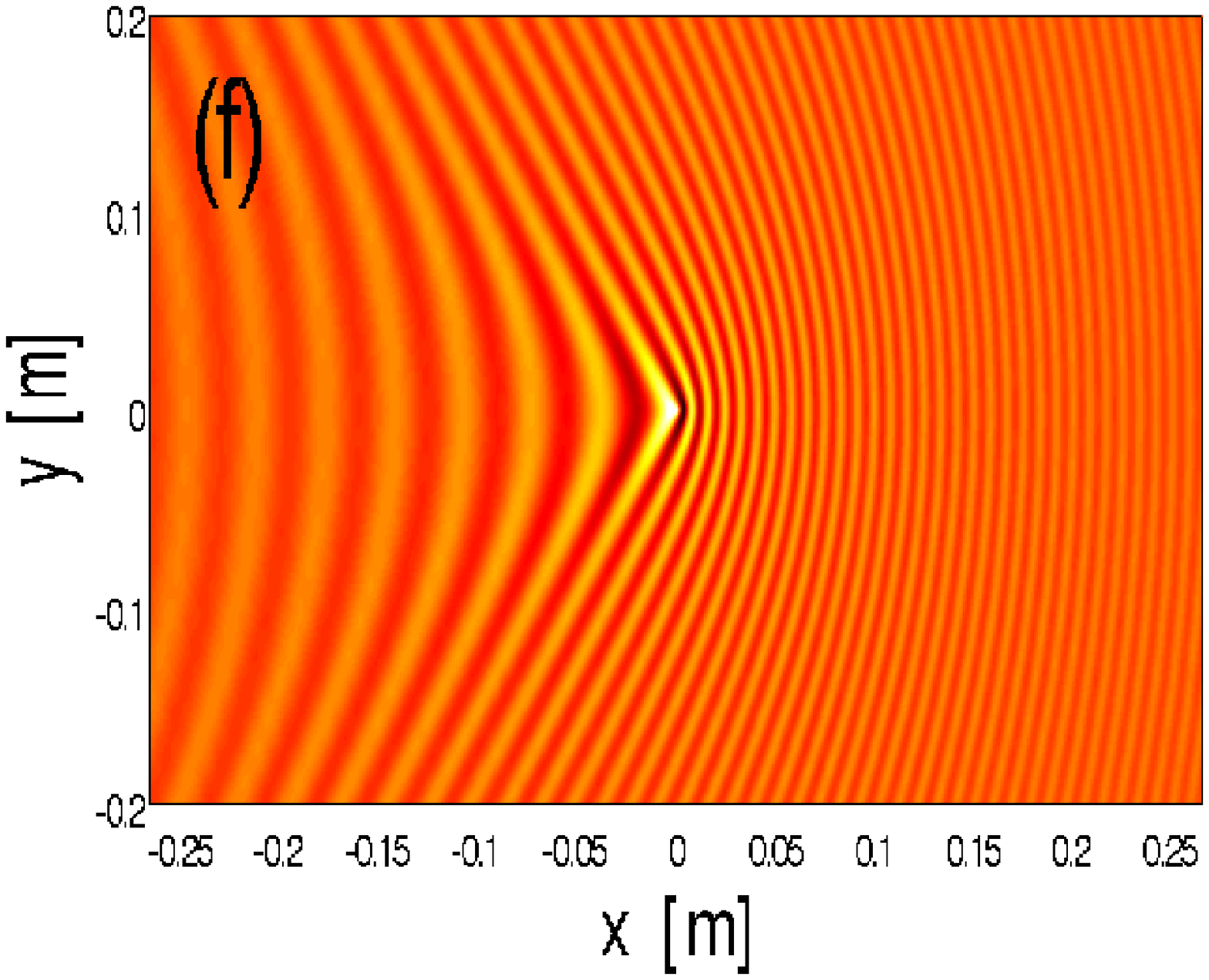} 
\end{center}
\caption{Top row: dispersion of surface wave in the $h=\infty$ deep water limit. The dashed line indicates the $\Omega=\kk\cdot \vv$ plane for generic particle speeds $v=2\,\textrm{m/s}$ (left (a) panel), $v=0.26\,\textrm{m/s}$ (right (d) panel).
Middle row: corresponding shapes of the $\kk$-space locus $\Sigma$ of resonantly excited modes (panels (b,e)); the green arrows indicate the normal to the locus $\Sigma$, that is the direction of the relative group velocity $\vv_g'$.
Bottom row: real space patterns of the surface height modulation (panels (c,f)). These patterns are numerically obtained via a fast Fourier transform of the $\kk$-space perturbation \eq{field_fourier2} using the density and surface tension values of water.
The (c) panel corresponds to the Kelvin's ship-wave pattern behind a duck swimming on a deep lake, a picture of which is shown in the upper left panel of Fig.\ref{fig:exp_surface}. An original sketch by Lord Kelvin is shown in the lower panel of the same figure.}
\label{fig:surface}
\end{figure*}

Let us start by investigating the deep fluid regime $h \to \infty$ for which the sonic speed $c_s \to \infty$.
The structure of the locus $\Sigma$ can be understood by looking at Fig.\ref{fig:surface}(d): for low speeds
\begin{equation}
v<v_{\rm min}=\left(\frac{4g \gamma}{\rho} \right)^{1/4},
\eqname{vmin}
\end{equation}
the locus $\Sigma$ is empty and there is no emission. This critical speed depends on the surface tension of the fluid: for the case of a water/air interface it is equal to $v_{\rm min}\simeq 0.23\,\textrm{m/s}$. The absence of emission corresponds to a vanishing {\em wave resistance} experienced by the slowly moving object which is able to slide with no friction on the surface of the fluid~\cite{Chevy}. Still, the localized deformation of the surface around the object is responsible for a renormalization of the mass of the object~\cite{Raphael_deGennes}.

An efficient emission of surface waves with the associated wave resistance~\cite{Chevy} is suddenly recovered as soon as $v>v_{\rm min}$. In this regime, the locus $\Sigma$ shown in Fig.\ref{fig:surface}(e) has a kind of oval shape~\cite{Raphael_deGennes}, with two intersections with the $k_x$ axis at respectively
\begin{equation}
k=k^{(1,2)}_x=k_\gamma\left[\frac{v^2}{v_{\rm min}^2}\pm 
\sqrt{\frac{v^4}{v_{\rm min}^4}-1} \right].
\eqname{k12}
\end{equation}
As expected, the two solutions merge to $k_x=k_\gamma$ for $v\gtrsim v_{\rm min}$, while at larger $v\gg v_{\rm min}$ they respectively tend to
\begin{eqnarray}
k_x^{(1)}&\simeq & g/v^2 \\
k_x^{(2)}&\simeq & 2k_\gamma v^2/v_{\rm min}^2 :
\end{eqnarray}
the former solution $k_x^{(1)}$ tends to zero in the large $v$ limit and corresponds to almost pure gravity waves, while the latter one $k_x^{(2)}$ quickly diverges as $v^2$ and corresponds to almost pure capillary waves. 

\subsubsection{Fast speed $v\gg v_{\rm min}$ (deep fluid, negligible surface tension)}
Within the $v\gg v_{\rm min}$ limit, we can start our discussion from the low wave vector region $k\ll k_\gamma$, where the waves have a mostly gravity nature, $\Omega(k)\simeq \sqrt{g k}$. Because of the fourth power of $v/v_{\rm min}$ that appears in \eq{k12}, this limit is achieved already for moderate values of $v/v_{\rm min}$ of the order of a few unities. In this region, the locus $\Sigma$ is approximately defined by the condition
\begin{equation}
k^2_y=\frac{v^4 k_x^2 }{g^2}\left(k_x^2-\frac{g^2}{v^4}\right),
\end{equation}
whose shape is plotted in the panel (b) of Fig.\ref{fig:surface}.
The locus $\Sigma$ extends in the $|k_x|\geq k_x^c=g/v^2$ regions: for $k_x\gtrsim k_x^c$, it has the form
\begin{equation}
k_y= \pm \sqrt{\frac{2g}{v^2}\,\left(k_x-k_x^c\right)},
\end{equation}
while for large $k_x\gg k_x^c$, one recovers an asymptotic behavior
\begin{equation}
k_y= \pm \frac{v^2}{g}\,k_x^2.
\end{equation}
A most remarkable feature are the inflection points at  $k^{\rm infl}_x=\pm \sqrt{3/2}\,g/v^2$ where the slope $dk_y/dk_x$ is minimum. At these points the normal to the locus $\Sigma$ makes the maximum angle to the $k_x$ axis, with a value $\phi_{\rm max}$ such that $\tan \phi_{\rm max}=1/\sqrt{8}\simeq 19\degree28'$. This angle determines the aperture of the wake cone behind the moving object: remarkably, this value is universal and does not depend on the speed of the object. A picture of the {\em Kelvin's ship-wave pattern} behind a swimming duck on a quiet lake is shown in the upper left panel of Fig.\ref{fig:exp_surface}; an original hand drawing by Lord Kelvin illustrating this physics is reproduced in the lower panel of the same figure.

Another, related feature that is worth noticing is that for each angle $|\phi|<\phi_{\rm max}$, there exist two points on the locus $\Sigma$ such that the normal to $\Sigma$ makes an angle $\phi$ with the negative $k_x$ axis: according to the theory discussed in Sec.\ref{sec:theory}, these two solutions are responsible for the two interpenetrating fringe patterns: the so-called {\em transverse} waves with a small $k_y$ and the so-called {\em diverging} waves with large $k_y$. 

The transverse waves are clearly visible as the long wavelength modulation right behind the object along the axis of motion:
their wavevector $k_x=k_x^c=g/v^2$ is determined by the intersection of the locus $\Sigma$ with the $k_x$ axis. Remarkably, the faster the object is moving, the smaller is the wavevector $k_x^c$. On Kelvin's hand drawing of Fig.\ref{fig:exp_surface}, the wavefront passing by point C belongs to the transverse wave pattern. 

The diverging waves are easily identified in the hand drawing as the wavefronts with opposite curvature connecting the source at A with the edge of the wake pattern where the two patterns collapse onto each other.
The direction of the peculiar fringe modulation of the edge of the wake [indicated by the blue dashed lines on Fig.\ref{fig:surface}(c)] is determined by the wavevector $\kk^{\rm infl}$ of the inflection point of the $\kk$-space locus $\Sigma$: the orientation of $\kk^{\rm infl}$ fixes the angle $\beta$ to a value such that $\tan \beta=\left.k_y/k_x\right|_{\rm infl}=1/\sqrt{2}$, i.e. $\beta \simeq 35\degree$.

Of course, a complete treatment of the wake would require including the capillary waves at very high wavevector $k\approx k_x^{(2)} \gg k_\gamma$, i.e. the part of the locus $\Sigma$ that closes the curve at large $\kk$'s outside the field of view of Fig.\ref{fig:surface}(b). However, the amplitude in these short-wavelength modes is quickly damped by viscous effects, so their contribution to the observable pattern turns out to be irrelevant in most practical cases.

\subsubsection{Moderate speed $v\gtrsim v_{\rm min}$ (deep fluid, significant surface tension)}

For moderate speeds $v\gtrsim v_{\rm min}$, surface tension effects are no longer negligible and all points of the $\kk$-space locus $\Sigma$ contribute to the real-space pattern. In particular, the locus $\Sigma$ is a closed curve that does not encircle the origin, as shown in Fig.\ref{fig:surface}(e): the two intersections with the $k_x$ axis at $k_x=k_x^{(1,2)}$, corresponding to gravity and capillary waves propagate with relative group velocities directed in opposite directions from the fishing line of the celebrated experiment by Thomson. The pattern of long-wavelength gravity waves is located downstream of the fishing line, while the short-wavelength capillary waves are located in the upstream region, see Fig.\ref{fig:surface}(f) and the drawing by J. S. Russell reproduced in the middle panel of Fig.\ref{fig:exp_surface}. 

An approximate analytical understanding of this pattern can be obtained by approximating the locus $\Sigma$ of Fig.\ref{fig:surface}(e) with a pair of circles analogously to the case of a parabolic dispersion discussed in Sec.\ref{sec:parabolic} and shown in Fig.\ref{fig:pol}(g-i): within this approximation, the shape of the wavefronts consists a system of hyperbolas, with a closer spacing ahead of the object. The qualitative agreement of the hyperbolic wavefronts of Fig.\ref{fig:pol}(i) with the full calculations shown in 
Fig.\ref{fig:surface}(f) is manifest.

\subsubsection{Effect of the source {\em structure factor}}
To complete the discussion, it is worth mentioning that the emission of waves can be hindered  by the source  {\em structure factor} $\tilde{S}(\kk)$ even at large $v>v_{\rm min}$. For example, the emission of surface waves will be strongly suppressed if the size $\ell$ of the source term (modelled as a Gaussian-shaped potential $S(\rr)$) is large enough to have $k \ell \gg 1$ for all points on $\Sigma$. For instance, for an object of typical size $\ell=30$~cm, the argument such that $k_x^{(1)} \ell \leq 1$ imposes a lower critical speed $v\geq v_c^{\rm size}=3\,\textrm{m/s}$ to the emission of gravity waves. A similar argument for capillary waves was mentioned to explain the characteristic swimming speed of some floating insects~\cite{insects}.

\subsection{Shallow fluid}
\label{sec:shallow}

\begin{figure*}[!htbp]
\begin{center}
\includegraphics[width=0.49\columnwidth]{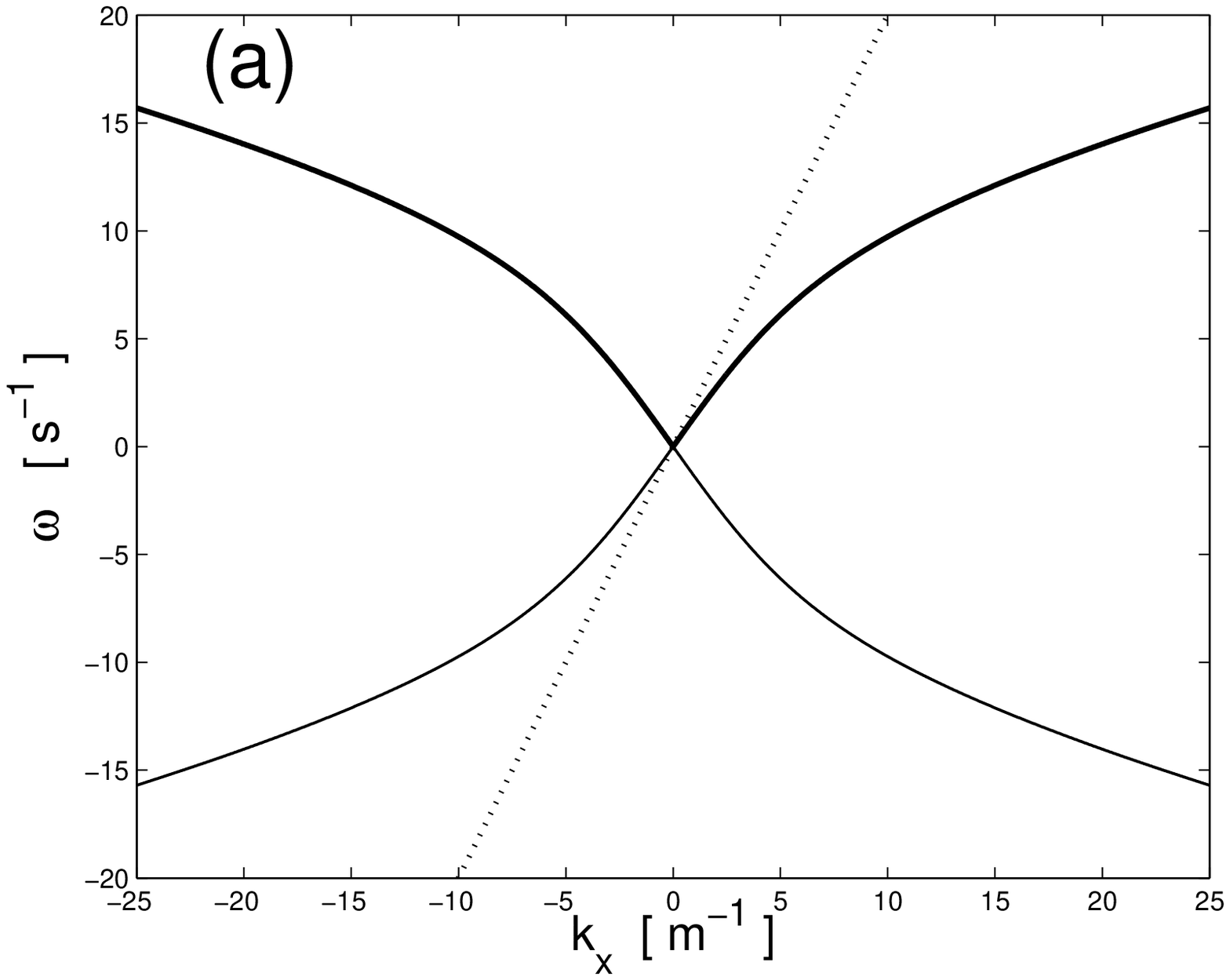}
\includegraphics[width=0.49\columnwidth]{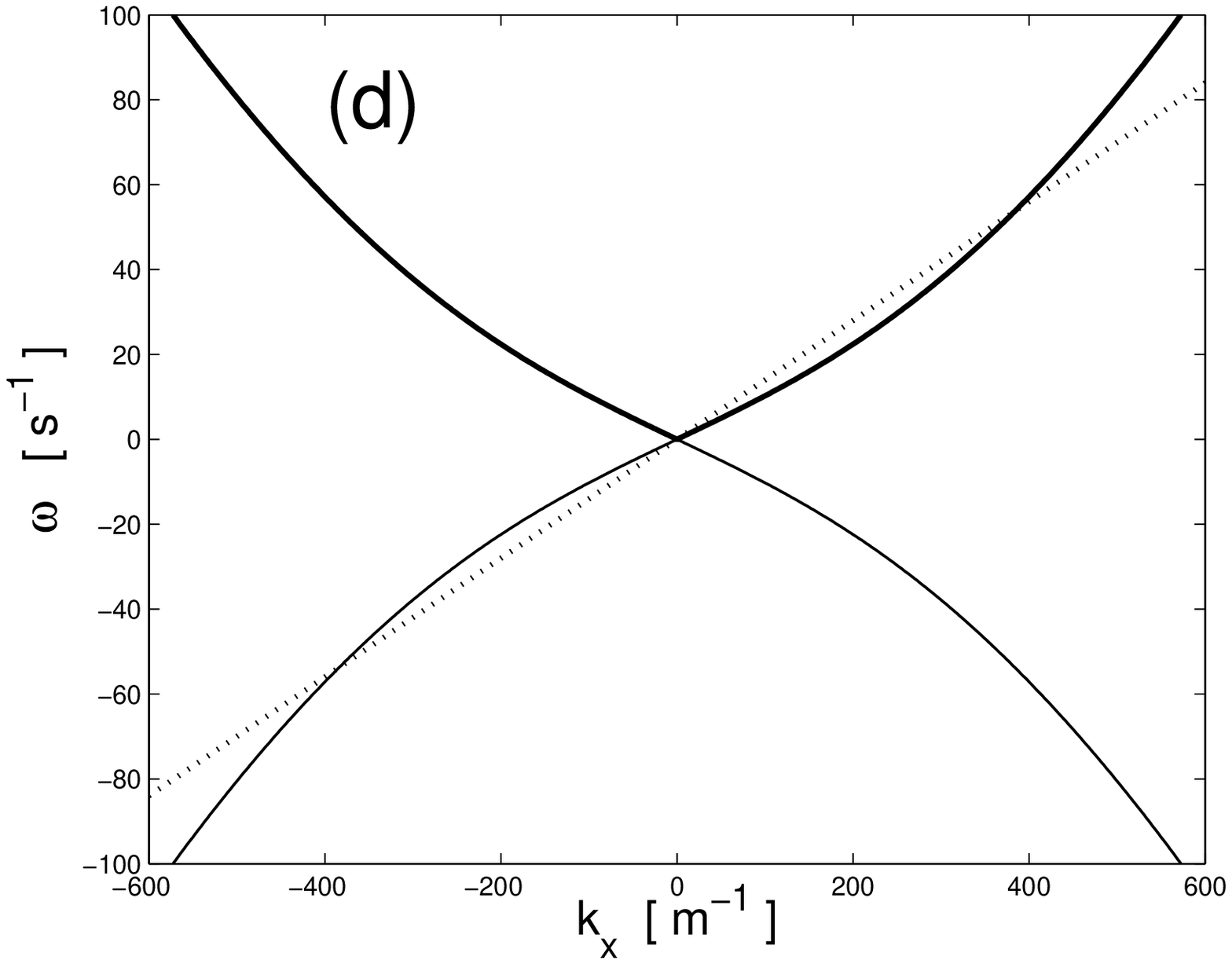} 
\\
\includegraphics[width=0.49\columnwidth]{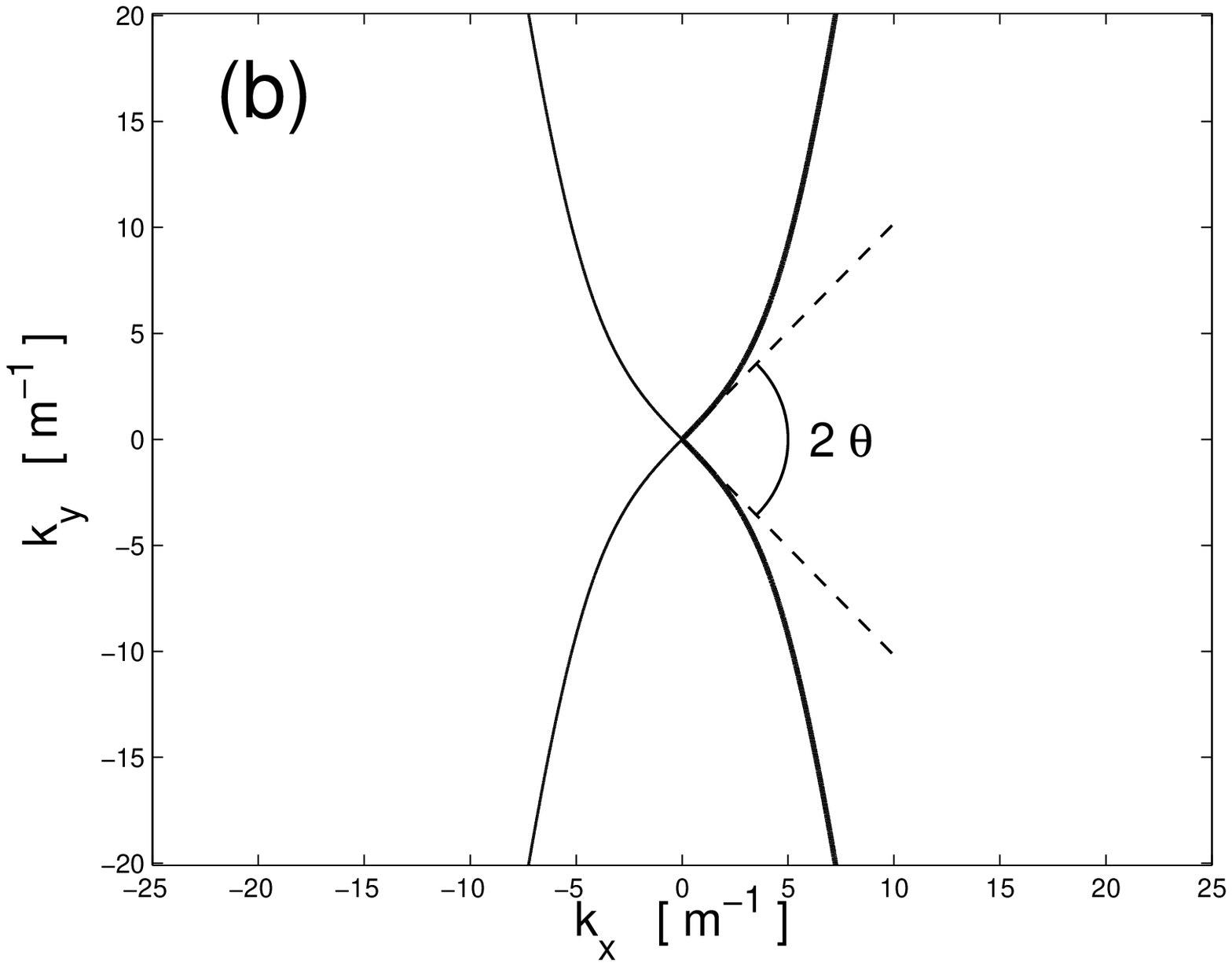}
\includegraphics[width=0.49\columnwidth]{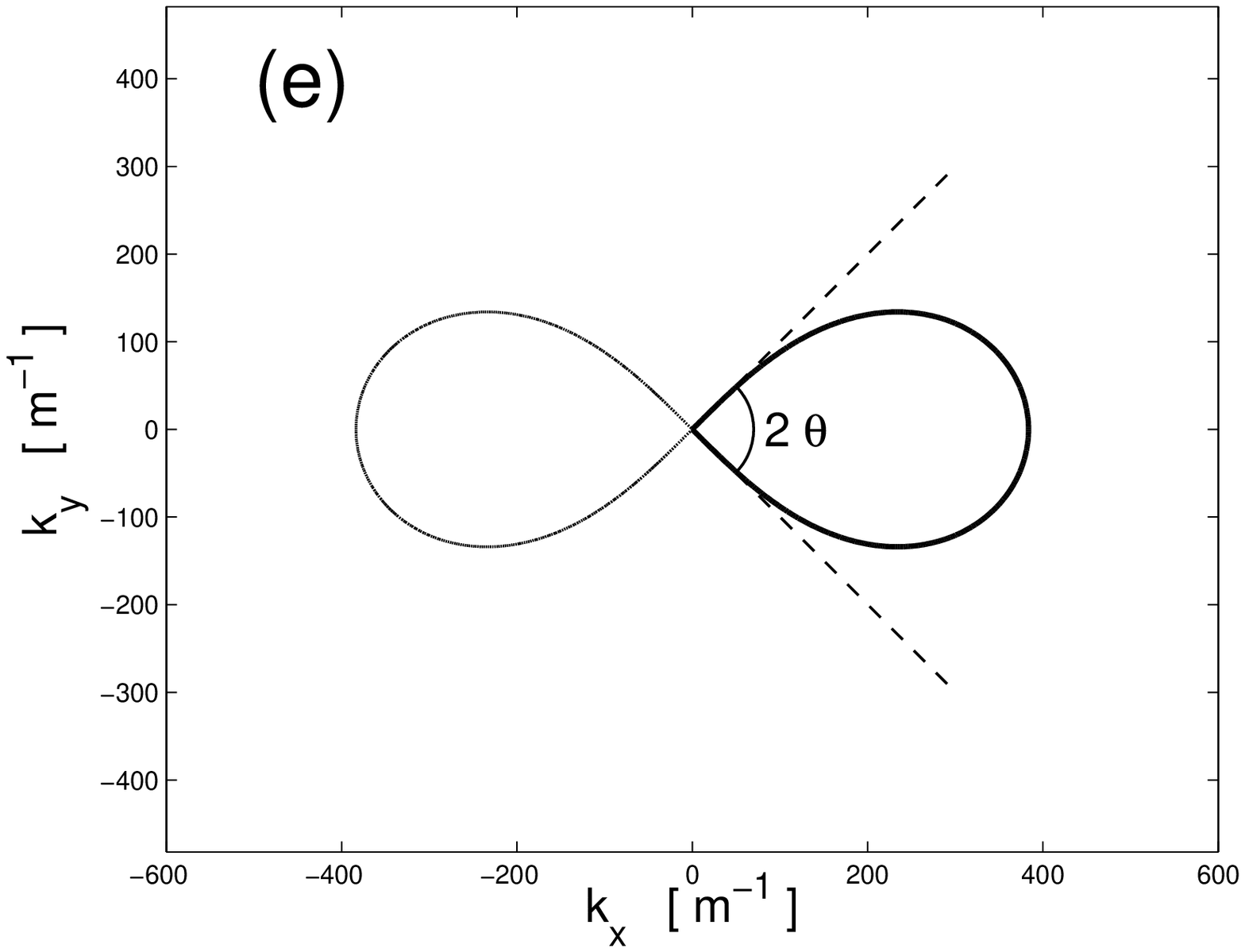} 
\\
\includegraphics[width=0.49\columnwidth]{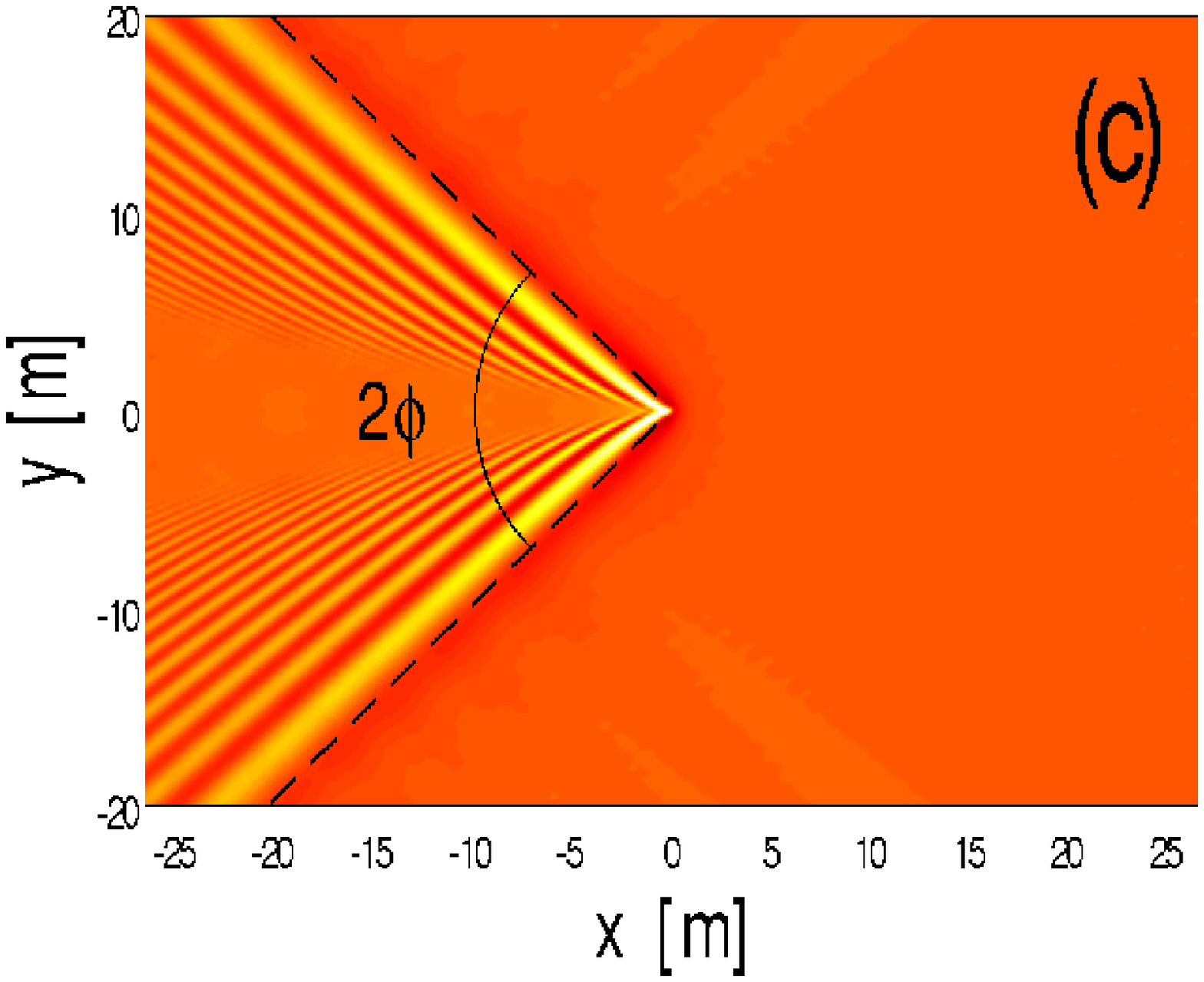} 
\includegraphics[width=0.49\columnwidth]{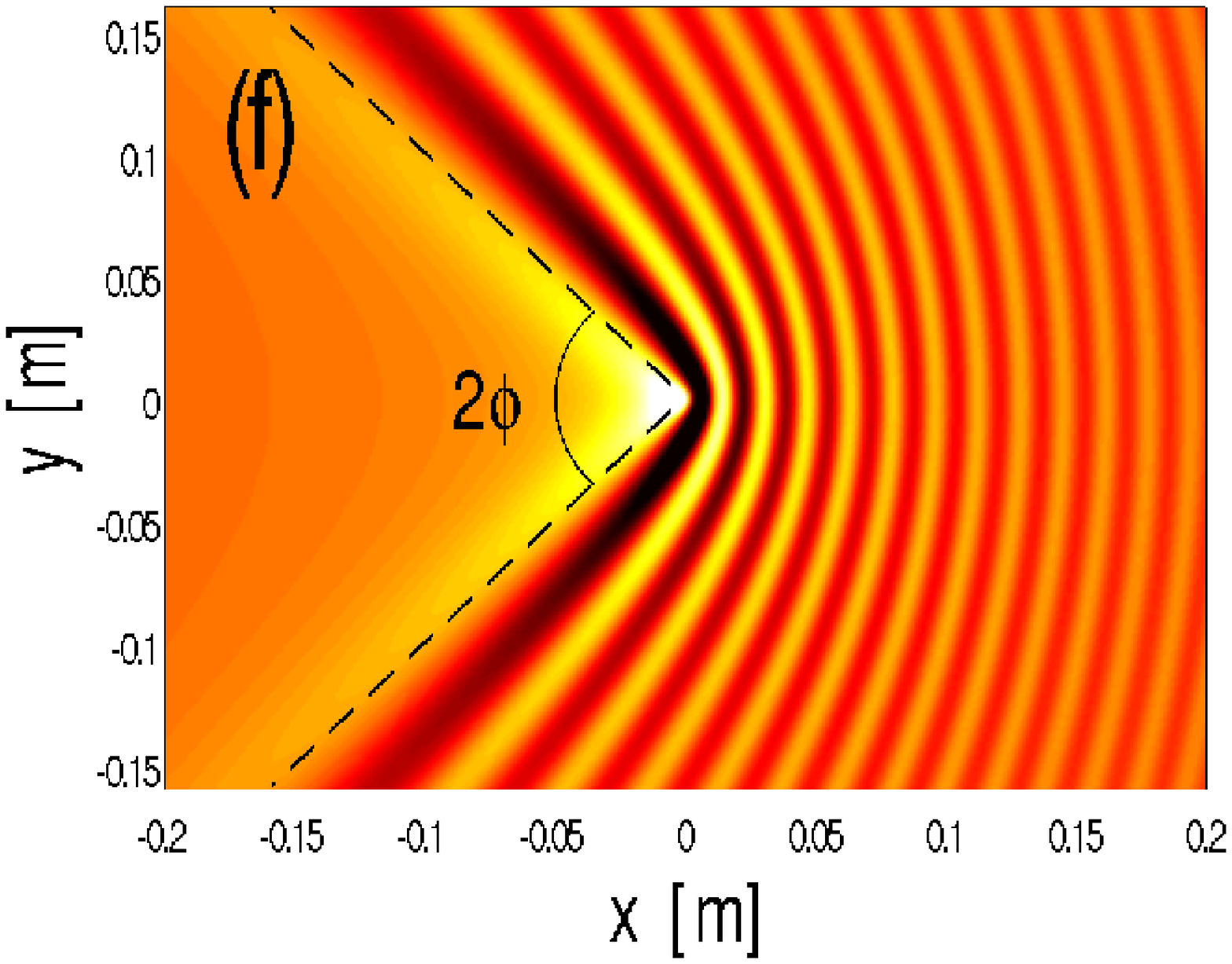}
\end{center}
\caption{
Top row: dispersion of surface wave on shallow water of height $h=0.2\,\textrm{m}$ (left (a) panel) and $h=0.001\,\textrm{m}$ (right (d) panel). The dashed line indicates the $\Omega=\kk\cdot \vv$ plane for generic particle speeds $v=2\,\textrm{m/s}$ (left (a) panel), $v=0.14\,\textrm{m/s}$ (right (d) panel).
Middle row: corresponding shapes of the $\kk$-space locus $\Sigma$ of resonantly excited modes (panels (b,e)). The dashed lines indicate the \u Cerenkov cone in the low wavevector region $k\xi \ll 1$.
Bottom row: real space patterns of the surface height modulation (panels (c,f)). The black dashed lines indicate the Mach cone. These patterns are numerically obtained via a fast Fourier transform of the $\kk$-space perturbation \eq{field_fourier2} using the density and surface tension values of water.
The left panels correspond to a case where $h>\sqrt{3}\ell_\gamma$ and the lowest-order correction to the sonic dispersion \eq{water_Bogo} is sub-linear. The right panels correspond to a case where $h<\sqrt{3}\ell_\gamma$ and the lowest-order correction has the same super-linear behavior as the Bogoliubov dispersion \eq{Bogo_disp} illustrated in Fig.\ref{fig:Bogo}.
}
\label{fig:surface2}
\end{figure*}

When the wavelength of the perturbation is longer than the depth $h$ of the fluid, the $\tanh(k h)$ term in the dispersion begins to be important and causes a radical change in the structure of the locus $\Sigma$. The left and right columns of Fig.\ref{fig:surface2} illustrate the two regimes $h > \sqrt{3}\,\ell_\gamma$ and $h < \sqrt{3}\,\ell_\gamma$ where the first correction to the sonic dispersion has either a sub-linear or a super-linear nature.

\subsubsection{Small surface tension (sub-luminal dispersion)}
\label{sec:small_surf}

We start here from the case where the surface tension is small enough to have $h > \sqrt{3}\,\ell_\gamma$. Depending on the speed $v$ of the object, several regimes can be identified.
For very low speeds $v<v_{\rm min}$, the locus $\Sigma$ is empty and there is no perturbation to the fluid. For intermediate speeds $v_{\rm min} < v \ll \sqrt{gh}$, the shape of the locus $\Sigma$ is determined by the high-$\kk$ capillary region of the dispersion and is almost unaffected by the finite height $h$ of the fluid:  as in the deep fluid limit, $\Sigma$ consists of closed, egg-shaped smooth curve and the real-space pattern again resembles a system of hyperbolas extending to the whole space, as illustrated in Fig.\ref{fig:surface}(d-f).  For increasing, yet sub-sonic speeds $v<c_s=\sqrt{gh}$, gravity recovers an important role, while the finite height $h$ keeps providing only a small correction to the deep water behavior illustrated in Fig.\ref{fig:surface}(a-c).

The situation is completely different for supersonic speeds $v>c_s$ [Fig.\ref{fig:surface2}(a-c)]: in this case, the locus $\Sigma$ starts from the origin, where it exhibits a conical singularity as a result of the sonic dispersion. At larger $\kk$, the locus $\Sigma$ recovers a shape similar to the infinitely deep fluid case: the sub-linear growth of the dispersion with $k$ is responsible for the fast increase of $k_y$ as a function of $k_x$. Of course, the super-sonic dispersion of capillary waves at very large $\kk$ [well outside the field of view of Fig.\ref{fig:surface2}(a,b)] makes the locus $\Sigma$ to close on itself. However, as already mentioned, these short-wavelength waves are quickly attenuated and hardly visible.

The singularity of $\Sigma$ at the origin $\kk=0$ is responsible for the Mach cone and the disappearance of the transverse wave pattern, as shown in Fig.\ref{fig:surface2}(c). As usual for sonic dispersions, the aperture of the Mach cone [indicated by the dashed line on Fig.\ref{fig:surface2}(c)] depends on the source speed $v$ as $\sin \phi=c_s/v$: on the $\Sigma$ locus shown in Fig.\ref{fig:surface2}(b), this corresponds to the fact that the normal to $\Sigma$ starts at a finite angle $\phi$ with the negative $k_x$ axis for $\kk=0$. For growing $\kk$'s, the angle monotonically decreases to $0$ meaning that the perturbation is restricted to the spatial region {\em inside} the Mach cone. The absence of the transverse wave pattern is clearly visible in Fig.\ref{fig:surface2}(c) as the absence of modulation along the negative $x$ direction right behind the object.

\subsubsection{Shallow one-dimensional channel}

The restriction of this model to a one-dimensional geometry provides interesting insight on the physics of long wavelength surface waves in a spatially narrow channel of width $W$. Spatial confinement along the orthogonal direction (say $y$) makes the corresponding wavevector  to be quantized in discrete values determined by the boundary conditions at the edges of the channel, $k_y=\pi p/W$ with the integer $p=0,1,2,\ldots$. For simplicity, we assume that the transverse shape of the source (e.g. a boat sailing along the channel) is broad enough to only excite the lowest mode at $k_y=0$, corresponding to a transversally homogeneous wave.

Neglecting for simplicity also capillarity effects, a generalized Landau criterion for one-dimensional gravity waves anticipates that a uniformly moving object can emit  $k_y=0$ gravity waves only if its velocity is slower than a maximum velocity 
\begin{equation}
v_{\rm max}=\max_{k_x} \left[\frac{\Omega(k_x,k_y=0)}{k_x}\right] = c_s=\sqrt{gh}.
\end{equation}
This feature is easily understood looking at the $k_y=0$ cut of the dispersion shown in Fig.\ref{fig:surface2}(a): for sub-sonic speed $v<c_s$, the $\Omega=k_x v$ straight line corresponding to the \u Cerenkov condition \eq{cerenkov} has a non-trivial intersection with the dispersion law, while the intersection reduces to the irrelevant $k_x=0$ point for supersonic speeds $v>c_s$. In this case, no modes are any longer available for the emission, which reflects into a marked decrease of the {\em wave drag} friction experienced by the object. The observation of an effect of this kind when a ship travels at sufficiently fast speed long a channel was first reported by Scott Russel and often goes under the name of Houston paradox in the hydrodynamics and naval engineering literature~\cite{Darrigol}. 

As compared to the standard Landau criterion for superfluidity, it is interesting to note that the condition on the object speed to have a (quasi-)frictionless flow is here reversed: friction is large at slow speeds and suddenly drops for $v>c_s$. This remarkable difference is due to the different sub-sonic rather than super-sonic shape of the gravity wave dispersion with respect to the Bogoliubov one. Of course, this suppression of friction is less dramatic when also higher $p>0$ transverse modes of the channel can be excited and a richer phenomenogy can be observed~\cite{channel}. As we have previously discussed at length, in a transversally unlimited geometry the transition from sub-sonic to super-sonic speeds  manifests itself as  the disappearance of the transverse wave pattern from Kelvin's wake and a corresponding sudden but only partial decrease of the friction force.

\subsubsection{Large surface tension (super-luminal dispersion)}

In the opposite regime of large surface tension $h < \sqrt{3}\, \ell_\gamma$, the super-linear form of the dispersion \eq{water_Bogo} makes the physics to closely resemble the behavior of impurities in a dilute superfluid discussed in Sec.\ref{sec:superfluid}.
For a slowly moving object at $v<c_s$, the locus $\Sigma$ is empty and the perturbation of the surface remains localized in the vicinity of the impurity. 
For a fast moving object at $v>c_s$, the locus $\Sigma$ and the wake pattern closely resemble the corresponding ones for the case of a supersonically moving impurity in a superfluid shown in Fig.\ref{fig:Bogo}: a Mach cone of aperture $\sin\phi=c_s/v$ located behind the object [indicated by the dashed line on Fig.\ref{fig:surface2}(f)] and a series of curved wavefronts ahead of the object. The most significant difference with the $h > \sqrt{3}\, \ell_\gamma$ case of Sec.\ref{sec:small_surf} is the position of the modulation with respect to the Mach cone: in the sub-linear case of panel (c), it lies within (i.e. behind) the Mach cone, while in the super-linear case of panel (f), it stays outside (i.e. in front of) the Mach cone.

\section{\u Cerenkov processes and the stability of analog black/white holes}
\label{sec:gravity}

The systems that were considered in the previous sections are presently among the most promising candidates for the realization of condensed matter analogs of gravitational black (or white) holes: the key idea of analog models is to tailor the spatial structure of the flow in a way to obtain a horizon surface that waves can cross only in one direction. Upon quantization, a number of theoretical works have predicted that a condensed matter analog of Hawking radiation should be emitted by the horizon. A complete review of this fascinating physics can be found in the other chapters of the book. In this last section, we shall review some consequences of \u Cerenkov processes that are most significant for the stability of analog black and white holes based on either flowing superfluids or surface waves on flowing water. 

The role of \u Cerenkov-like emission processes in the dynamics of the strong optical pulses that are used in optical analog models based on nonlinear optics~\cite{faccio} is still in the course of being elucidated and interesting experimental observations in this direction have recently appeared~\cite{faccio2}. Here, it is important to remind that, differently from the all-optical \u Cerenkov radiation experiments of~\cite{dipolar_nonlopt_cherenk,fs_nonlopt_cherenk} where an effective moving dipole was generated by $\chi^{(2)}$ nonlinearity, the analog models of~\cite{faccio} are based on the time- and space-dependent effective refractive index profile due to a $\chi^{(3)}$ optical nonlinearity: given the centro-symmetric nature of the medium under examination, no effective moving dipole can in fact appear unless the medium shows some material imperfection.

\subsection{Superfluid-based analog models}

\begin{figure}[!htbp]
\begin{center}
\includegraphics[width=0.9\columnwidth,clip]{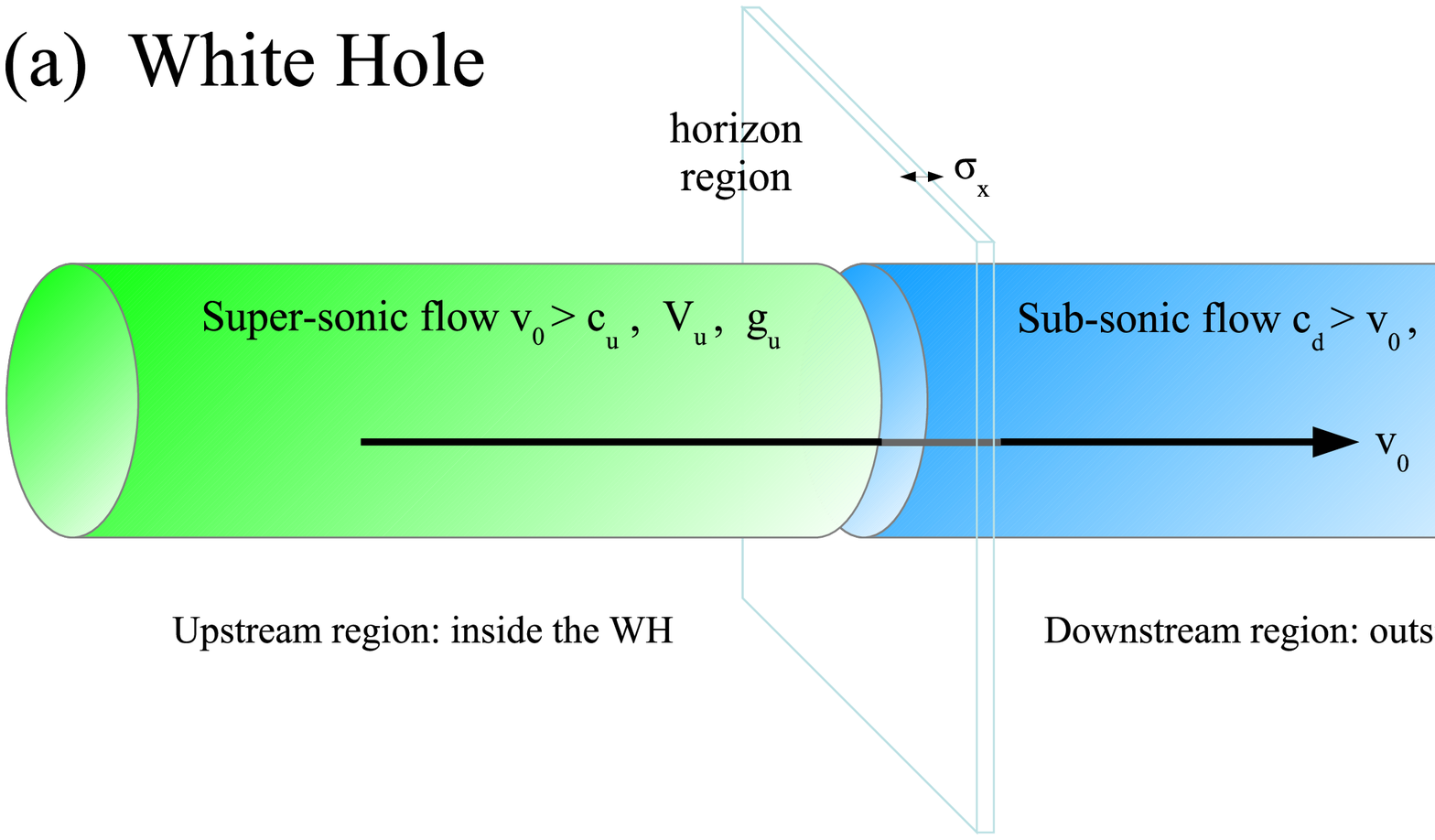}
\includegraphics[width=0.9\columnwidth,clip]{figura_WH_dispersion.eps}\\
\vspace{1cm}
\includegraphics[width=0.9\columnwidth,clip]{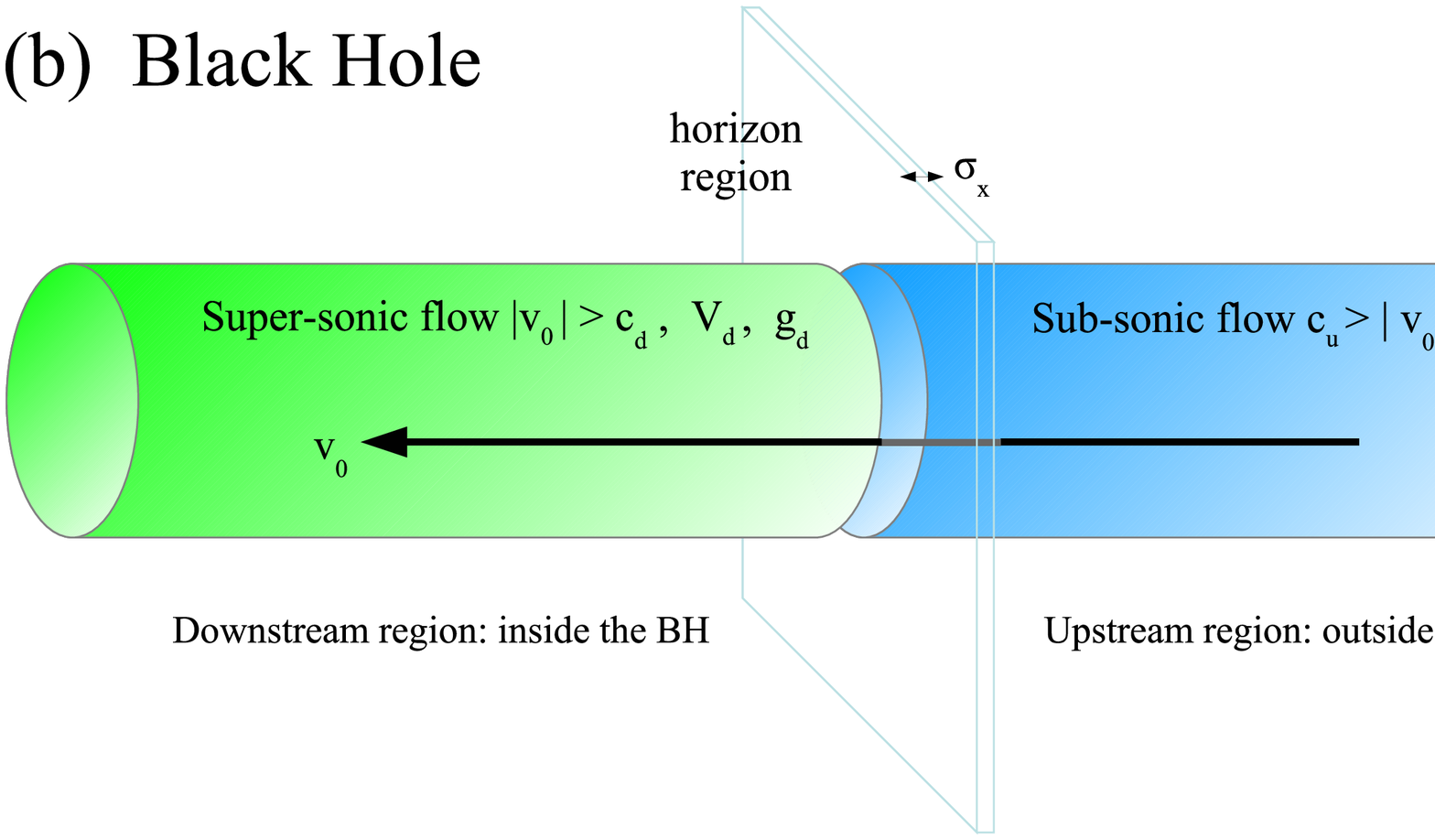}
\includegraphics[width=0.9\columnwidth,clip]{figura_WH_dispersion_BH.eps}
\end{center}
\caption{Main panels: sketch of the flow geometry for white (a) and black (b) hole configurations based on superfluids. Smaller panels (a1,a2,b1,b2): dispersion of Bogoliubov excitations in the asymptotic regions far from the horizon. \u Cerenkov emission is only possible for white hole configurations: the corresponding {\em zero mode} is indicated in blue in (a1).
Figure from~\cite{noiWH}.}
\label{fig:BHWH_superfluid}
\end{figure}

Let us start from the simplest case of analog black/white holes based on flowing superfluids for which a complete theoretical understanding is available~\cite{noiWH,noiBH,ParentaniBHWH}. In a one dimensional geometry, the horizon consists of a point separating a region of sub-sonic flow from a region of super-sonic flow. In a black hole the sub-sonic region lies upstream of the horizon, while in a white hole the sub-sonic region lies downstream of the horizon.  A sketch of both configurations is reproduced in Fig.\ref{fig:BHWH_superfluid} together with the Bogoliubov dispersion of excitations as observed from the laboratory frame: in the most common configurations, the flow has a non-trivial structure only in a small region around the horizon and recovers a homogeneous shape with space-independent density and speed farther away from the horizon. In the laboratory reference frame (corresponding to the rest frame of the impurity), the \u Cerenkov emission occurs in the zero-frequency Bogoliubov modes, the so-called {\em zero modes}.

As we have discussed in detail in Sec.\ref{sec:superfluid}, the super-linear nature of Bogoliubov dispersion restricts \u Cerenkov emission processes to super-sonic flows, where they generate waves that propagate in the upstream direction. In the geometry under consideration here, the flow is everywhere smooth exception made for the horizon region. Combining these requirements immediately rules out the possibility of \u Cerenkov emission in black hole configurations: the group velocity of the zero mode waves emitted at the horizon points in the direction of the sub-sonic region, where it can no longer be supported. This simple kinematic argument contributes to explaining the remarkable dynamical stability of acoustic black hole configurations, as observed in numerical simulations of their formation starting from a uniformly moving fluid hitting a localized potential barrier~\cite{Pavloff_Kamchatnov}.

In contrast, white hole configurations are much more sensitive to the dissipation of energy via \u Cerenkov processes: Bogoliubov excitations can appear in the supersonic region upstream of the horizon and give rise to significant modulations of the density and flow speed, the so-called {\em undulation} patterns. Several reasons make such processes to be potentially harmful to the study of quantum features of the white hole radiation.
To the best of our knowledge, the only known realistic scheme to generate a white hole configuration in a flowing superfluid is the one of~\cite{noiWH} using a simultaneous spatial and temporal modulation of both the atom-atom interaction strength and the external confining potential.
The main difficulty of this configuration is that it requires a very precise tuning of the system parameters to eliminate unwanted \u Cerenkov emission processes that may mask the quantum vacuum radiation.
\begin{figure}[!htbp]
\begin{center}
\includegraphics[width=0.8\columnwidth,clip]{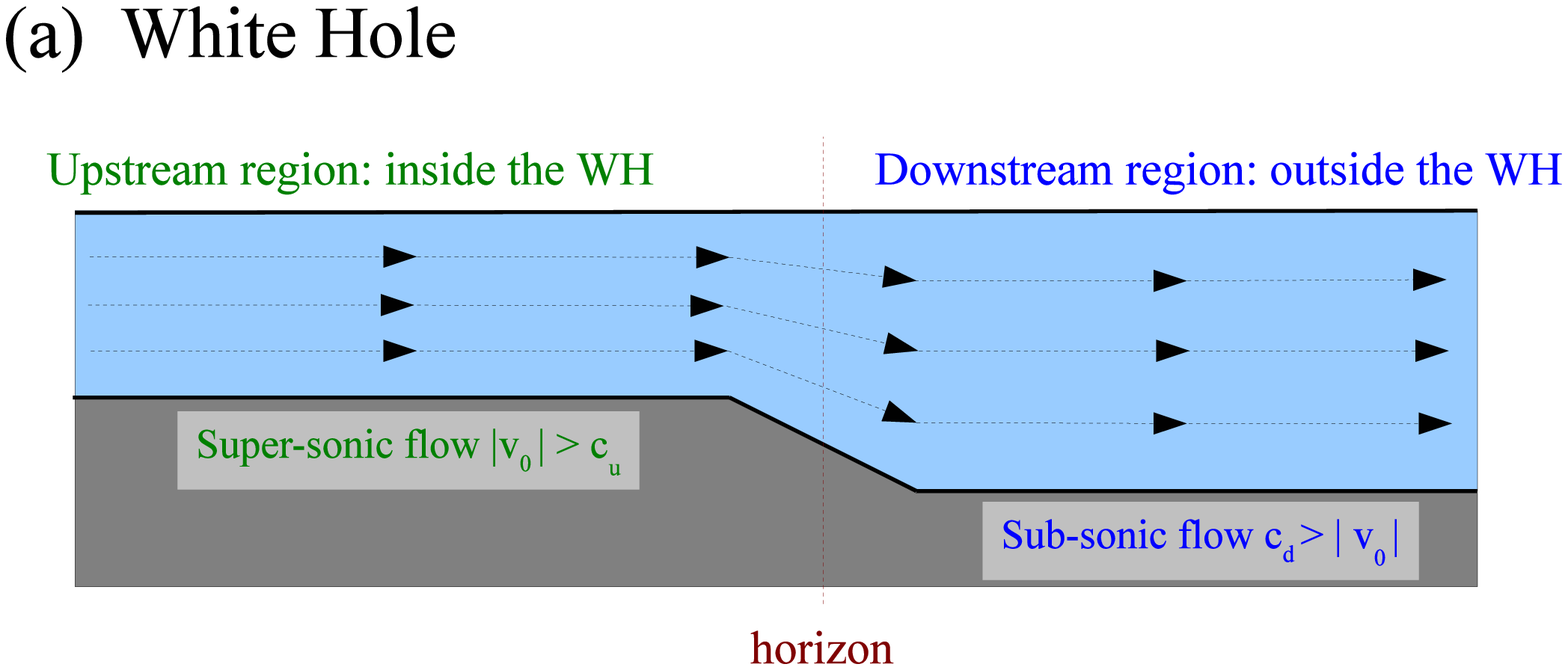}
\includegraphics[width=0.8\columnwidth,clip]{figura_WH_surface.eps}\\
\vspace*{1cm}
\includegraphics[width=0.8\columnwidth,clip]{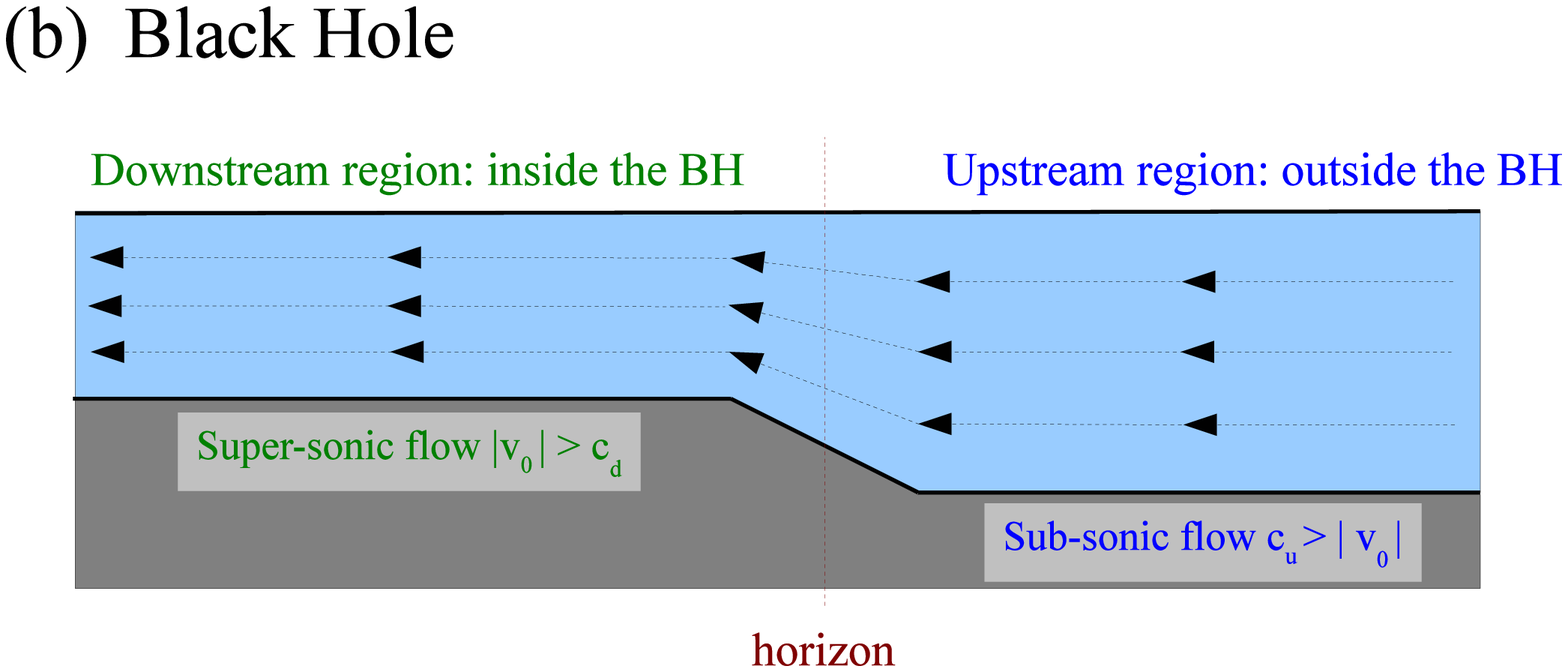}
\includegraphics[width=0.8\columnwidth,clip]{figura_BH_surface.eps}
\end{center}
\caption{
Main panels: sketch of the flow geometry for white (a) and black (b) hole configurations based on surface waves on a fluid. Smaller panels (a1,a2,b1,b2): surface wave dispersion in the asymptotic regions far from the horizon. \u Cerenkov emission is only possible for white hole configurations: the corresponding {\em zero mode} is indicated in blue in (a2). Differently from the superfluid case of Fig.\ref{fig:BHWH_superfluid}, the zero mode now has a group velocity in the downstream direction into the subsonic region.
Parameters: flow speed $v=2$~m/s, fluid depth $h=0.1$~m [white hole, upstream inner region, panel (a1)], $v=0.666$~m/s, fluid depth $h=0.3$~m [white hole, downstream outer region, panel (a2)];  flow speed $v=-2$~m/s,  fluid depth $h=0.1$~m [black hole, downstream inner region, panel (b1)], $v=-0.666$~m/s, fluid depth $h=0.3$~m [black hole, upstream outer region, panel (b2)]. For the chosen parameters, the effect of surface tension at the water/air interface is negligible.
}
\label{fig:BHWH_surface}
\end{figure}

Even if a perfect preparation of the white hole is assumed, \u Cerenkov emission processes may still be triggered by nonlinear effects in the horizon region. As it was shown in~\cite{noiWH}, an incident classical Bogoliubov wavepacket is able to induce a distortion of the horizon proportional to the square of its amplitude, which then results in the onset of a continuous wave \u Cerenkov emission from the horizon and the appearance of a spatially oscillating modulation in the density profile upstream of the horizon. Of course, a similar mechanism is expected to be initiated by quantum fluctuations when back-reaction effects are included in the model, i.e. the non-linear interaction of quantum fluctuations with the underlying flow.
A third, more subtle mechanism of instability of a white hole configuration was unveiled in~\cite{noiWH}: the $1/\sqrt{\omega}$ divergence of the matrix elements of the $\mathbf{S}$-matrix for low-frequency outgoing modes in the neighborhood of the finite wavevector zero mode is responsible for a steady growth of the density fluctuation amplitude in time since the formation of the white hole. Even if the temporal growth of fluctuations follows a slow logarithmic (linear at a finite initial temperature $T>0$) law, still it is expected to strongly affect the properties of the horizon at long times.

The situation is expected to be different if fully three dimensional systems without transverse confinement are considered. In this case, \u Cerenkov emission can take place also in a black hole configuration: a distortion of the horizon by classical or quantum fluctuations with a non-trivial transverse structure is in fact able to excite Bogoliubov modes with a finite transverse component $k_y\neq 0$. As we have seen in Fig.\ref{fig:Bogo}, there exist such modes that can propagate in the downstream direction into the supersonic region inside the black hole. The effect of these zero modes on the finite-$k_y$ $\mathbf{S}$-matrix for the black hole is presently under investigation~\cite{fabbri}.

\subsection{Analog models based on surface waves}

The different dispersion of surface waves is responsible for dramatic qualitative differences in the wave propagation from the horizon of analog black and white holes configurations. As it is sketched in Fig.\ref{fig:BHWH_surface}(a,b), the trans-sonic interface is generally created in these systems by means of a spatial variation of the fluid depth $h$~\cite{NJP,Weinfurtner}. Depending on the detailed shape of the transition region and/or on the presence of fluctuations, \u Cerenkov emission by the horizon can occur into the zero modes of zero energy in the laboratory frame, as observed e.g. in~\cite{unruh_0}: a quantitative estimate of the amplitude of the resulting stationary undulation pattern for specific configurations of actual experimental interest requires however a complete solution of the hydrodynamic equations, which goes beyond the scope of the present work.

With an eye to the experiments of~\cite{NJP,Weinfurtner}, we can restrict our attention to low-wavevector gravity waves and neglect surface tension effects in a simplest one-dimensional geometry.
In this case, \u Cerenkov emission processes only occur for flow speeds lower than the sonic speed $c_s=\sqrt{gh}$ and result in an emission in the downstream direction. From the surface wave dispersions shown in Fig.\ref{fig:BHWH_surface}(a1,a2,b1,b2), one can easily see that \u Cerenkov emission can again only occur in white hole configurations, which are then expected to be again less stable than black hole ones. As in the superfluid case, \u Cerenkov emission of surface waves with a finite $k_y\neq 0$ become possible also in the black hole case as soon as a two dimensional geometry is considered and Kelvin's diverging waves are allowed. In spite of its importance in view of on-going experiments, we are not aware of any comprehensive work having studied in full detail the dynamical stability of surface wave-based analog white/black holes as in the case of superfluid-based ones~\cite{noiWH}.

\section{Conclusions}

\label{sec:conclu}

In this chapter we have presented a review on some most significant aspects of the \u Cerenkov effect from a modern and interdisciplinary point of view. In our perspective the same basic process of  generalized \u Cerenkov emission encompasses all those emission processes that take place when a uniformly moving source is coupled to some excitation field: as soon as the source velocity exceeds the phase velocity of some mode of the field, this gets continuously excited. Simple geometrical arguments are presented that allow to extract the shape of the emission pattern in real and $\kk$ space from the dispersion law $\Omega(\kk)$ of the field. Application of the general concepts to some most illustrative cases is discussed, from the standard \u Cerenkov emission of relativistically moving charged particles in non-dispersive media, to the Mach cone behind a supersonically moving impurity in a  superfluid, to the wake of gravity and capillary waves behind a duck swimming on the surface of a quiet lake. The impact of \u Cerenkov emission processes on condensed-matter analog models of gravitational physics is finally discussed. Open questions in this direction are reviewed.


\begin{acknowledgement}
IC is grateful to G. C. La Rocca and M. Artoni for triggering his interest in the rich physics of the \u Cerenkov effect, to C. Ciuti, M. Wouters, A. Amo, A. Bramati, E. Giacobino, and A. Smerzi for a long-lasting collaboration on (among other) the Bogoliubov-\u Cerenkov emission in atomic and polaritonic superfluids, and to R. Balbinot, A. Fabbri, C. Mayoral, R. Parentani, and A. Recati for fruitful collaboration on analog models.  \\ 
IC acknowledges financial support from ERC through the QGBE grant. GR is grateful to Conseil G\'en\'eral 06 and r\'egion PACA (HYDRO Project) for financial support.
\end{acknowledgement}

\end{document}